\documentclass[letterpaper,twocolumn,10pt]{article}
\usepackage{zhanggroup}

\usepackage[absolute]{textpos}
\usepackage{tikz}
\usepackage{multirow}
\usepackage{booktabs}
\usepackage{graphicx}
\usepackage{float}
\usepackage{cite}
\usepackage{amsmath,amssymb,amsfonts}
\usepackage{textcomp}
\usepackage[ruled,linesnumbered]{algorithm2e}
\newcommand{\mypara}[1]{\medskip\noindent{\bf {#1}.}}

\newcommand{\namedref}[2]{\hyperref[#2]{#1~\ref*{#2}}\xspace}
\newcommand{\appendixref}[1]{\namedref{Appendix}{#1}}
\newcommand{\subfig}[1]{\namedref{Figure}{#1}}
\newcommand{\equationref}[1]{\namedref{Equation}{#1}}
\newcommand{\tabincell}[2]{\begin{tabular}{@{}#1@{}}#2\end{tabular}} 
\newcommand{\system}{\texttt{UnGANable}\xspace}
\newcommand{\advG}{G_{\text{t}}}
\newcommand{\advE}{E_{\text{t}}}
\newcommand{\advI}{I_{\text{o}}}
\newcommand{\userG}{G_{\text{s}}}
\newcommand{\userE}{E_{\text{s}}}
\newcommand{\targetX}{\mathbf{x}}
\newcommand{\cloakedX}{\hat{\mathbf{x}}}
\newcommand{\code}{\mathbf{z}}
\newcommand{\Lrec}{\mathcal{L}_{\text{rec}}}
\newcommand{\Lper}{\mathcal{L}_{\text{percept}}}
\newcommand{\Lcos}{\mathcal{L}_{\text{cos}}}
\newcommand{\Lmse}{\mathcal{L}_{\text{mse}}} 

\begin{document}
\begin{textblock}{15}(3.9,1)
To Appear in the 32nd USENIX Security Symposium, August 9–11, 2023.
\end{textblock}
\date{}
\title{\bf UnGANable: Defending Against GAN-based Face Manipulation}
\author{
{\rm Zheng Li\textsuperscript{1}}\ \ \
{\rm Ning Yu\textsuperscript{2}}\ \ \
{\rm Ahmed Salem\textsuperscript{3}}\ \ \
{\rm Michael Backes\textsuperscript{1}}\ \ \
{\rm Mario Fritz\textsuperscript{1}}\ \ \
{\rm Yang Zhang\textsuperscript{1}}
\\
\\
\textsuperscript{1}\textit{CISPA Helmholtz Center for Information Security} \ \ \ 
\\
\textsuperscript{2}\textit{Salesforce Research} \ \ \
\textsuperscript{3}\textit{Microsoft Research}
}

\maketitle
\begin{abstract}
Deepfakes pose severe threats of visual misinformation to our society. One representative deepfake application is face manipulation that modifies a victim's facial attributes in an image, e.g., changing her age or hair color. 
The state-of-the-art face manipulation techniques rely on Generative Adversarial Networks (GANs).
In this paper, we propose the first defense system, namely \system, against GAN-inversion-based face manipulation. 
In specific, \system focuses on defending GAN inversion, an essential step for face manipulation. Its core technique is to search for alternative images (called cloaked images) around the original images (called target images) in image space. 
When posted online, these cloaked images can jeopardize the GAN inversion process. 
We consider two state-of-the-art inversion techniques including optimization-based inversion and hybrid inversion, and design five different defenses under five scenarios depending on the defender's background knowledge. 
Extensive experiments on four popular GAN models trained on two benchmark face datasets show that \system achieves remarkable effectiveness and utility performance, and outperforms multiple baseline methods.
We further investigate four adaptive adversaries to bypass \system and show that some of them are slightly effective.\footnote{See our code at \url{https://github.com/zhenglisec/UnGANable}.}
\end{abstract}

\section{Introduction}
Nowadays, machine learning (ML) models have become a core component for many real-world applications, ranging from image classification~\cite{HZRS16,KSH12} to recommendation systems~\cite{HZKC16,WSFHWW19}.
One major advancement of ML techniques in the image domain is deep generative models.
The resolution and quality of generated images have been improved exponentially since the introduction of Generative Adversarial Networks (GANs)~\cite{GPMXWOCB14}. 
Although realistic synthetic images can be used for various applications, e.g., virtual reality, avatars, and games, and detrimental uses also emerge, such as deepfakes.

\begin{figure}[!t]
\centering
\includegraphics[width=0.8\columnwidth]{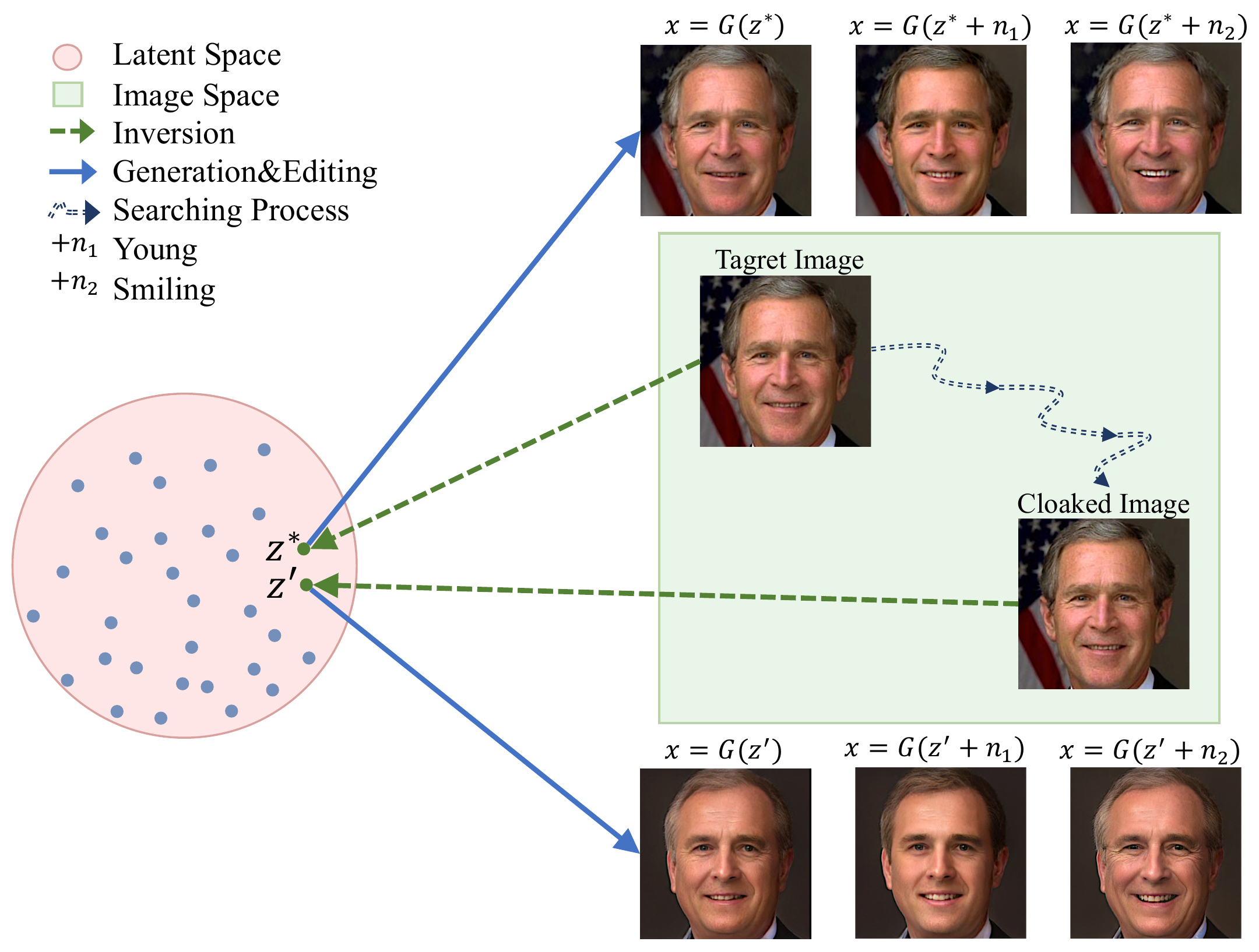}
\caption{An illustration of GAN inversion and latent code manipulation, as well as a high-level overview of \system.}
\label{fig:intro}
\end{figure}

One major example of deepfakes is face manipulation with GANs, which has been an emerging topic in very recent years~\cite{XYHZHLDB21,DHMG19,ZKS21,SGTZ20,JCI20,HHLP20,SZ21,YSEY21,PWSCL21,CVB20,GAOI19,CCKHKC18,TZSTN16,PAMSM20,WP21,YZTG17}. 
As face manipulation systems can change the target face with respect to certain attributes, such as hairstyle or facial expression, and considering that the manipulated results become increasingly more realistic, these techniques can easily be misused for malicious purposes, such as misinformation generation.
In detail, the malicious manipulator may edit the portrait image of any person without his/her permission. Moreover, the manipulator is able to forge the expression (e.g. lip shape) of political figure’s speech video, which might seriously mislead the public.
Therefore, heavy concerns on such risks are raised, and we believe that individuals need tools to protect their facial images from being misused by malicious manipulators. 

To leverage GANs to manipulate facial images, the manipulator/adversary needs to perform a two-step operation.
The first step is \emph{GAN inversion}~\cite{ZKSE16,AQW19,AQW20,ZSZZ20,BBYM20,WCZLZYHY21} which inverts a victim's facial image to a latent code.
The second step is \emph{latent code manipulation}~\cite{XYHZHLDB21,ZKS21,SGTZ20,JCI20,HHLP20,SZ21,YSEY21,PWSCL21,CVB20,GAOI19} which manipulates the latent code to get the modified image, such as adding a pair of glasses on the victim's face.
See \autoref{fig:intro} for an illustration of the two-step operation.

\subsection{Our Contributions}
In this paper, we propose the first defense system, namely \system, against GANs-inversion-based face manipulation.
In particular, \system focuses on defending against GAN inversion.
Once an image is successfully inverted to its accurate latent code, it is extremely hard (if not possible) to defend the following manipulation step as the adversary can perform any operation on the latent code. 
Therefore, we believe the most effective defense is to reduce the performance of GAN inversion - the adversary can only obtain an inaccurate latent code that is far from the accurate one, thus the following latent code manipulation step will not achieve the ideal result.
See \autoref{fig:intro} for an illustration of our defense.

\system searches for cloaked images in the image space which are indistinguishable from the target images but can cause the adversary's GAN inversion to obtain an inaccurate latent code.
In this way, any individual can use \system to protect their images by sharing only the cloaked images online.
Further, we focus on two state-of-the-art GAN inversion techniques, i.e., optimization-based inversion~\cite{AQW19,AQW20} and hybrid inversion~\cite{ZKSE16,ZSZZ20,WCZLZYHY21}, and consider five scenarios to characterize the defender's background knowledge along multiple dimensions.
By considering what knowledge the defender has, we obtain a taxonomy of five different types of methods (called ``cloaks'' throughout the paper) to disenable GAN inversion. 
More concretely, two cloaks are designed against optimization-based inversion, while the other three cloaks are designed against hybrid inversion. 

We evaluate all our five cloaks on four popular GAN models that are constructed on two benchmark face datasets of different sizes and complexity. 
Extensive experiments show that \system in general achieves remarkable performance with respect to both effectiveness and utility. 
We also conduct a comparison of our \system with thirteen baseline image distortion methods.
The results show that our defenses can outperform all these methods.
We also explore four adaptive adversaries to bypass \system and conduct sophisticated studies. 
Empirical results show that Spatial Smoothing~\cite{SpatialSmoothing} and more iterations of inversion are slightly effective.

In summary, we make the following contributions.
\begin{itemize}
    \item We take the first step towards defending against malicious face manipulation by proposing \system, a system that can jeopardize the process of GAN inversion.
    \item We consider five scenarios to comprehensively characterize a defender's background knowledge along multiple dimensions, and propose five different defenses for each scenario.
    Extensive evaluations on four popular GAN models show that \system can achieve remarkable performance with respect to both effectiveness and utility.
    \item We conduct a comparison of our defenses with thirteen baseline image distortion methods. 
    The results show that our defenses can outperform all these methods.
    \item We further explore four adaptive adversaries to bypass \system and show that some of them are slightly effective.
\end{itemize}

\section{Background and Related Work}
\label{sec2:preli}
In this section, we first introduce the two-step of GAN-based face manipulation, namely GAN inversion and latent code manipulation. Then we discuss other face manipulation techniques and existing defenses.
For presentation purposes, we summarize the notation throughout the paper in Appendix \autoref{tab:notions}. 
In particular, we emphasize that the adversary-controlled generator is marked as the target generator $\advG$ and the adversary-controlled encoder is marked as the target encoder $\advE$.

\subsection{GAN Inversion}
\label{sec:gan_inversion}
In this paper, we consider two representative and most widely-used techniques of GAN inversion, i.e., optimization and hybrid formulations, as shown in \autoref{fig:inversions}. The algorithms can be found in \autoref{appendix:inversion_algorithms}.

\mypara{Optimization-based Inversion}
Existing optimization-based inversions~\cite{AQW19,AQW20} typically reconstruct a target image by optimizing the latent vector
\begin{equation}
\label{equ:opt}
\textbf{z}^{*}=\underset{\textbf{z}}{\arg\min } \Lrec\big(\mathbf{x}, \advG(\mathbf{z})\big)
\end{equation}
where $\mathbf{x}$ is the target image and $\advG$ is the target generator. Starting from a Gaussian initialization $\mathbf{z}$, we search for an optimized vector $\mathbf{z}^{*}$ to minimize the reconstruction loss $\Lrec$ which measures the similarity between the given image $\targetX$ and the image generated from $\mathbf{z}^{*}$. $\Lrec$ is a weighted combination of the perceptual loss~\cite{JAF16} and MSE loss:
\begin{equation}
\Lrec=\Lper\big(\advG(\code), \targetX\big)+\Lmse\big(\advG(\code), \targetX\big)\nonumber
\end{equation}
where  $\Lper$ measures the similarity of features extracted from a pretrained neural network, such as VGG-16~\cite{SZ15}, and $\Lmse$ measures the pixel-wise similarity. 

\mypara{Hybrid Inversion}
An important issue for optimization-based inversion is initialization. Since \autoref{equ:opt} is highly non-convex, the reconstruction quality strongly relies on a good initialization of $\mathbf{z}$. 
Consequently, researchers~\cite{ZKSE16,ZKSE16,WCZLZYHY21,TANPC21} propose to use an encoder to provide better initialization $\mathbf{z}$ for optimization, namely hybrid inversion.

Hybrid inversion first predicts $\code$ of a given image $\targetX$ by training a separate encoder, then uses the obtained $\code$ as the initialization for optimization. The learned predictive encoder serves as a fast bottom-up initialization for the non-convex optimization problem \autoref{equ:opt}. 

\begin{figure}[!t]
\centering
\includegraphics[width=0.8\columnwidth]{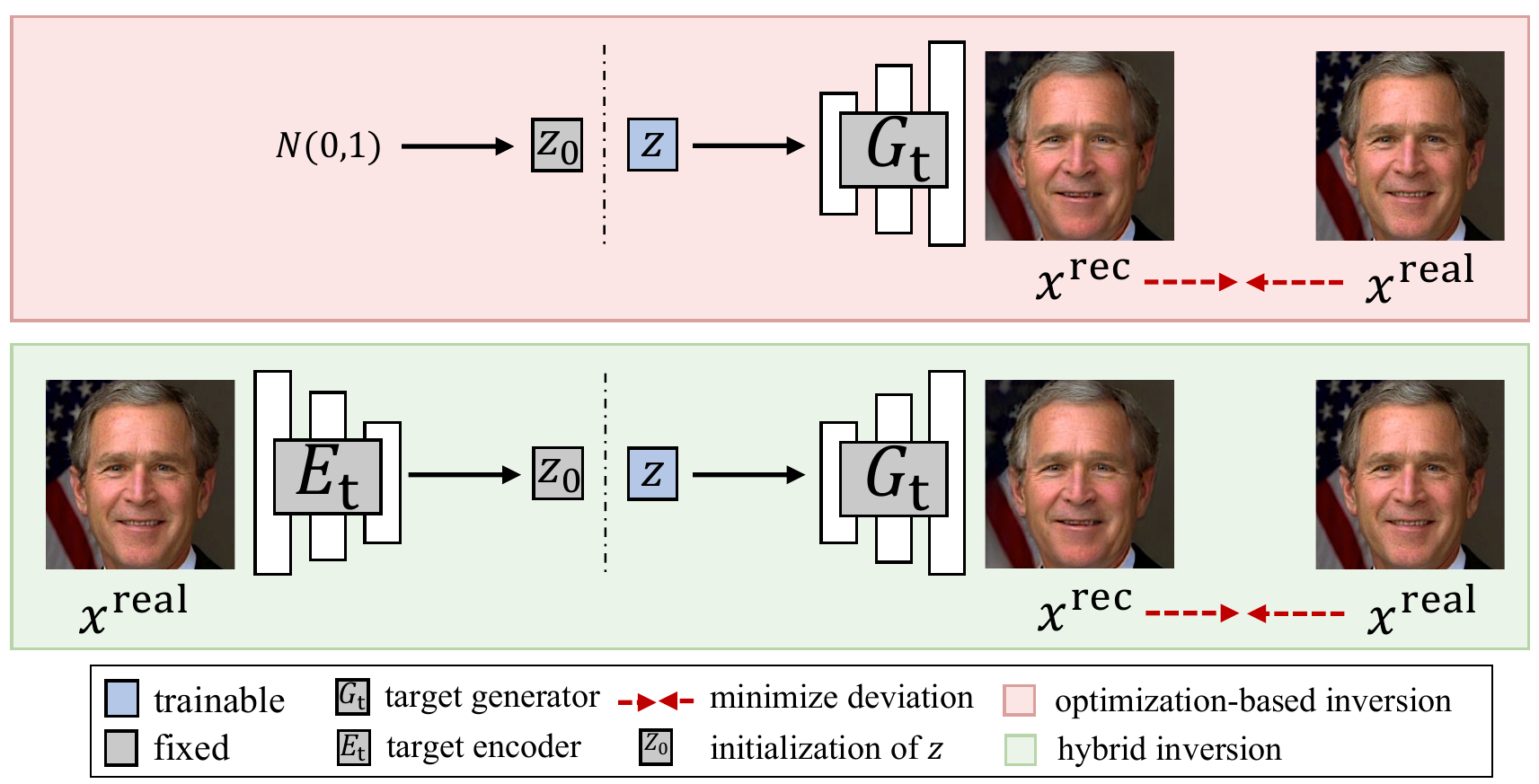}
\caption{Illustration of GAN inversion methods. The upper is the optimization-based inversion. The bottom is the hybrid inversion.}
\label{fig:inversions}
\end{figure} 

\subsection{Latent Code Manipulation}
Considering that a given image has been successfully inverted into the latent space, the editing of the image can be easily executed.
There are multiple methods~\cite{XYHZHLDB21,ZKS21,SGTZ20,JCI20,HHLP20,SZ21,YSEY21,PWSCL21,CVB20,GAOI19} to manipulate the latent code, most of them are based on algebraic operations on the latent code.
For instance, in InterFaceGAN~\cite{SGTZ20}, the authors move the latent code $\code$ along a certain semantic direction $n$ to edit the corresponding attribute of the image ($\code+n$).
See \autoref{appendx:face_edit} for how to edit faces by algebraic operations.
As the adversary has full control over the manipulation step, it is extremely difficult (if not possible) to defend this step.
Therefore, we only focus on defending against the GAN inversion step - the adversary can only obtain a misleading latent code that is already far from its exact one. 
In this way, the latent code manipulation step will not achieve its ideal result.

\subsection{Other Related Work}
\mypara{Image-Translation-Based Face Manipulation} 
This face manipulation ~\cite{CCKHKC18,TZSTN16,PAMSM20,YZTG17,IZZE17,ZPIE17,ZZCYW20}, also known as Image-to-Image Translations (I2I), represented by StarGANv2~\cite{CUYH20} and AttGAN~\cite{HZKSC19}, has received increasing attention in recent years.
More concretely, I2I builds an end-to-end neural network as the backbone, to translate source images into the target domain with many aligned image pairs for training. 
When editing images, I2I uses the backbone network to accept the target image and output a new style of it without GAN-inversion process. 
Considering that the defense against I2I has been well studied~\cite{RBS20,YCTW20,HZZZY21,LYWL19}, the defense against GAN-inversion-based is still an open research problem. 
Our work is therefore well-motivated to complete this puzzle map.
We also provide a more in-depth discussion of I2I in \autoref{sec:discussion}.

\mypara{Existing Defenses Against Face Manipulation}
As face manipulation causes a great threaten to individual privacy even political security, it is of paramount importance to develop countermeasures against it. 
To mitigate this risk, many defenses have been proposed, and these defenses can be broadly divided into two categories: detection~\cite{LBZYCWG20, RCVRTN19,ANYE18,ZHMD17,MRS19,NYE19} and disrupting I2I~\cite{RBS20,YCTW20,HZZZY21,LYWL19}.
However, the former defense is designed in a passive manner to detect whether face images have been tampered with after wide propagation.
The latter defense can only mitigate image-translation-based face manipulation by spoofing the backbone network. 
However, there is still no approach to defend against GAN-inversion-based face manipulation in a proactive manner.
In this paper, we propose \system of initiative defense to degrade the performance of GAN inversion, which is an essential step for subsequent face manipulation.
See more discussion about limitations of existing defenses in \autoref{limitation_exiting_defense}.

\begin{table*}[!t]
\centering
\caption{
An overview of assumptions.
``\checkmark'' means the defender needs the knowledge and ``-'' indicates the knowledge is not necessary.
``Target'' means the adversary-controlled entities, and ``Shadow'' means the defender-built entities locally.
}
\scalebox{0.75}
{
\begin{tabular}{cccccccc} 
\toprule
\multirow{2}{*}{Inversion Category} &\multirow{2}{*}{Cloaks} & Target & Shadow &Target& Shadow& Feature& Inversion \\
& & Generator &  Generator&Encoder &Encoder & Extractor & Technique \\
\midrule
\multirow{2}{*}{Optimization-based}& White-box
& \checkmark & - & - &  \checkmark & \checkmark  & \checkmark \\
\cmidrule(lr){2-8}
& Black-box & - & -& - & - &  \checkmark  & - \\
\cmidrule(lr){1-8}
\multirow{3}{*}{Hybrid}& White-box & - &  - & \checkmark &- &\checkmark & - \\
\cmidrule(lr){2-8}
& Gray-box & - &  \checkmark & - & \checkmark &\checkmark & - \\
\cmidrule(lr){2-8}
& Black-box & - & - & - & - &  \checkmark & -\\
\bottomrule
\end{tabular}
}
\label{tab:defenseoverview}
\end{table*}

\section{Overview of \system}
\label{sec3:overview}
In this section, we provide an overview of \system.

\subsection{Intuition}
\label{sec3:intuition}
We derive the intuition behind our \system from the basic pipeline of how inversion works. 
Since the optimization-based inversion is part of the hybrid inversion, here we focus only on the former. 
As described in \autoref{sec:gan_inversion}, the inversion employs a loss function that is a weighted combination of the perceptual loss~\cite{JAF16} and the pixel-wise MSE loss, to guide the optimization into the correct region of the latent space.
This methodology leads to the following observations.
\begin{itemize}
    \item The pixel-wise MSE loss works in the pixel space, i.e., the image space.
    \item The perceptual loss measures the similarity of features extracted from different images using a pretrained model, which works in the feature space.
    \item The optimization aims to search for the optimal latent code, which works in the latent space.
\end{itemize}
Thus, GAN inversion actually works in at least three spaces, i.e., the image space, the feature space, and the latent space. 
These observations motivate our \system, which aims to maximize deviations in both latent and feature spaces with the cloaked images, meanwhile maintain the image indistinguishable in the image space.

\subsection{Threat Model}
\label{sec3:threatmodels}
The goal of the face manipulator (i.e., adversary) is to manipulate the face without any authorization from the owner of the face image to serve its own purposes, such as violating individual privacy or even misleading political opinions,  
The face manipulator could be a commercial company or even an individual. 
We assume the face manipulator has access to advanced GANs (e.g., via GitHub), and can apply two advanced GAN inversion techniques, namely optimization-based inversion and hybrid inversion, to invert the images into the latent space. 
These two inversion methods are shown in \autoref{fig:inversions}.

\subsection{System Model}
Any user (also called defender) can use \system to search for cloaked images, which are around the target images in the image space.
The design goals for these cloaks are:
\begin{itemize}
    \item cloaked images should be indistinguishable from the target images;
    \item when inverting the cloaked image, the adversary can only get a misleading latent code, which is far from its accurate one in the latent space (see \autoref{equ:Lrec}).
\end{itemize}
Generally, \system aims to maximize the deviations in the latent space and feature space, while keeping the images indistinguishable in image space.
Therefore, the challenge for \system is to obtain the representation in each space.
To this end, we make different assumptions for \system in different scenarios where \system can use different methods to search for invisible images.
The overview of background knowledge is introduced in \autoref{tab:defenseoverview}

\section{\system Against Optimization-based Inversion}
\label{sec4:cloak-v01}
In this section, we present \system against the first type of GAN inversion, i.e., optimization-based inversion. 

\subsection{Defender’s Knowledge}
For optimization-based inversion, we consider two different scenarios to characterize a defender's background knowledge. 
See more detailed explanation about background knowledge in \autoref{assumption_defender}.

\mypara{White-Box (Cloak v0)} 
To maximize the deviation in the latent space, a defender has white-box access to the target generator $\advG$, and knows the adversary's inversion techniques $\advI$, thus he/she can obtain the accurate latent code of the original image.
Besides, the defender trains a shadow encoder $\userE$ to embed interim cloaked images to obtain the cloaked latent code.
Then, the adversary can maximize the deviation between them.
To maximize deviation in the feature space, we further assume that the defender has access to a feature extractor $F$, which can map both original image and cloaked image to feature space. 
Here, the feature extractor can be different from feature extractor used in perceptual loss.

\mypara{Black-Box (Cloak v1)}
In this scenario, we assume the defender has no knowledge of the target generator or inversion techniques. 
Here, the defender only has access to a feature extractor $F$.
\begin{figure}[!t]
\centering
\includegraphics[width=0.8\columnwidth]{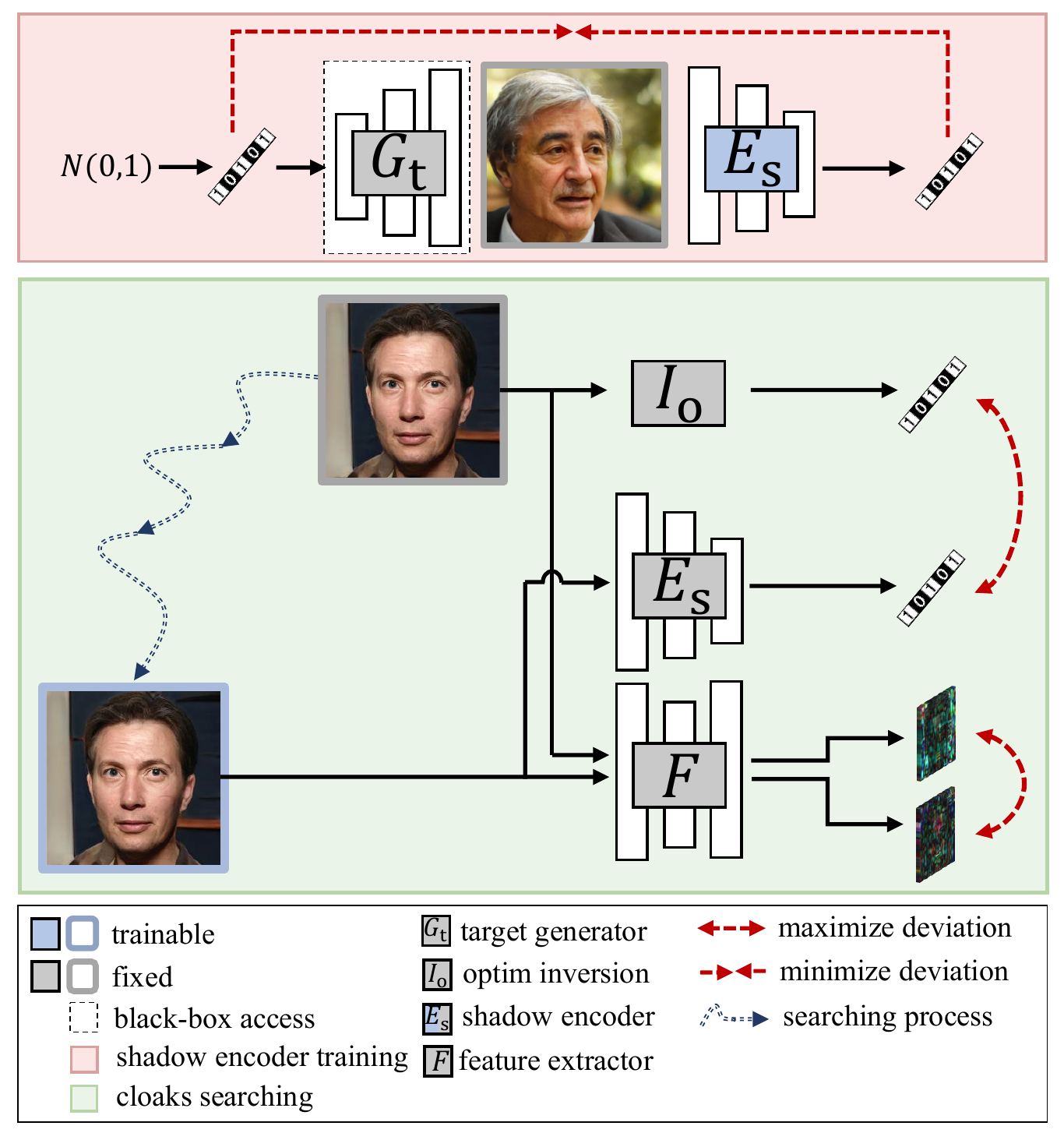}
\caption{An illustration of white-box (Cloak v0) and black-box (Cloak v1) defenses against optimization-based inversion.}
\label{fig:cloak-0}
\end{figure}

\subsection{Methodologies}\label{optim_method}
From a high-level overview, the defense can be divided into three simultaneous components, namely maximizing latent deviation, maximizing feature deviation, and searching for cloaked images in the image space.
The algorithms can be found in \autoref{appendix:unganable_algorithms}.

\mypara{White-Box (Cloak v0)} 
The defender first leverages optimization-based inversion $\advI$ to invert a target image $\targetX$ to obtain its exact latent code $\advI(\targetX)$.\footnote{This process requires white-box access to the target generator $\advG$, as shown in \autoref{fig:inversions}.}
For maximizing latent deviation, the defender needs to build an end-to-end model, namely shadow encoder $\userE$, to invert the cloaked image $\cloakedX$ of each step to obtain its latent code.\footnote{The reason is that when iteratively searching in the image space, the defender needs to compute the cloaked image's gradient of each step with respect to the latent deviation by backpropagation, which is intractable through optimization-based inversion. The optimization-based inversion is just an inverted process, not an end-to-end model.}
To train $\userE$, as shown in the pink part of \autoref{fig:cloak-0}, the defender leverages the target generator $\advG$ to create a dataset of generated images $\advG(\code)$ and their latent codes $\code$, then minimize a similarity reconstruction loss $\Lrec$ between these latent codes $\userE\big(\advG(\code)\big)$ and $\code$.
\begin{equation}
\begin{array}{c}
\Lrec=-\Lcos\Big(\userE\big(\advG(\code)\big), \code\Big)+\Lmse\Big(\userE\big(\advG(\code)\big), \code\Big)
\end{array}
\label{equ:Lrec}
\end{equation}
where both $\Lcos$ and $\Lmse$ measure the element-wise similarity of latent codes. Here, $\Lcos$ is cosine similarity loss, and $\Lmse$ is MSE similarity loss.

For maximizing feature deviation, the defender uses a third-party pre-trained model (e.g., via GitHub) as the feature extractor $F$ to obtain the features $F(\targetX)$ and $F(\cloakedX)$.
Once the defender obtains $\advI(\targetX)$, $\userE$ and $F$, the defender iteratively searches for $\cloakedX$ in the image space by modifying $\targetX$, to maximize the latent and feature deviations between $\targetX$ and $\cloakedX$.
\begin{equation}
\begin{array}{c}
\max _{\cloakedX} \kappa \Big( \Lrec\big(E_{\text{s}}(\cloakedX), \advI(\targetX)\big) \Big)  +(1-\kappa )\Big( \Lrec\big(F(\cloakedX), F(\targetX)\big) \Big) \\
\text { s.t. } |\cloakedX - \targetX|_\infty < \epsilon \nonumber\\
\kappa \in [0,1] 
\end{array}
\end{equation}
where $\mathcal{L}_{\text{rec}}(.)$ introduced in \equationref{equ:Lrec} measures the element-wise similarity of two feature vectors or latent vectors, $|\cloakedX - \targetX|_\infty$ measures the distance between $\cloakedX$ and $\targetX$, $\varepsilon$ is the distance budget in image space, and $\kappa$ is a trade-off hyper-parameter between latent and feature spaces. 

\mypara{Black-Box (Cloak v1)} 
The defender can only produce significant alterations to images’ feature space, i.e., searching for $\cloakedX$ in the image space by modifying $\targetX$, to maximize the feature deviation between $\cloakedX$ and $\targetX$. 
\begin{equation}
\begin{array}{c}
\max _{\cloakedX} \mathcal{L}_{\text{rec}}(F(\cloakedX), F(\targetX)) \nonumber \\
\text { s.t. } |\cloakedX - \targetX|_\infty < \epsilon 
\end{array}
\end{equation}

\begin{table}[!t]
    \centering
    \caption{Target GANs, datasets and resolutions used to evaluate defense performance.}
    \scalebox{0.75}
    {
    \begin{tabular}{c|c|c|c}
    \toprule
        Model Zoo & $Z$ dims & Dataset & Resolution\\
         \midrule
         DCGAN (2016)\cite{RMC16}& 100 & CelebA~\cite{LLWT15} & 64$\times$64\\
         \midrule
         WGAN (2017)\cite{GAADC17}& 128 & CelebA~\cite{LLWT15}  & 128$\times$128\\
         \midrule
         StyleGANv1 (2019)\cite{KLA19}& 512 & FFHQ~\cite{KLA19,KLAHLA20}  & 256$\times$256\\
         \midrule
         StyleGANv2 (2020)\cite{KLAHLA20}& 512& FFHQ~\cite{KLA19,KLAHLA20}  & 256$\times$256\\
         \bottomrule
    \end{tabular}
    }
    \label{tab:GAN_dataset}
\end{table}

\subsection{Experimental Setup}
\label{setup_optim}
\mypara{GAN Models and Datasets} 
Without losing representativeness, we focus on four generative applications in recent years - DCGAN~\cite{RMC16}, WGAN~\cite{GAADC17}, StyleGANv1~\cite{KLA19}, and StyleGANv2~\cite{KLAHLA20}.
These GAN models are built with different architectures, losses and training schemes.
Each generation application benchmarks its own dataset. As summarized in \autoref{tab:GAN_dataset}, we considered two benchmark datasets of different sizes and complexities, including CelebA~\cite{LLWT15} and FFHQ~\cite{KLA19,KLAHLA20}, to construct different GAN models.  
Details of GAN models and datasets can be found in \appendixref{appendix:GAN_dataset}.

\mypara{Manipulator/Adversary}
For face manipulator/adversary, we follow the original configurations of optimization-based inversion (Image2StyleGAN~\cite{AQW19}). 
More specifically, we set up 500 iterations for the optimization step of inversion. 
In addition, we use perceptual loss and pixel-level MSE loss to reconstruct the target image in the optimization step.
Though StyleGANv1~\cite{KLA19} and StyleGANv2~\cite{KLAHLA20} also work on $\mathbf{w}$ space that is converted from $\mathbf{z}$ space, $\mathbf{z}$ space is applicable to all GAN models, thus we only consider $\mathbf{z}$ space in this work.

\mypara{Defender} 
For the defender, we use a random initialized ResNet-18~\cite{HZRS16} as the shadow encoder $\userE$ in the white-box scenario (Cloak v0). 
Besides, for both white- and black-box scenarios (Cloak v0/v1), we adopt the easy-to-download, widely-used, and pre-trained ResNet-18 as the feature extractor. 
Further, we set up 500 iterations to iteratively search for the cloaked image in the image space by modifying the target image. 

\mypara{Target Samples} 
We first evaluate \system on generated images from each GAN model.
The reason is that, as stated in previous works~\cite{AQW19,AQW20,ZSZZ20}, and also shown in our experimental results, the generated images are more easily inverted into accurate latent codes.
In other words, in the competition between attackers and defenders, we actually make a very strong advantageous assumption for the former. 
We investigate whether \system can achieve acceptable or even superior performance in such a worst-case scenario.
Thus, for each GAN model, we evaluate the performance of \system on 500 randomly selected generated images that can be successfully reconstructed. 

\begin{figure}[t]
    \centering
    \subfloat[StyleGANv1\label{fig:optim-styleganv1}]{\includegraphics[width=0.5\linewidth]{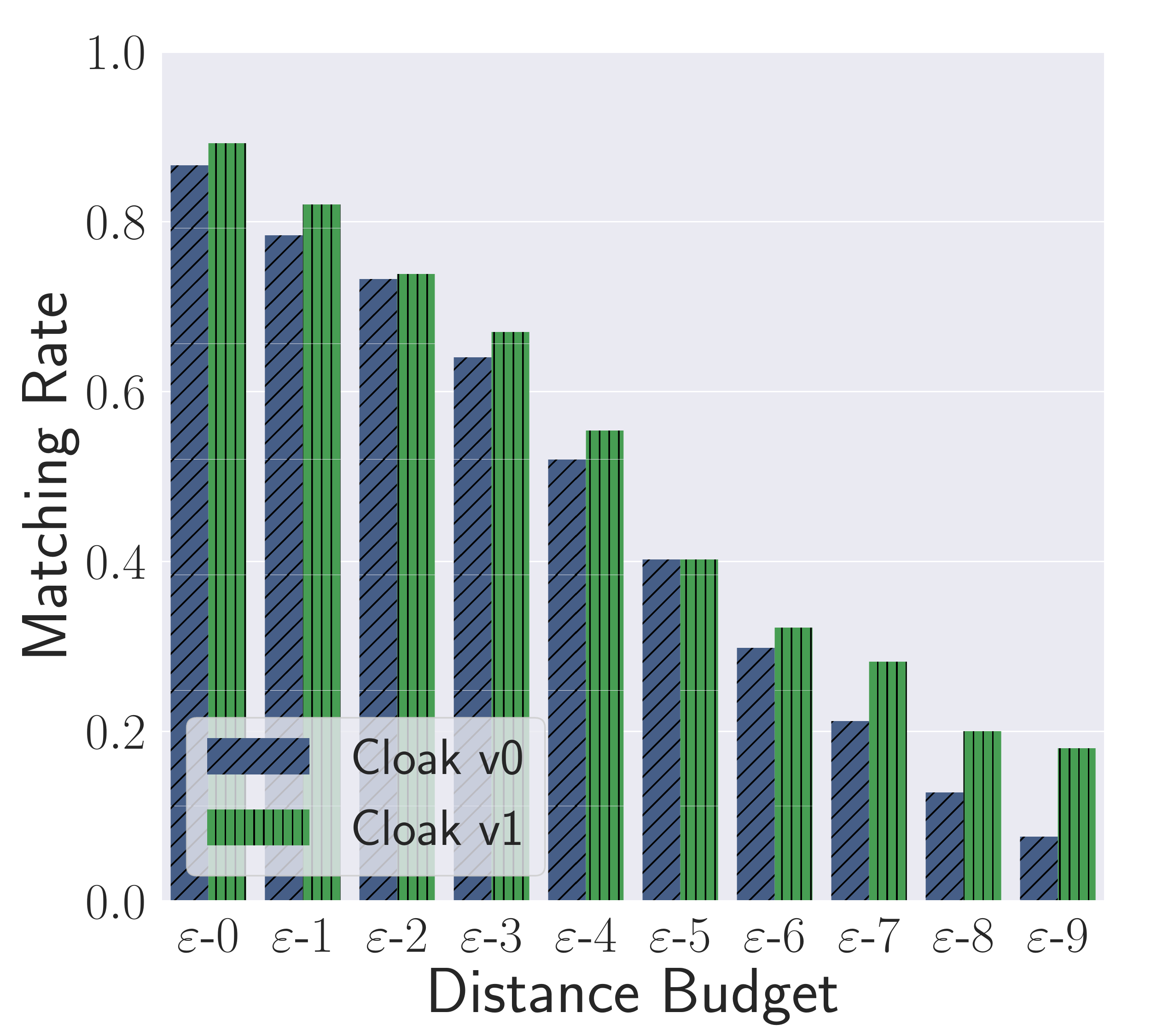}}
    \subfloat[StyleGANv2\label{fig:optim-styleganv2}]{\includegraphics[width=0.5\linewidth]{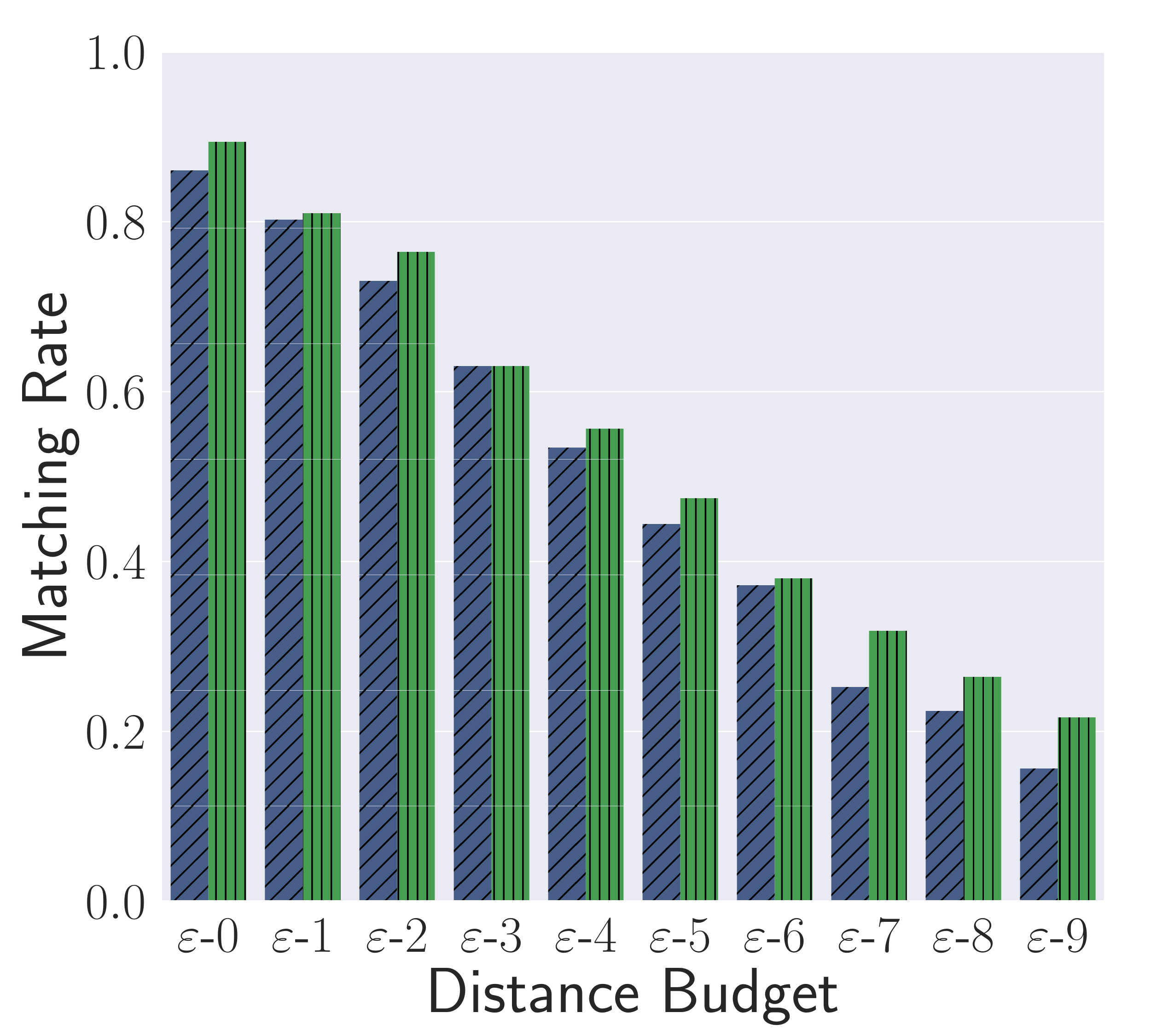}}
    \caption{The effectiveness performance of Cloak v0 and Cloak v1.}
    \label{fig:effectiveness-0}
\end{figure}

\begin{table*}[!t]
    \centering
    \caption{Some visual examples of reconstructed images based on StyleGANv2. The defense method is Cloak v1.}
    \scalebox{0.75}
    {
    \begin{tabular}{c|c|c|c|c|c|c}
    \toprule
         Target Image& No cloak  & $\varepsilon$-$1$ & $\varepsilon$-$3$ & $\varepsilon$-$5$& $\varepsilon$-$7$ & $\varepsilon$-$9$\\
         \midrule
         \raisebox{-.5\height}{\includegraphics[width=0.15\linewidth]{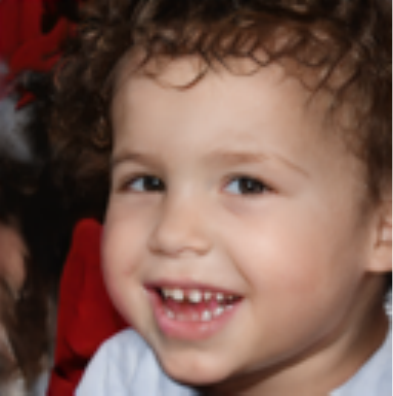}} & 
         \raisebox{-.5\height}{\includegraphics[width=0.15\linewidth]{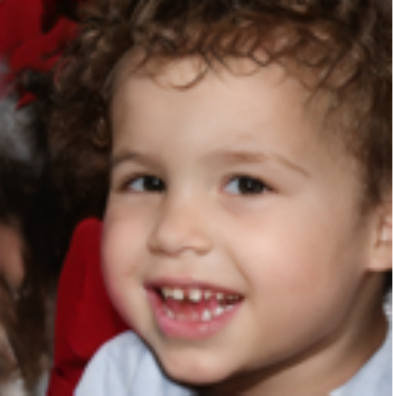}} & 
        \raisebox{-.5\height}{\includegraphics[width=0.15\linewidth]{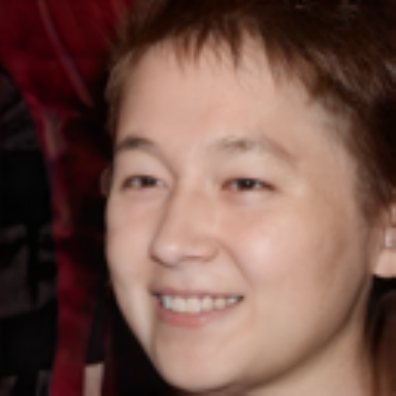}} & 
        \raisebox{-.5\height}{\includegraphics[width=0.15\linewidth]{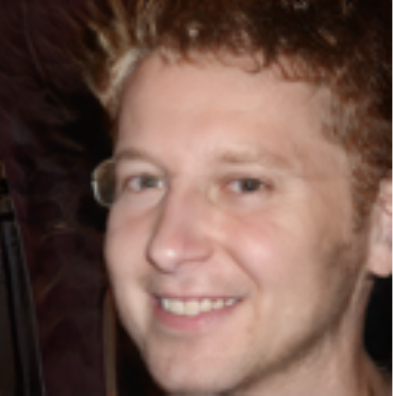}} & \raisebox{-.5\height}{\includegraphics[width=0.15\linewidth]{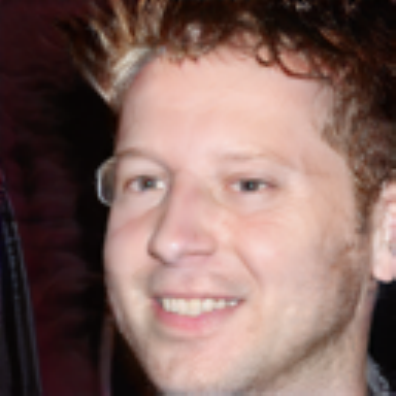}} &
        \raisebox{-.5\height}{\includegraphics[width=0.15\linewidth]{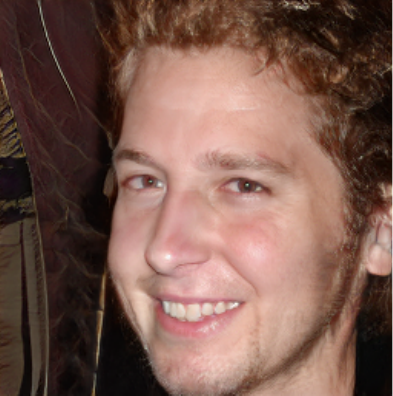}} &
        \raisebox{-.5\height}{\includegraphics[width=0.15\linewidth]{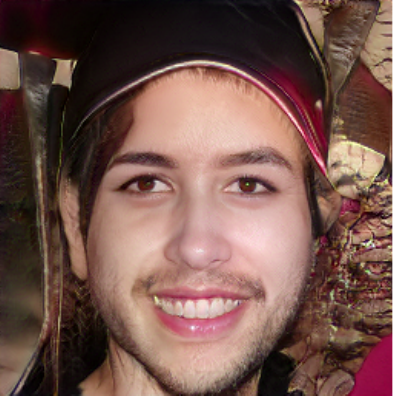}}
        \\
         \midrule
         \raisebox{-.5\height}{\includegraphics[width=0.15\linewidth]{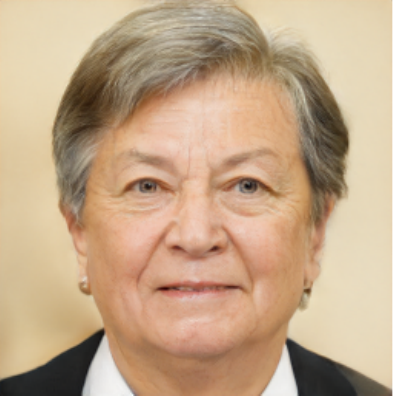}} & 
          \raisebox{-.5\height}{\includegraphics[width=0.15\linewidth]{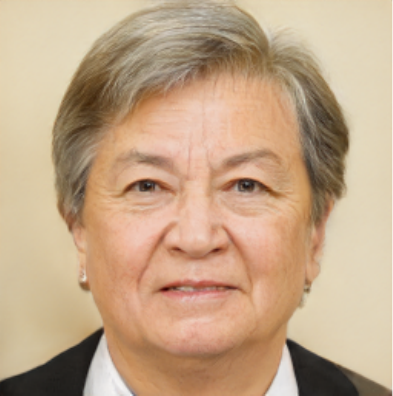}} & 
        \raisebox{-.5\height}{\includegraphics[width=0.15\linewidth]{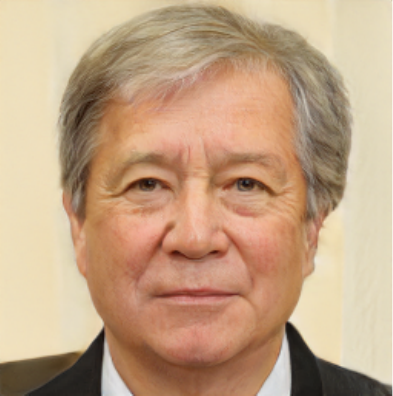}} & 
        \raisebox{-.5\height}{\includegraphics[width=0.15\linewidth]{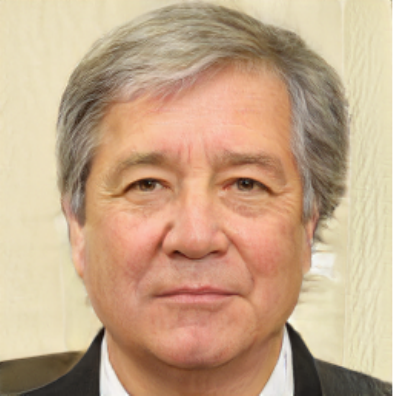}} & \raisebox{-.5\height}{\includegraphics[width=0.15\linewidth]{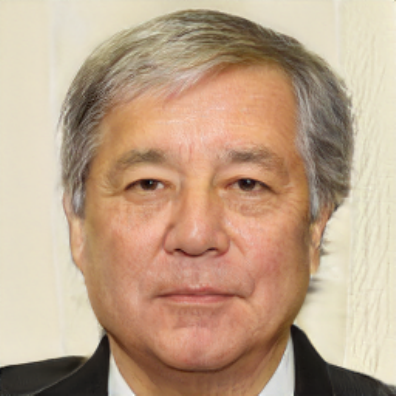}} &
       \raisebox{-.5\height}{ \includegraphics[width=0.15\linewidth]{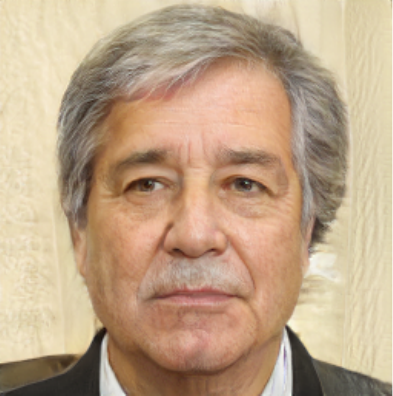}} &
        \raisebox{-.5\height}{\includegraphics[width=0.15\linewidth]{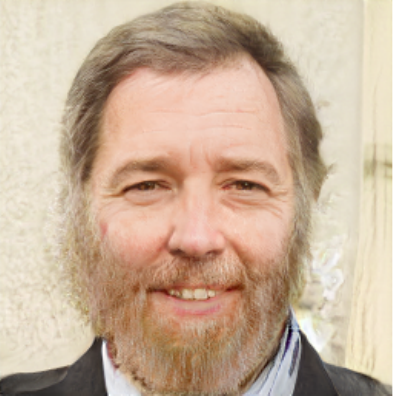}}
        \\
         \bottomrule
    \end{tabular}
    }
    \label{fig:quali-0}
\end{table*}

\mypara{Evaluation Metrics}
For evaluation metrics, we consider two perspectives: effectiveness and utility.
Effectiveness measures the extent to which \system jeopardizes the GAN inversion process. 
Given a target image, the sign of successful defense is a change in the identity of the reconstructed image, as shown in \autoref{fig:intro}. 
The reason is that once the identity of the reconstructed image changes, the defender no longer cares about the manipulation of the reconstructed image because the reconstructed image does not belong to the defender.
To this end, we use \textit{Matching Rate} to evaluate effectiveness:

\begin{equation}
    \textit{Matching Rate} = \frac{\# \text{successful reconstructed images}}{\# \text{total images}} \nonumber
\end{equation}
Therefore, the lower the matching rate is, which means the more reconstructed images with changed identity, the better effectiveness \system achieves. 
In our implementation, we utilize a popular open-source face verification/comparison tool FaceNet~\cite{SKP15} to compute the defense success rate. 
Given the embedding distance of a pair of two face images, a pre-calibrated threshold is used to determine the classification of \emph{same} and \emph{different}, i.e., the two face images belong to the same person if the embedding distance is less than the threshold, otherwise different person.
See more details on threshold selection in \appendixref{appendix:threshold}.

Utility measures whether the cloaked images searched by \system is indistinguishable from the target images.
To measure the utility, we use a variety of most widely-used similarity metrics, including mean squared error (MSE), structural similarity (SSIM)~\cite{WBSS04}, and peak signal-to-noise ratio (PSNR).
Here, the lower the MSE is, the higher the SSIM and PSNR are, then the better utility \system achieves. More details about these metrics are presented in \appendixref{appendix:metrics}. 

\begin{table}[!t]
    \centering
    \caption{The utility performance of \system against optimization-based inversion.}
    \tabcolsep 3pt
    \scalebox{0.75}
    {
    \begin{tabular}{c|c|c|c|c|c|c|c}
    \toprule
          Budget  &Metric&Cloak v0& Cloak v1&Budget  &Metric&Cloak v0& Cloak v1 \\
         \midrule
         \multirow{3}{*}{$\varepsilon$-1} & MSE & 7.3e-05 & 7.2e-05 &\multirow{3}{*}{$\varepsilon$-7} & MSE & 0.0010 & 0.0014\\
                                    & SSIM & 0.9889 & 0.9891&& SSIM & 0.8802 & 0.8431\\
                                    & PSNR & 41.376 & 41.408&& PSNR & 30.118 & 28.532\\
         \midrule
        \multirow{3}{*}{$\varepsilon$-3} & MSE & 0.0003 & 0.0003&\multirow{3}{*}{$\varepsilon$-9} & MSE & 0.0014 & 0.0022\\
                                    & SSIM & 0.9612 & 0.962&& SSIM & 0.8347 & 0.7820\\
                                    & PSNR & 35.684 & 35.716&& PSNR & 28.423 & 26.637\\
        \midrule
        \multirow{3}{*}{$\varepsilon$-5} & MSE & 0.0006 & 0.0006\\
                                    & SSIM & 0.9228 & 0.9245\\
                                    & PSNR & 32.419 & 32.455\\
         \bottomrule
    \end{tabular}
    }
    \label{tab:utility_v0/1}
\end{table}

\subsection{Results}
\label{sec:optim_eval}
\mypara{Effectiveness Performance}
In our \system, we adopt a budget $\varepsilon$ to limit distance between the cloaked and target image, aiming to ensure that the cloaked image is indistinguishable from the target image. 
Here, we first investigate the effectiveness of \system by reporting matching rate under the effects of the distance budget $\varepsilon$.
More concretely, we set 10 different distance budgets $\varepsilon$-0, $\varepsilon$-1, ... , $\varepsilon$-9 (uniformly ranging from 0.01 to 0.07 for DCGAN and WGAN, and from 0.01 to 0.1 for StyleGANv1 and StyleGANv2.\footnote{We conducted a pre-experiment and showed that only a small distance can jeopardize DCGAN and WGAN inversions, so we set the maximum magnitude of the distance budget to 0.07 for DCGAN and WGAN, and 0.1 for StyleGANv1/v2.}).
Under each distance budgets, we perform grid search to find the optimum trade-off hyper-parameter $\kappa$. The exact settings for $\varepsilon$ and $\kappa$ can be found in Appendix \autoref{tab:eps_kappa}.

\autoref{fig:effectiveness-0} depicts the effectiveness performance of Cloak v0 and Cloak v1 (see more results on DCGAN and WGAN in Appendix \autoref{fig:effectiveness-appendix}). 
As we can see, with the increase of the budget $\varepsilon$, both Cloak v0 and Cloak v1 can significantly reduce matching rate.
For example, in \autoref{fig:effectiveness-0} (Cloak v0, StyleGANv2), the matching rate of $\varepsilon$-0 is 0.86, and that of $\varepsilon$-9 is 0.156, which drops sharply.
These results imply that if we set a relatively high distance budget, \system can achieve significant effectiveness against optimization-based inversion.

Besides above quantitative results, we further provide random qualitative examples to demonstrate the effectiveness of \system performed on StyleGANv2.
As shown in \autoref{fig:quali-0}, we can observe that as $\varepsilon$ increases, more and more facial attributes cannot be successfully reconstructed. The difference between the reconstructed image and the target image becomes more extensive, which implies the effectiveness is getting better.

\begin{table}[!t]
    \centering
    \caption{Some visual examples of cloaked images searched by Cloak v1 performed on StyleGANv2 under different perturbation budgets.}
    \scalebox{0.75}
    {
    \begin{tabular}{c|c|c|c}
    \toprule
       Target Image & $\varepsilon$-$1$ &  $\varepsilon$-$5$& $\varepsilon$-$9$\\
         \midrule
        \raisebox{-.5\height}{\includegraphics[width=0.25\linewidth]{plots/qualitative/0_0.pdf}} & 
        \raisebox{-.5\height}{\includegraphics[width=0.25\linewidth]{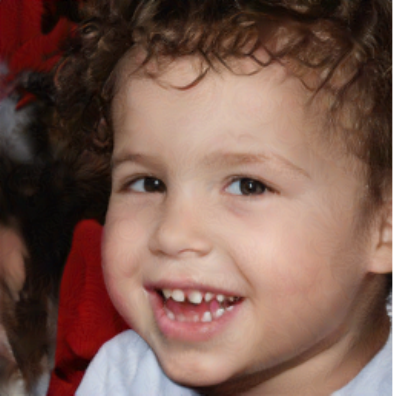}} & 
        \raisebox{-.5\height}{\includegraphics[width=0.25\linewidth]{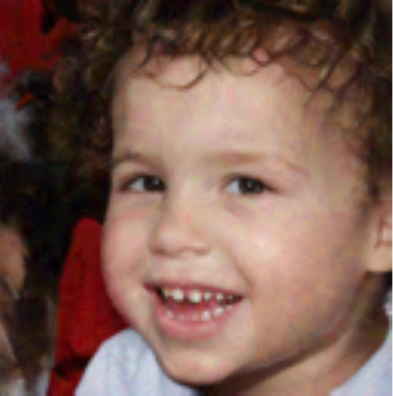}} &
        \raisebox{-.5\height}{\includegraphics[width=0.25\linewidth]{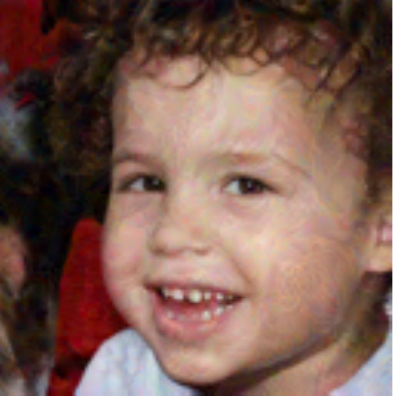}}
        \\
         \midrule
          \raisebox{-.5\height}{\includegraphics[width=0.25\linewidth]{plots/qualitative/2_0.pdf}} &
            \raisebox{-.5\height}{\includegraphics[width=0.25\linewidth]{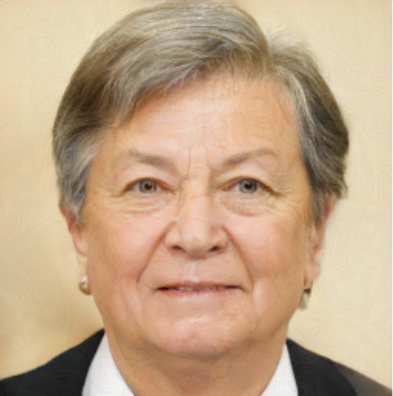}} & 
            \raisebox{-.5\height}{\includegraphics[width=0.25\linewidth]{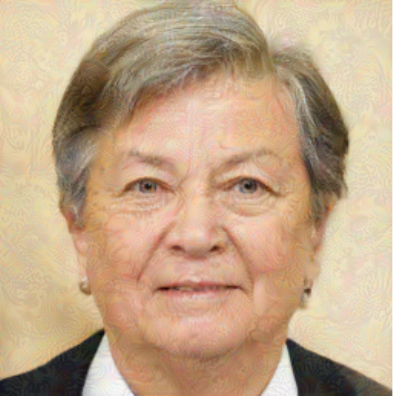}} &
            \raisebox{-.5\height}{\includegraphics[width=0.25\linewidth]{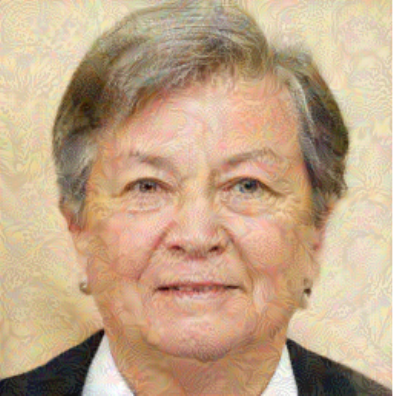}}
        \\
         \bottomrule
    \end{tabular}
    }
    \label{fig:quali-utility}
\end{table}

\mypara{Utility Performance}
To evaluate the utility performance, we first quantitatively report a variety of similarity metrics (MSE/SSIM/PSNR) in \autoref{tab:utility_v0/1}.
Typically, a SSIM value greater than 0.9 or a PSNR greater than 35 means a good quality of cloaked images.
To elaborate more on utility performance, we show in \autoref{fig:quali-utility} some qualitative samples of cloaked images searched by \system performed on StyleGANv2. 
We can observe that when distance budget is set as $\varepsilon$-1 (0.02) and $\varepsilon$-3 (0.04), which represents a completely imperceptible perturbation, \system can achieve acceptable effectiveness performance (see qualitative reconstructed examples in \autoref{fig:quali-0}).
In addition, we acknowledge that some perturbations are perceptible to our naked eye when the distance budget is set to $\varepsilon$-7 (0.08) or $\varepsilon$-9 (0.1).
But note that these visual results are performed on the images generated by their corresponding GAN models.
In the following \autoref{eval_real}, we further conduct experiments on real images.
It is encouraging that \system can apply a much lower distance budget to obtain excellent effectiveness performance while guaranteeing the visual quality of the cloaked image.

\begin{table*}[!t]
    \centering
    \caption{Visual examples of different baseline distortion methods.}
    \scalebox{0.75}
    {
    \begin{tabular}{c|c|c|c|c|c|c}
    \toprule
       Target Image & ShearX & ShearY & TranslateX & TranslateY & Rotate & Brightness\\
       \midrule
        \raisebox{-.5\height}{\includegraphics[width=0.15\linewidth]{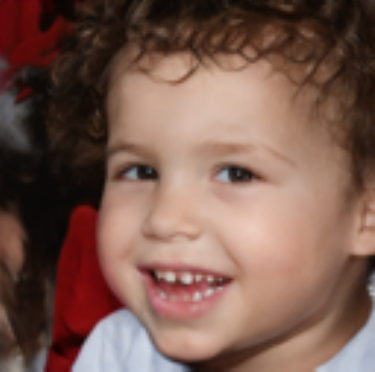}} & 
        \raisebox{-.5\height}{\includegraphics[width=0.15\linewidth]{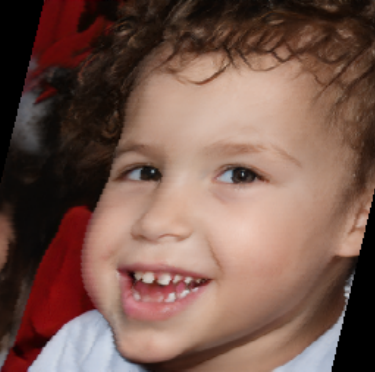}} & 
        \raisebox{-.5\height}{\includegraphics[width=0.15\linewidth]{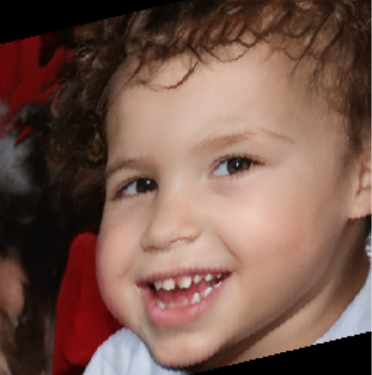}} & \raisebox{-.5\height}{\includegraphics[width=0.15\linewidth]{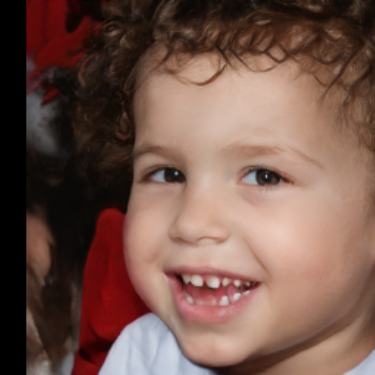}} &
        \raisebox{-.5\height}{\includegraphics[width=0.15\linewidth]{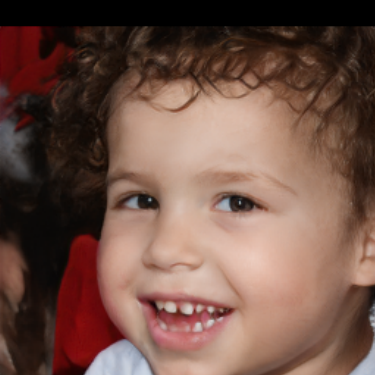}} &
        \raisebox{-.5\height}{\includegraphics[width=0.15\linewidth]{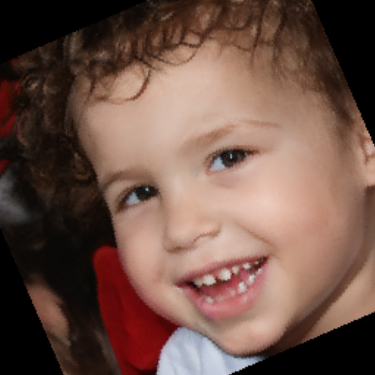}} &
        \raisebox{-.5\height}{\includegraphics[width=0.15\linewidth]{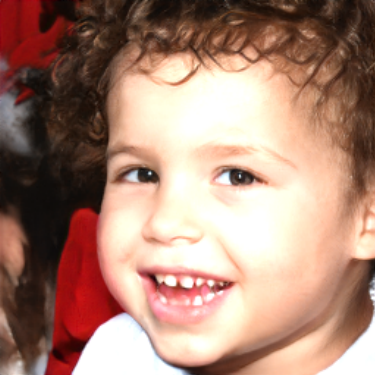}} 
        \\
        \midrule
        Color & Contrast & Solarize & CenterCrop & GaussianBlur & GaussianNoise & JPEGCompression\\
        \midrule
        \raisebox{-.5\height}{\includegraphics[width=0.15\linewidth]{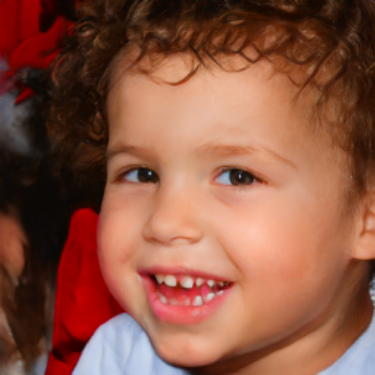}} & 
        \raisebox{-.5\height}{\includegraphics[width=0.15\linewidth]{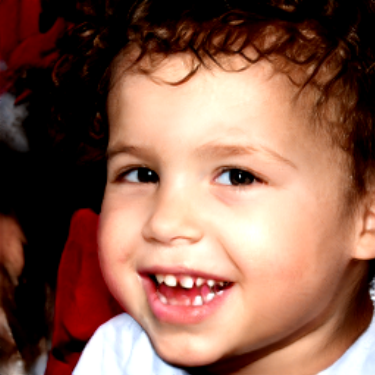}} & 
        \raisebox{-.5\height}{\includegraphics[width=0.15\linewidth]{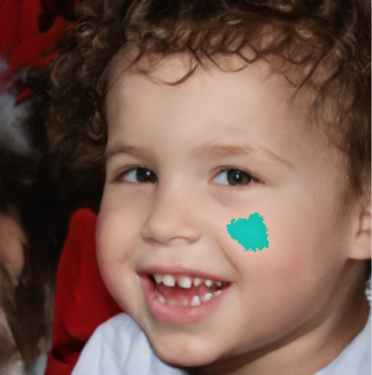}} & \raisebox{-.5\height}{\includegraphics[width=0.15\linewidth]{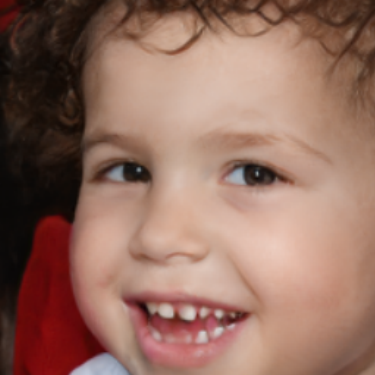}} &
        \raisebox{-.5\height}{\includegraphics[width=0.15\linewidth]{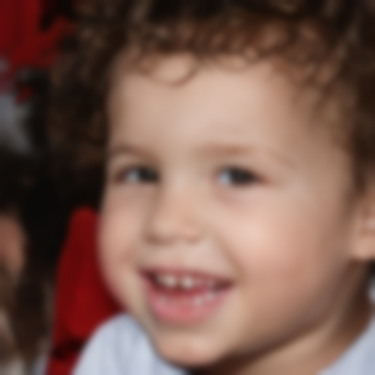}} & 
        \raisebox{-.5\height}{\includegraphics[width=0.15\linewidth]{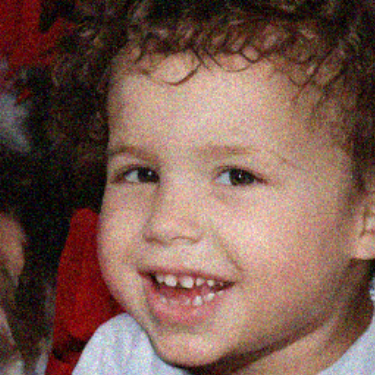}}&
        \raisebox{-.5\height}{\includegraphics[width=0.15\linewidth]{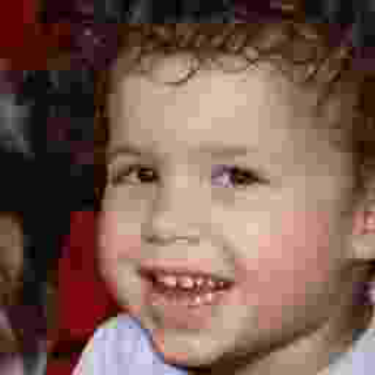}}
        \\
         \bottomrule
    \end{tabular}
    }
    \label{tab:baseline}
\end{table*}

\begin{figure*}[t]
    \centering
    \includegraphics[width=0.30\linewidth]{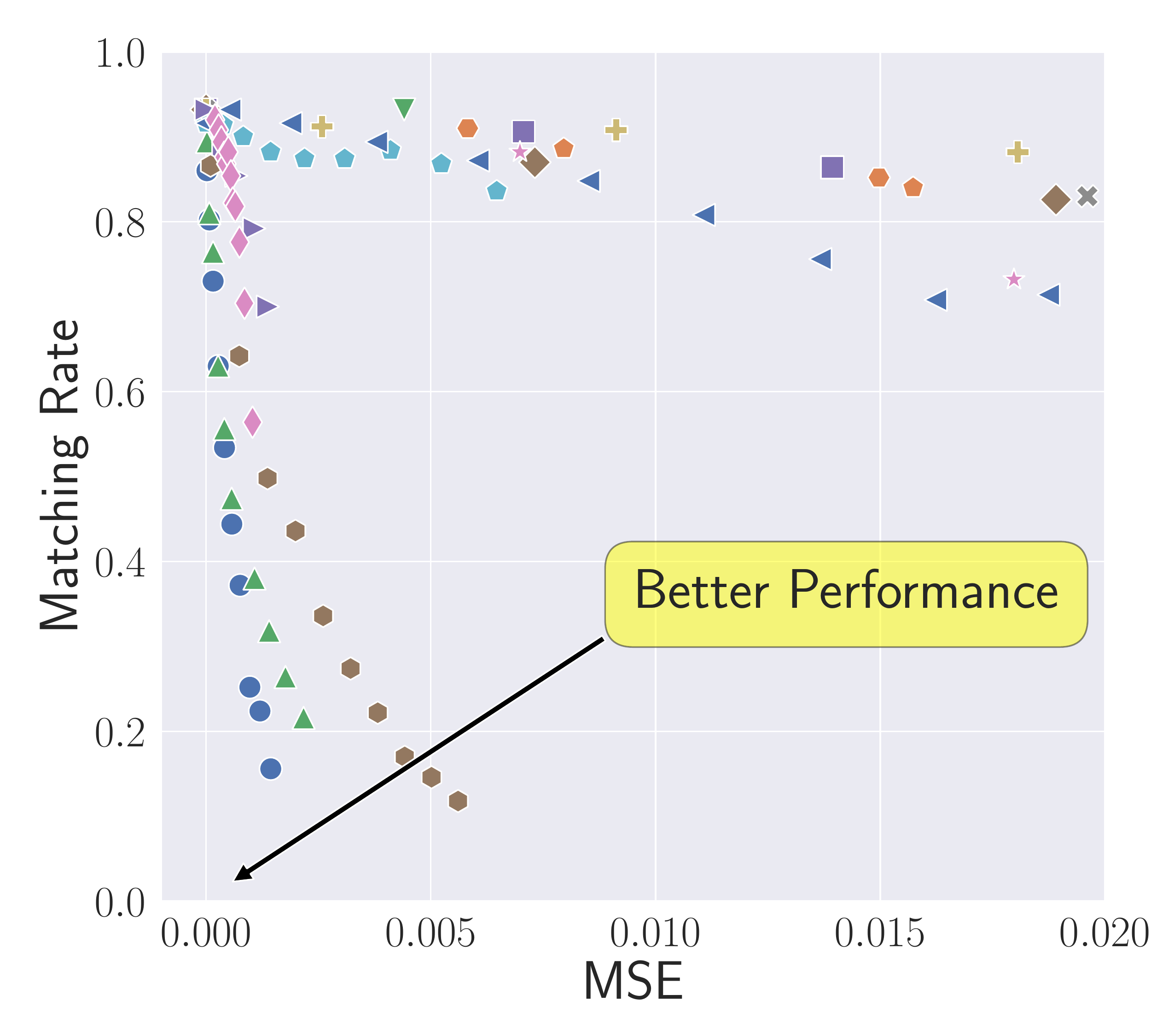}
\includegraphics[width=0.30\linewidth]{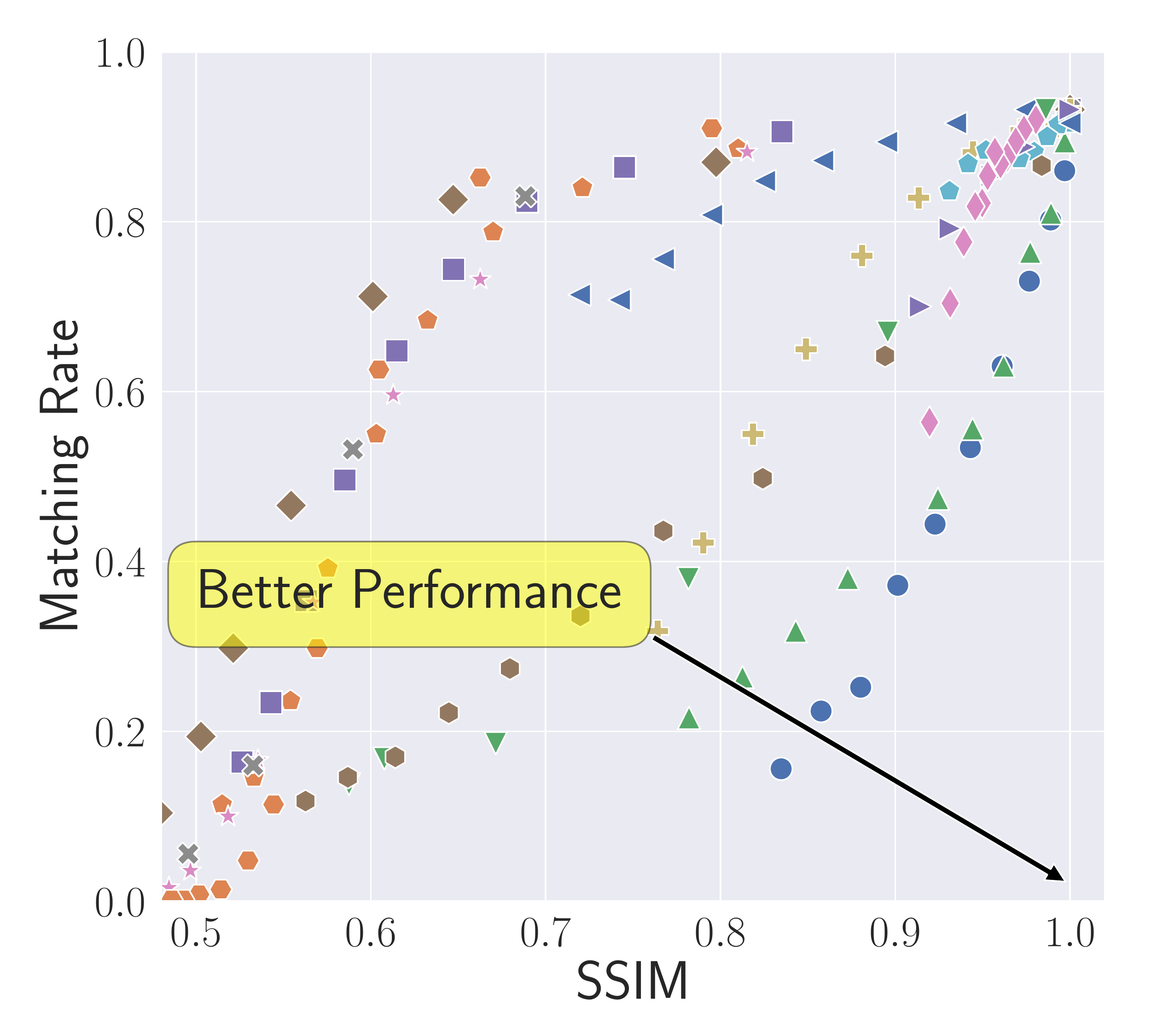}
\includegraphics[width=0.30\linewidth]{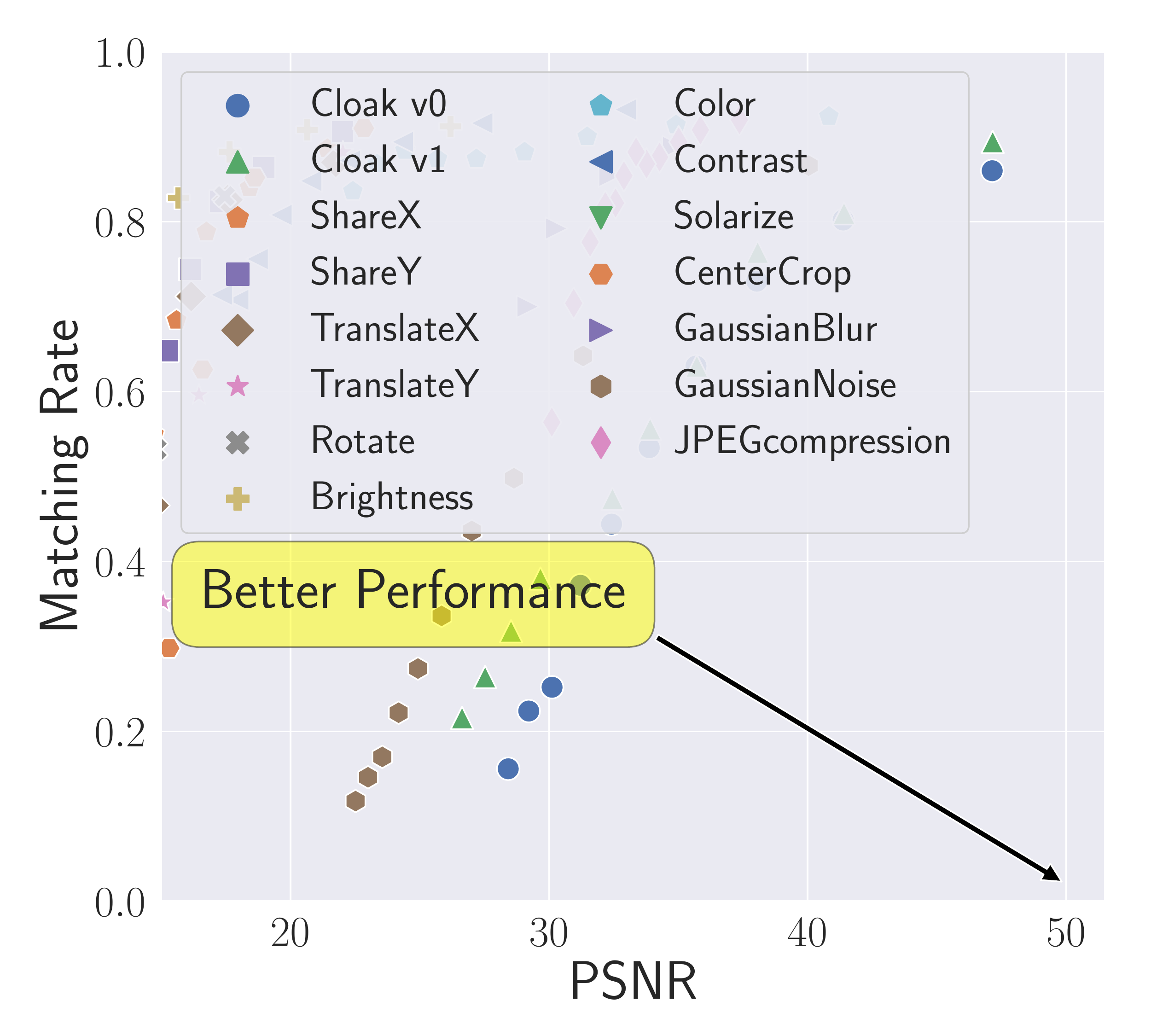}
    \caption{Comparison between all baseline methods and Cloak v0/v1 on generated images and StyleGANv2. The different points of each method represent different budgets.}
    \label{fig:scatter_opti}
\end{figure*}

\mypara{The Effect of Latent/Feature Deviation}
We further investigate the effect of latent/feature deviation on the performance of \system.
In the white-box scenario (Cloak v0), \system search for the cloaked images which can maximize both latent and feature deviations, while in the black-box scenario (Cloak v1) only feature deviations are maximized. 
As shown in \autoref{fig:effectiveness-0}, we can obverse that Cloak v0 achieve better effectiveness performance than Cloak v1 under each distance budget. 
However, we cannot prematurely claim that Cloak v0 is better because we need to consider whether Cloak v0 is at least as good as Cloak v1 in terms of utility performance. \autoref{tab:utility_v0/1} reports the utility performance of \system on the StyleGANv2. 
First, we can observe that Cloak v0 performs at least on-par with Cloak v1 under budget $\varepsilon$-1, $\varepsilon$-3, and $\varepsilon$-5. More encouragingly, under budget $\varepsilon$-7 and $\varepsilon$-9, Cloak v0 achieves better utility performance than Cloak v1. 
These results show that Cloak v0 outperforms Cloak v1 in terms of both effectiveness and utility, and convincingly demonstrate that the additional latent deviation we introduce for Cloak v0 does improve performance.

\mypara{Comparison with Baselines}
To elaborate on \system's performance in a more convincing manner, we compare \system extensively with thirteen baseline distortion methods, as shown in \autoref{tab:baseline}. 
For each baseline method, we evaluate both effectiveness and utility performance with a wide variety of different magnitude of the budget. More detailed descriptions of each method are presented in Appendix \autoref{tab:alldistortions}.
\autoref{fig:scatter_opti} displays the relationship between each baseline method’s matching rate and MSE/SSIM/PSNR score (see more results in Appendix \autoref{appfig:scatter_opti}). Thus, we can make the following observations.

First, as the budget increases (i.e., MSE becomes larger and SSIM/PSNR becomes smaller), all baseline methods can significantly reduce the matching rate, meaning that baseline methods that work only in image space can also achieve good effectiveness performance.
 
More encouragingly, the plot also clearly indicates the benefits of latent and feature deviations: among baseline methods with similar utility performance levels (similar MSE/SSIM/PSNR), our Cloak v0 and Cloak v1 consistently achieve better effectiveness (lower matching rate), as they benefit from maximizing latent and feature deviations. 
In other words, searching for cloaked images to maximize latent and feature deviations can further disenable GAN inversions at nearly no cost in utility.
Another interesting finding is that when \system is not an option, \textit{GaussianNoise}, \textit{GaussianBulr}, and \textit{JPEGCompression} appear to perform better.

\section{\system Against Hybrid Inversion}
\label{sec4:cloak-v02}
We now present \system against the second GAN inversion technique, i.e., hybrid inversion. 

\subsection{Defender’s Knowledge}
For hybrid inversion, we consider three different scenarios to characterize a defender's background knowledge.
See more detailed explanation about background knowledge in \autoref{assumption_defender}.

\mypara{White-Box (Cloak v2)} 
Hybrid inversion actually adopts an encoder to provide a better initialization $\code$ for the following optimization step. 
Here, we assume that a defender has complete knowledge of the target encoder $\advE$ to mislead the encoder, i.e., provide a worse initialization latent code $\code$ for the optimization. 
We give a quantitative illustration of this intuition in \autoref{sec:hybrid_method}. Besides that, we also assume that the defender has access to a feature extractor $F$. 
Note that the defender does not need to have white-box access to the target generator $\advG$ due to the design of this defense (see \autoref{sec:hybrid_method} more details).

\mypara{Grey-Box (Cloak v3)} 
Here, we relax the assumption that the defender has complete knowledge of the target encoder $\advE$. 
In particular, we assume that the defender can send many queries to the target encoder $\advE$ and train a shadow encoder $\userE$ to mimic the behavior of the target encoder $\advE$, and relies on the shadow encoder to act as the target encoder. 
Besides that, we assume that the defender has access to a feature extractor $F$ for feature deviation. 

\mypara{Black-Box (Cloak v4)} 
Here, we assume the defender has no knowledge of the adversary's generator or encoder. 
Here, the defender only has access to a feature extractor $F$.

\subsection{Methodologies}
\label{sec:hybrid_method}
Here the defenses are also divided into three simultaneous components, namely maximizing latent deviation, maximizing feature deviation, and searching for cloaked images in the image space. 
In particular, we introduce a new novel method to maximize the latent deviation. 
The algorithms can be found in \autoref{appendix:unganable_algorithms}.

\mypara{New Perspective of Latent Deviation} 
As aforementioned in \autoref{sec:gan_inversion}, an important issue for optimization-based inversion is initialization. 
Recent research~\cite{KALL18,BDS19,KLA19,RMC16} shows that using different initializations leads to a significant perceptual difference in generated images. 
Here, we conduct a pre-experiment on using different initializations to perform the optimization-based inversion, including Gaussian, zeros, etc (see~\cite{Distribution-is-all-you-need} for each distribution). 
In particular, hybrid inversion adopts an encoder to provide initialization for optimization. 

\begin{figure}[t]
    \centering
    \subfloat[Perceptual Loss]{\includegraphics[width=0.5\linewidth]{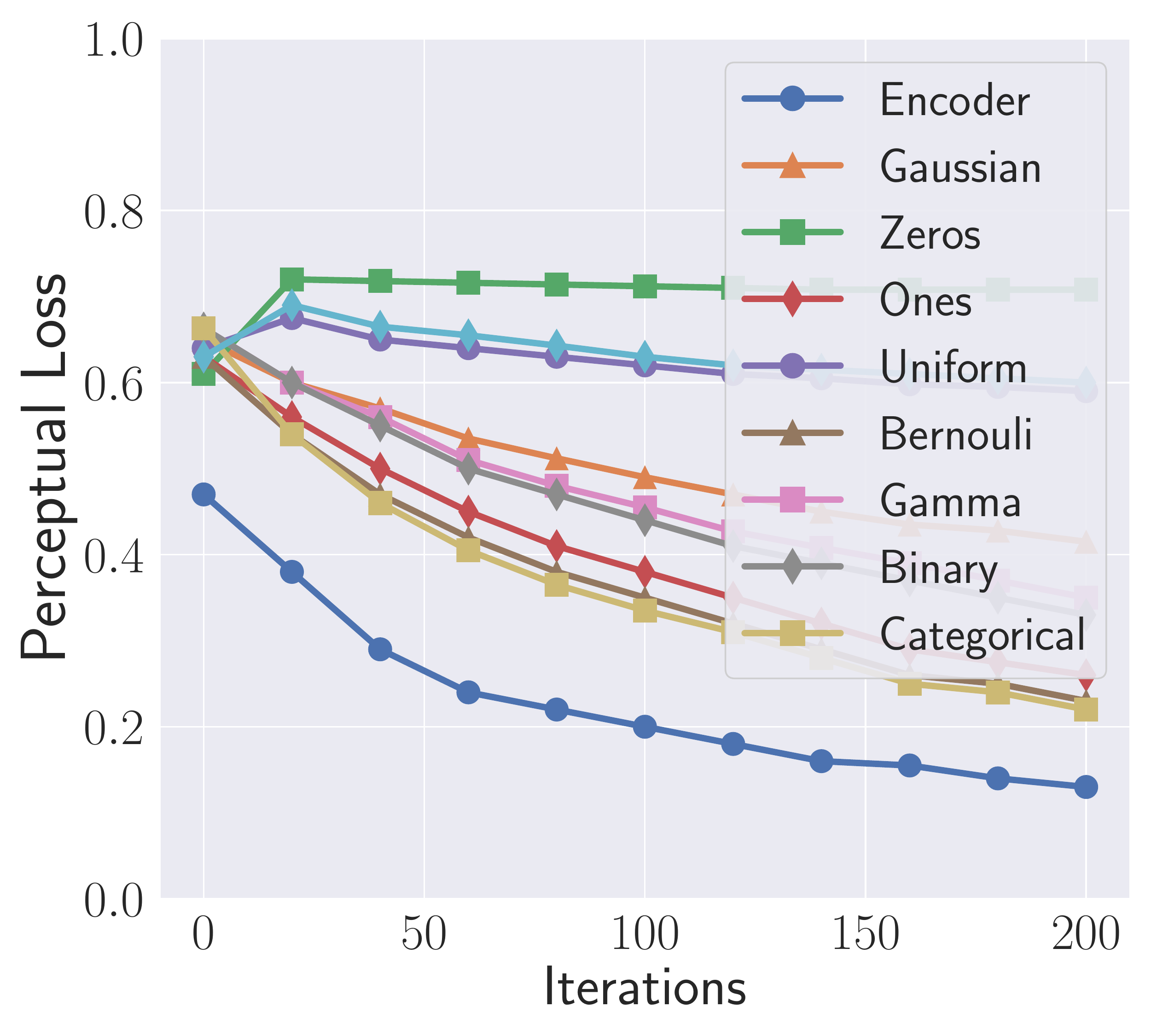}}
    \subfloat[MSE Loss]{\includegraphics[width=0.5\linewidth]{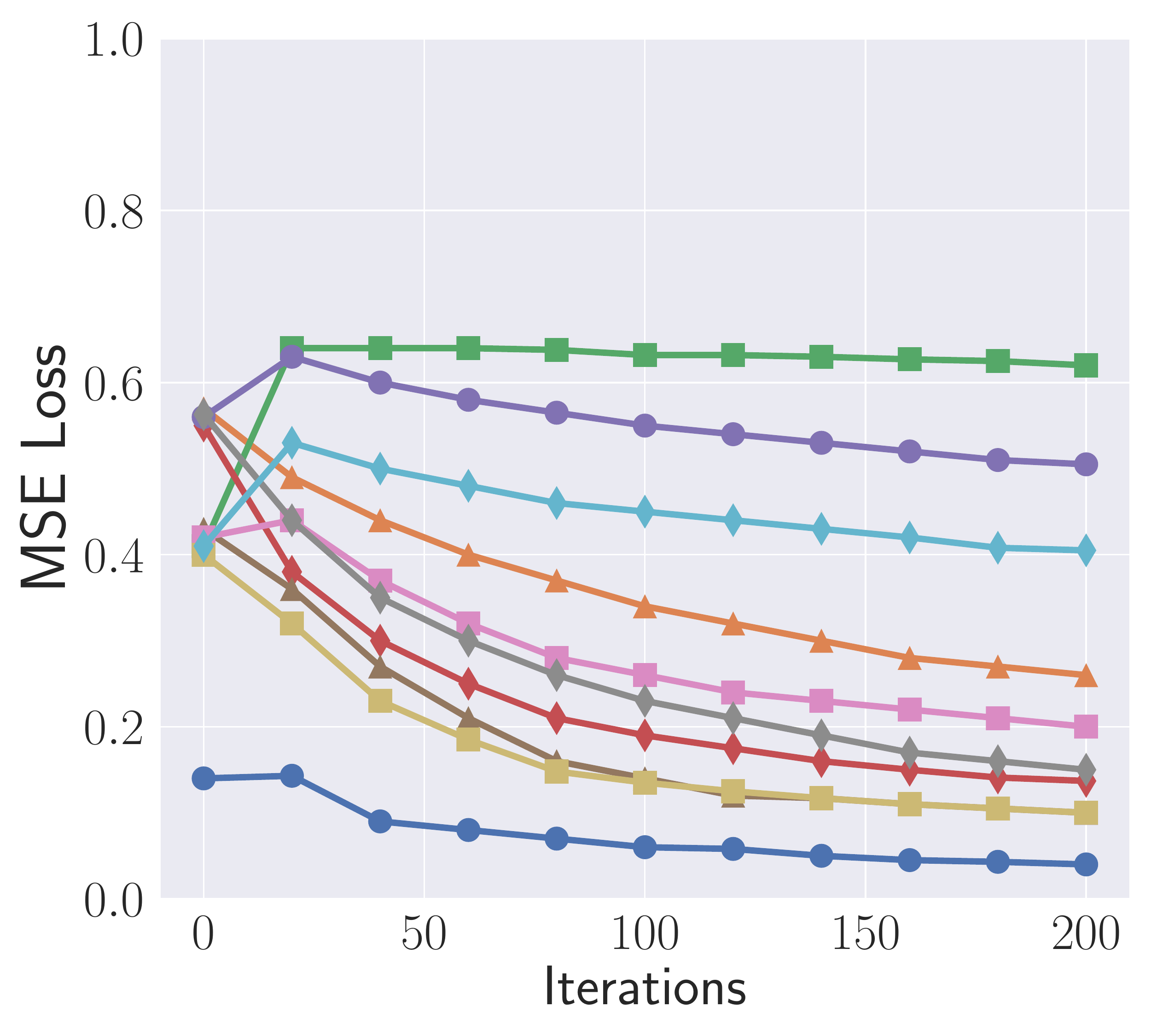}}
    \caption{The loss trend under the effect of different initialization for optimization.}
    \label{fig:distri}
\end{figure}

\autoref{fig:distri} shows the trend of perceptual and MSE loss, respectively.
First, the encoder indeed provides better initialization, which leads to better initial and final performance. 
Second, the trend of loss remains constant when the initialization is set to zero, which means it is quite difficult to invert the target image into the latent space. 
This observation suggests a new perspective on the latent deviation -- misleading the encoder to provide zero initialization, or close to zero.
In other words, our defense's goal against hybrid inversion should be to force the output of the encoder to zero.
This is actually a special case of maximizing latent deviation, which provides the movement direction of the cloaked image in the latent space, i.e., towards zero. 

\begin{figure}[!t]
\centering
\includegraphics[width=0.8\columnwidth]{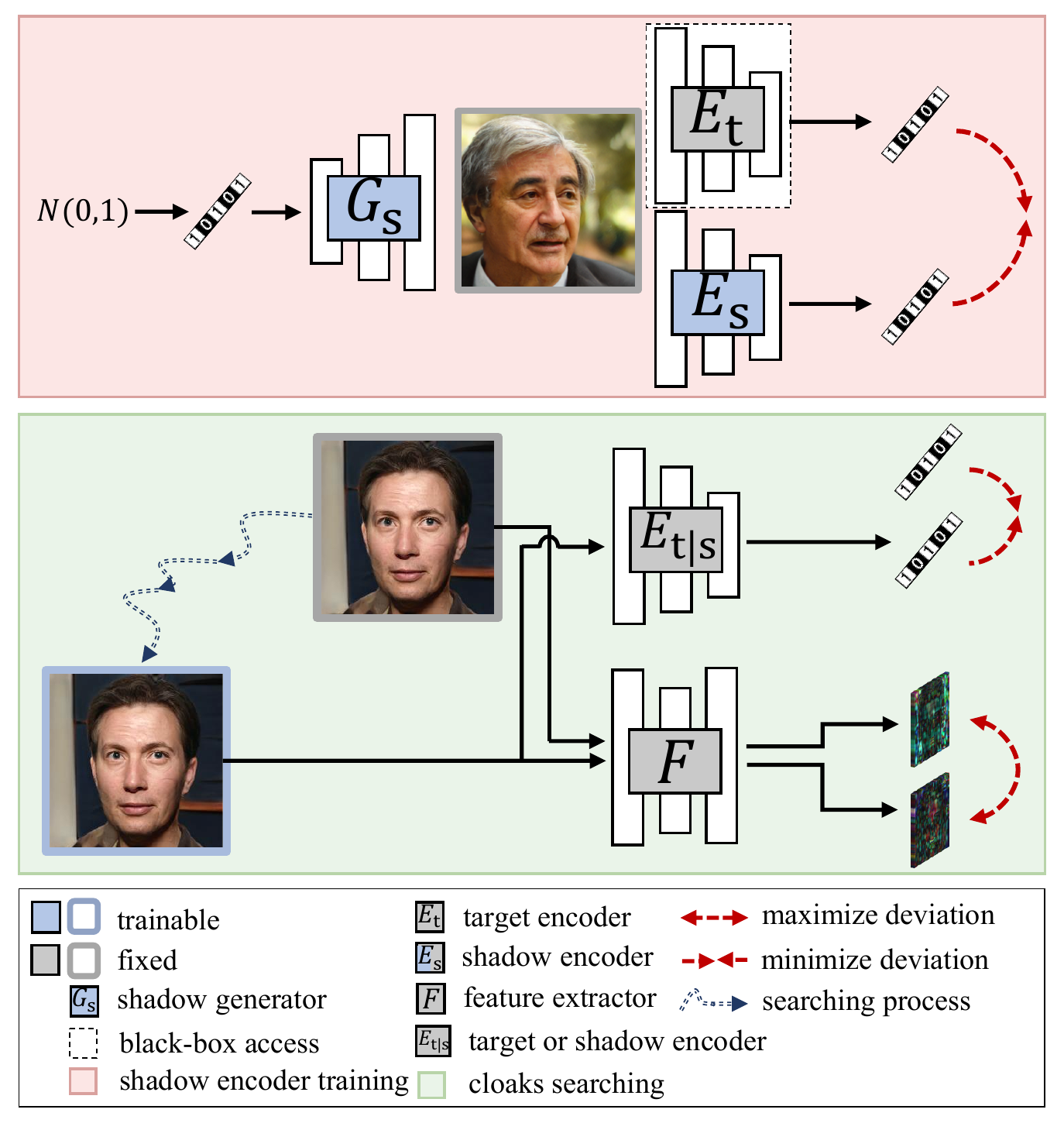}
\caption{An illustration of white-box (Cloak v2), grey-box (Cloak v3) and black-box (Cloak v4).}
\label{fig:cloak-23}
\end{figure}

\mypara{White-Box (Cloak v2)} 
In this scenario, we assume that the defender has full knowledge of the target encoder $\advE$, as well as an additional feature extractor $F$.
As shown in the green part of \autoref{fig:cloak-23}, the defender iteratively searches for $\cloakedX$ in the image space by modify $\targetX$, in order to minimize the deviation between $\advE(\cloakedX)$ and zero, and maximize the deviation between $F(\cloakedX)$ and $F(\targetX)$.
\begin{equation}
\begin{array}{c}
\max _{\cloakedX} \kappa \Big( -\Lrec\big(E_{\text{t}}(\cloakedX), 0\big) \Big)  +\left(1-\kappa \right)\Big( \Lrec\big(F(\cloakedX), F\left(\targetX\right)\big) \Big) \\
\text { s.t. } |\cloakedX - \targetX|_\infty < \epsilon \nonumber\\
\kappa \in [0,1] 
\end{array}
\end{equation}

\begin{figure*}[t]
    \centering
    \includegraphics[width=0.30\linewidth]{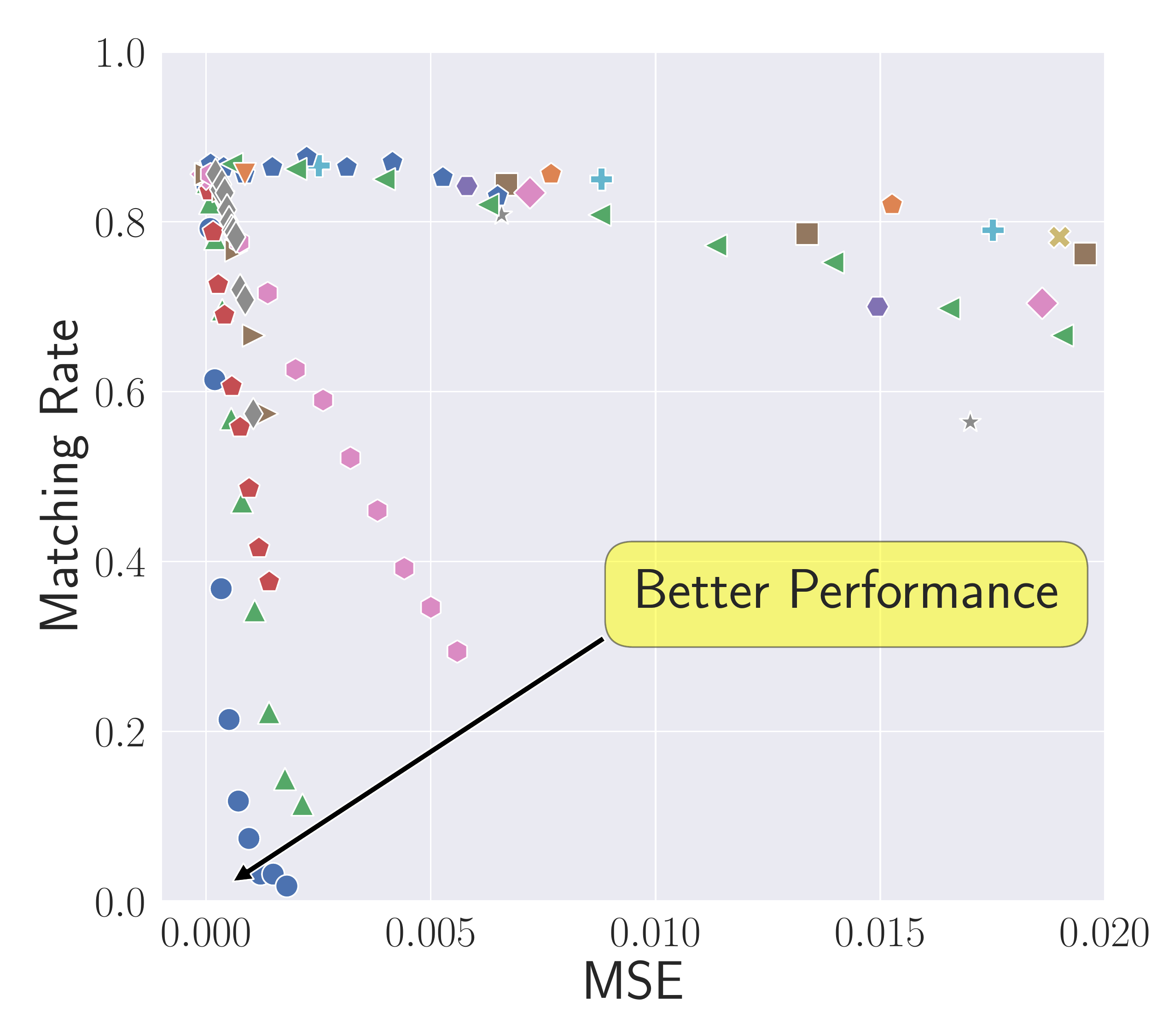}
    \includegraphics[width=0.30\linewidth]{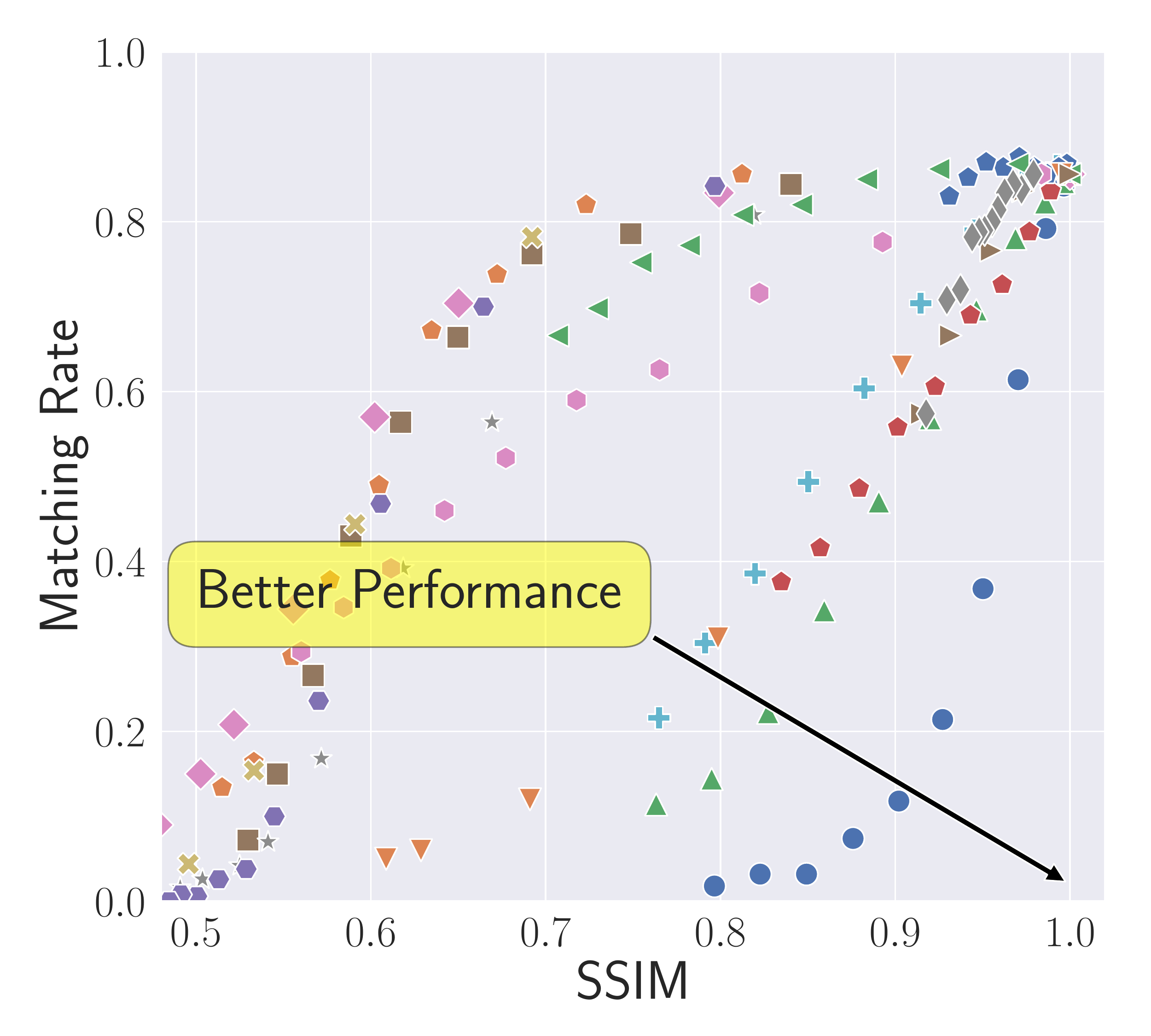}
    \includegraphics[width=0.30\linewidth]{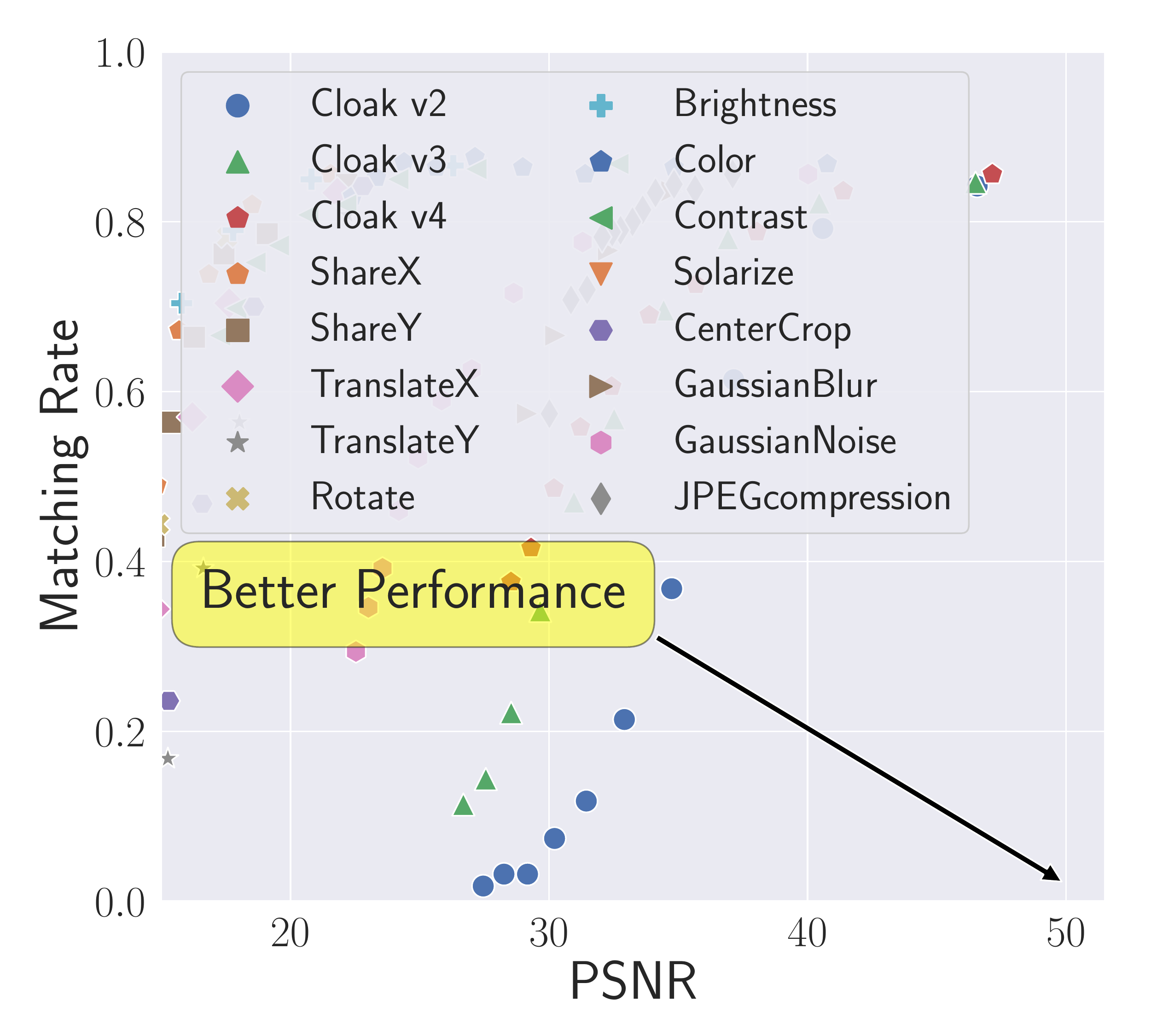}
    \caption{Comparison between all baseline methods and Cloak v2/v3/v4 on generated images and StyleGANv2. The different points of each method represent different budgets.}
    \label{fig:scatter_hybrid}
\end{figure*}

\mypara{Grey-Box (Cloak v3)} 
Here,  we relax the assumption that the defender has complete knowledge of the target encoder $\advE$. 
The defender needs to build a shadow encoder $\userE$ to match the predictions of $\advE$, i.e., find the shadow encoder's parameters that minimize the probability of errors between the shadow and target predictions. 

As shown in the pink part of \autoref{fig:cloak-23}, the defender builds a shadow generator $\userG$ which is responsible for crafting some input images, and $\userE$ serves as a discriminator while being trained to match target encoder's predictions on these images. 
In this setting, the two adversaries are $\userE$ and $\userG$, which try to minimize and maximize the deviation between $\userE$ and $\advE$ respectively. Then, shadow encoder $\userE$ becomes a functionally equivalent copy of target encoder $\advE$. 

Finally, the defender iteratively searches for $\cloakedX$ in the image space by modify $\targetX$, in order to minimize the deviation between $\userE(\cloakedX)$ and zero, and maximize the deviation between $F(\cloakedX)$ and $F(\targetX)$.
\begin{equation}
\begin{array}{c}
\max _{\cloakedX} \kappa \Big( -\Lrec\big(\userE(\cloakedX), 0\big) \Big)  +\left(1-\kappa \right)\Big( \Lrec\big(F(\cloakedX), F\left(\targetX\right)\big) \Big) \\
\text { s.t.\ } |\cloakedX - \targetX|_\infty < \epsilon \nonumber\\
\kappa \in [0,1] 
\end{array}
\end{equation}

\mypara{Black-Box (Cloak v4)} 
In this scenario, the defender has no knowledge of the target generator or target encoder or inversion techniques. 
The defender can only search for $\cloakedX$ in the image space by modifying $\targetX$, to maximize the feature deviation between $\cloakedX$ and $\targetX$.

\begin{equation}
\begin{array}{c}
\max _{\cloakedX} \Lrec(F(\cloakedX), F(\targetX)) \\
\text { s.t. } |\cloakedX - \targetX|_\infty < \epsilon \nonumber\\
\end{array}
\end{equation}

\subsection{Experimental Setup}
\label{sec5:setup}
For the manipulator/adversary, we follow the configurations of hybrid inversion (Zhu et al.~\cite{ZSZZ20}). 
Here, we again only consider the $\mathbf{z}$ space for all GAN models.
We set up 100 iterations for the optimization step of inversion, and use perceptual loss and pixel-level MSE loss to reconstruct the target image in the optimization step.

As a defender, for Cloak v3, we build the shadow generator by using 1 linear layer to accept Gaussian noise, followed by five convolutional layers and five Batch Normalization~\cite{IS15} layers. 
Furthermore, we again use a random initialized ResNet-18 as the shadow encoder. 
For all Cloaks (v2/v3/v4), we again use a pretrained ResNet-18~\cite{HZRS16} as the feature extractor. 
Besides, we fix the number of iterations as 500, to search for cloaked images.
In addition, all other experimental settings are the same as described in \autoref{setup_optim}.

\subsection{Results}
\label{sec:hybrid_eval}
\begin{figure}[t]
    \centering
    \subfloat[StyleGANv1\label{fig:hybrid-styleganv1}]{\includegraphics[width=0.5\linewidth]{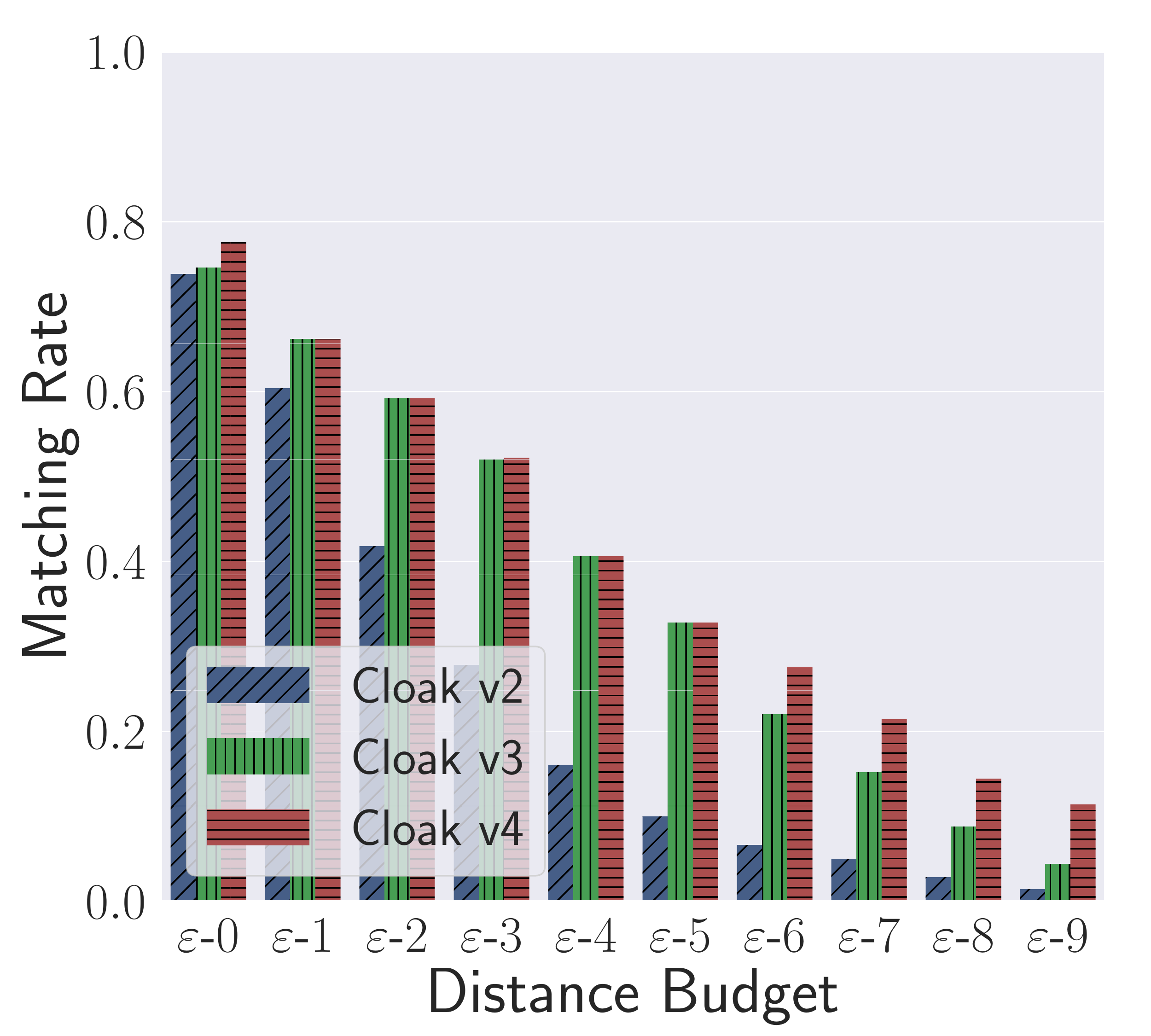}}
    \subfloat[StyleGANv2\label{fig:hybrid-styleganv2}]{\includegraphics[width=0.5\linewidth]{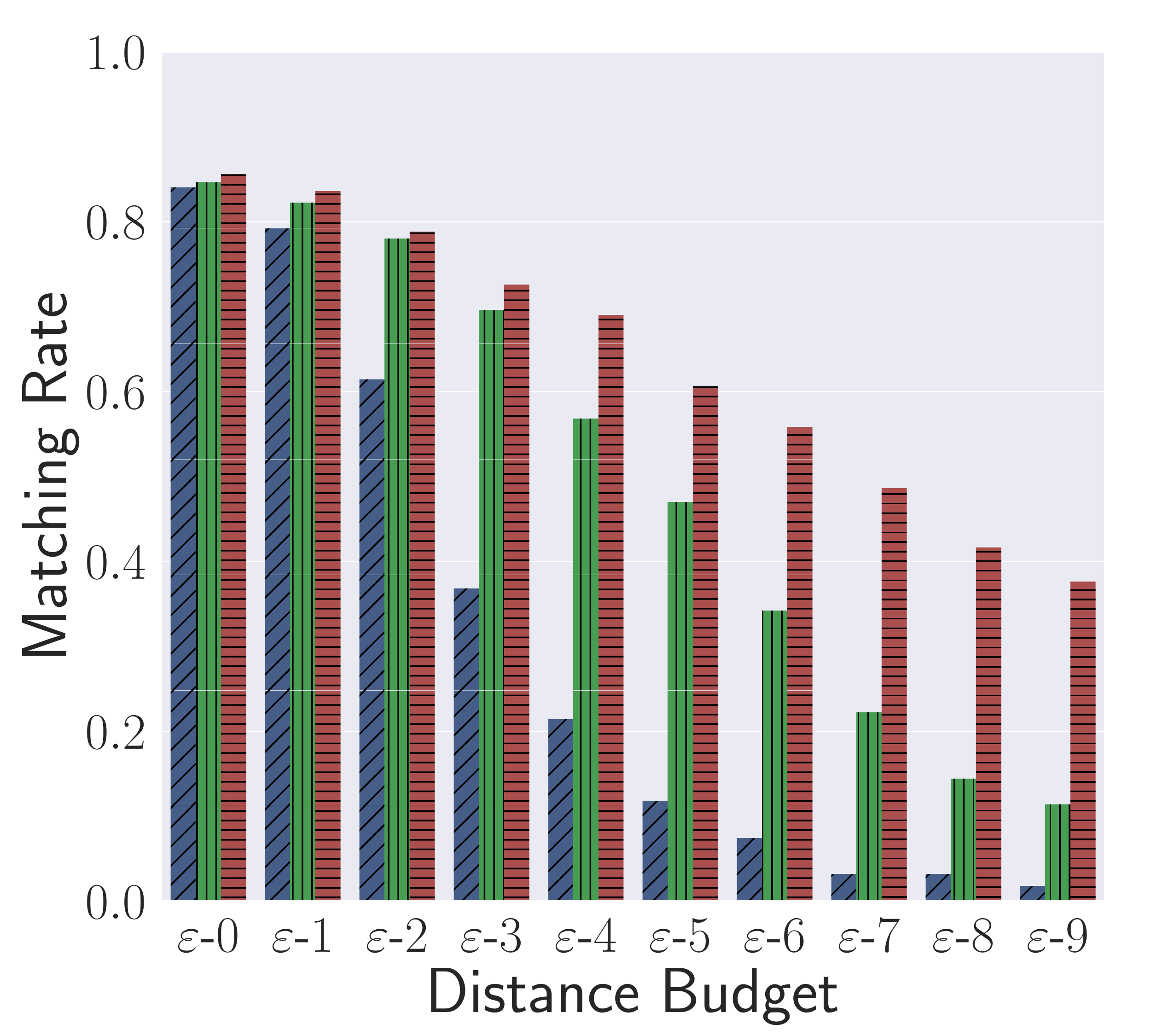}}
    \caption{The effectiveness performance of Cloak v2, Cloak v3 and Cloak v4.}
    \label{fig:effectiveness-234}
\end{figure}

\mypara{Effectiveness Performance}
To evaluate the effectiveness performance quantitatively, we use the same evaluation setup as presented in \autoref{sec:optim_eval}.
\autoref{fig:effectiveness-234} depicts the effectiveness performance of Cloak v2/v3/v4 (See more results on DCGAN and WGAN in Appendix \autoref{fig:effectiveness-appendix}).
First, we again observe that as the budget increases, all Cloak v2/v3/v4 can significantly reduce the matching rate. 
These results indeed imply that UnGANable can achieve significant effectiveness against hybrid inversion.
For qualitative results, the same perturbation budget will lead to similar reconstructed results, as shown in \autoref{fig:quali-0}. 

\mypara{Utility Performance}
Similarly, since we set the same distance budgets as adopted against optimization-based inversion, thus for the same perturbation budget will lead to similar quantitative and qualitative utility performance, as shown in \autoref{fig:quali-0} and \autoref{tab:utility_v0/1}. 

\mypara{The Effect of Latent/Feature Deviation}
In \subfig{fig:hybrid-styleganv1} and \subfig{fig:hybrid-styleganv2}, we can observe that searching for cloaked images to mislead the target encoder controlled by adversary (Cloak v2) leads to much better effectiveness performance. Furthermore, the larger the distance budget, the larger the gap between Cloak v2 and both Cloak v3 and Cloak v4, reflecting the fact that zero initialization can significantly jeopardize the process of GAN inversion.
This convincingly verifies our new perspective of latent deviation--misleading the adversary's encoder to provide zero initialization, or close to zero. 

\mypara{Comparison with Baselines}
We compare \system extensively with thirteen baseline methods, as shown in \autoref{tab:baseline}.  
We use the same experimental setup as described in \autoref{sec:optim_eval}, such as the perturbation budget setting strategy and the result reporting metrics. 
We report comparisons between baseline methods and \system in \autoref{fig:scatter_hybrid}, and we can make the similar observations as mentioned in \autoref{sec:optim_eval}.
See more results on DCGAN/WGAN/StyleGANv1 in Appendix \autoref{appfig:scatter_hybrid}.
Here, we again emphasize that Cloak v2/v3/v4 achieves consistently better effectiveness (lower matching rate) and utility (lower MSE, higher SSIM and PSNR) performance than all baselines.

\section{Evaluation on Real Images}
\label{eval_real}
To elaborate on \system's performance, here we investigate the performance of \system on real facial images.
Concretely, we consider the strictest setting in which the defender has no knowledge of the adversary-controlled entities. 
Thus, we only consider the black-box scenario against optimization-based and hybrid inversion, i.e., Cloak v1 and Cloak v4.
In addition, the adversary-controlled GAN model is the state-of-the-art deepfake generative model StyleGANv2.
We collect 200 real images from the FFHQ dataset, and these images are the most successfully inverted into the latent space among the whole FFHQ dataset.

\begin{figure}[!t]
\centering
    \subfloat[Optimization-based \label{}]{
    \includegraphics[width=0.5\linewidth]{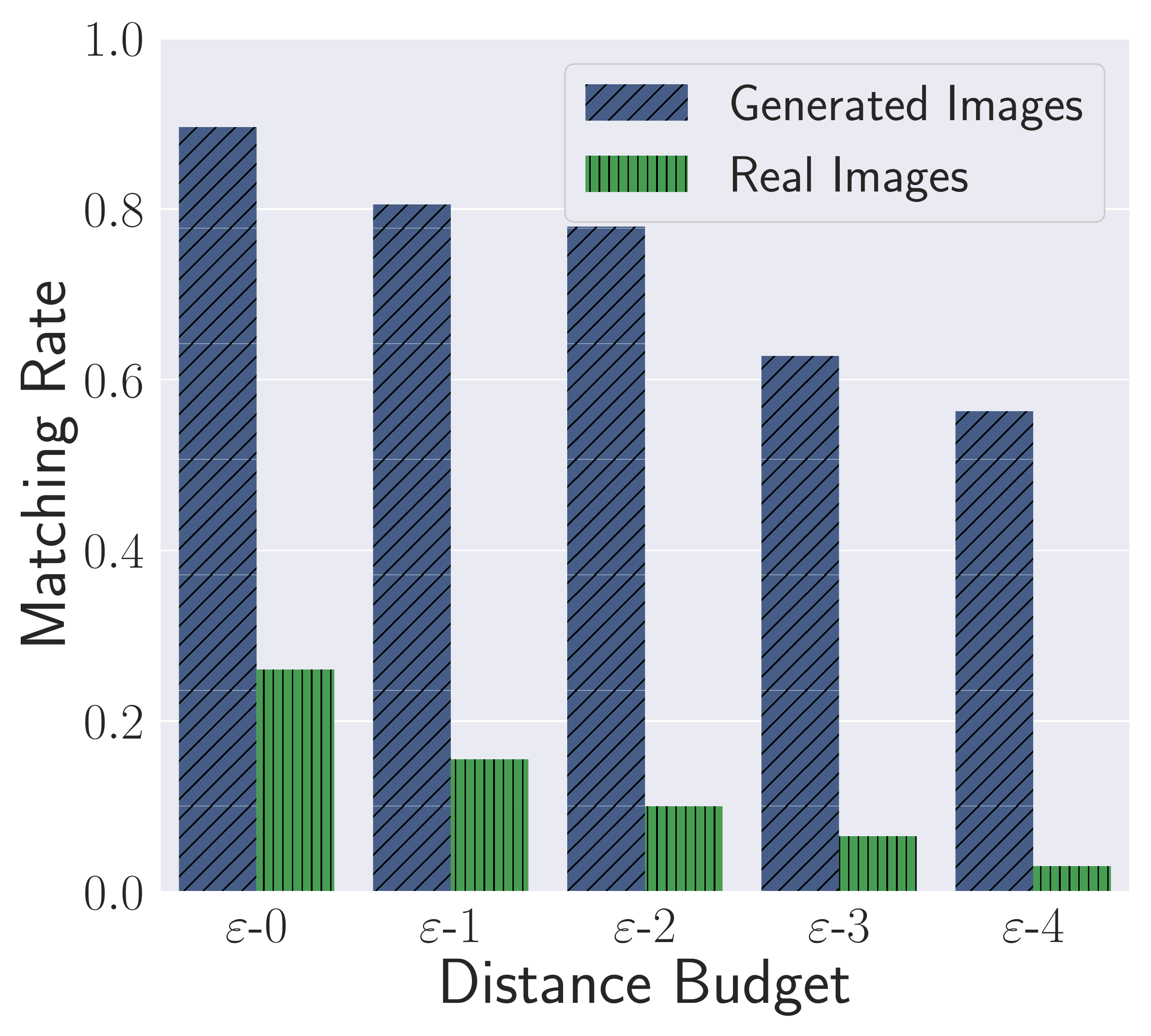}}
    \subfloat[Hybrid \label{real_hybrid_real}]{
    \includegraphics[width=0.5\linewidth]{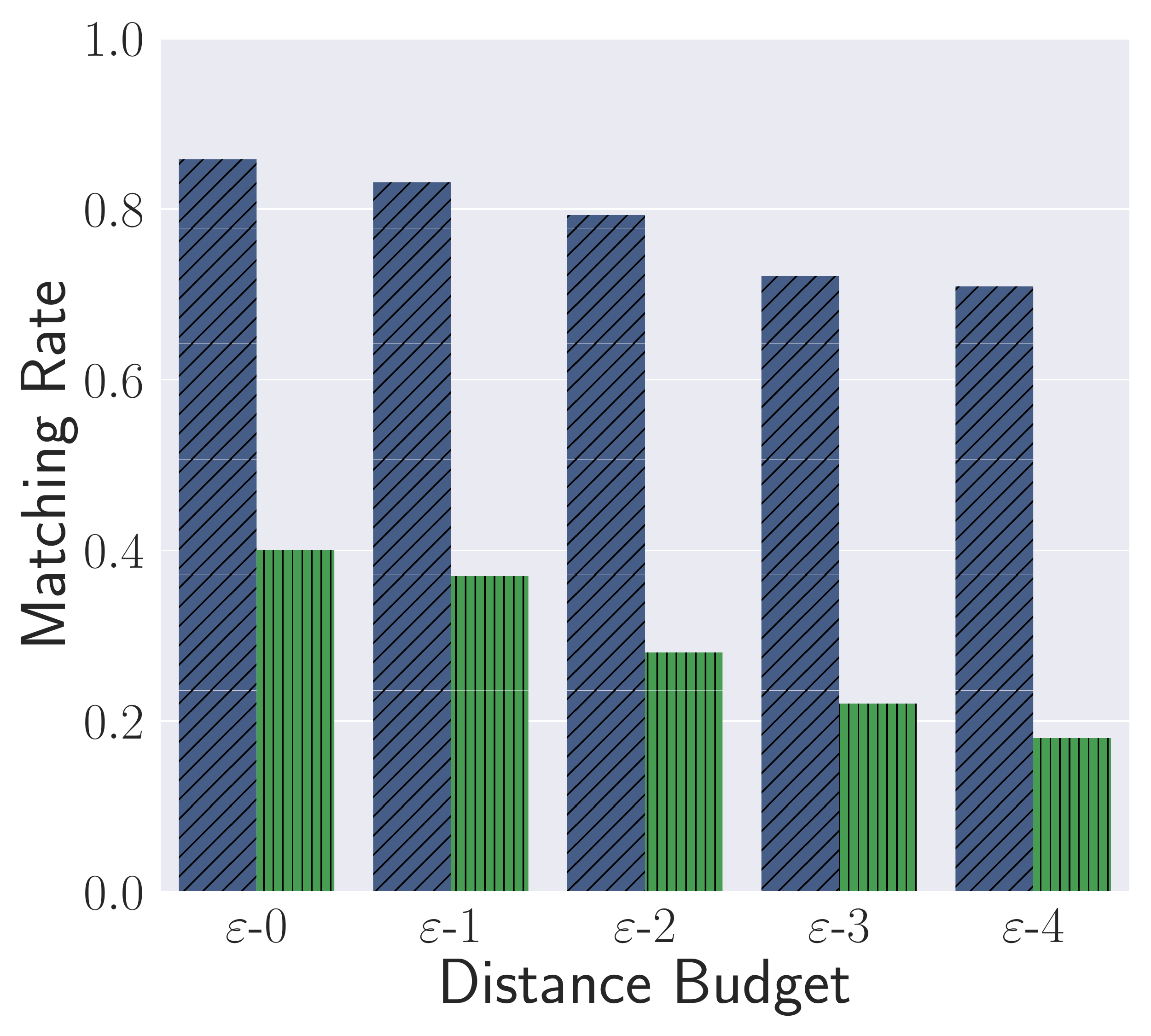}}
\caption{The effectiveness performance of Cloak v1/v4 on generated and real images, respectively.}
\label{fig:v1_4_reak_fake}
\end{figure}

\mypara{Effectiveness Performance}
We first present the effectiveness performance of \system.
We use the same evaluation setup as presented in \autoref{sec:optim_eval}.
We set 5 different distance budgets $\varepsilon$-0/1/2/3/4, the same as adopted in previous evaluations. 
\autoref{fig:v1_4_reak_fake} depicts the effectiveness performance of Cloak v1 and Cloak v4. 
First, we again observe that as the budget $\varepsilon$ increases, both Cloak v1 and Cloak v4 can significantly reduce the matching rate. 
Then we can see that the matching rate of Cloak v4 is clearly higher than that of Cloak v1, which verifies that the encoder of hybrid inversion indeed leads to better reconstruction performance.

What is more encouraging is that \system can achieve better effectiveness performance compared to that on generated images.
For example, when the distance budget is set as $\varepsilon$-4 (0.05), the matching rate of Cloak v1/v4 on the real image is about 0.072/0.191, while that on the generated image is about 0.474/0.606.  
We also provide a more detailed analysis of the reasons behind this observation in \autoref{reason_real_low_per}.
The results clearly show that \system can apply a much lower perturbation budget to obtain better effectiveness performance, and this lower distance budget further benefits utility performance.

\mypara{Utility Performance}
For utility performance, we conduct the evaluations both quantitatively and qualitatively.
we first quantitatively report a variety of similarity metrics (MSE/SSIM/PSNR) in \autoref{tab:utility_v1/4}.
Generally, SSIM values of 0.97, 0.98, and 0.99 imply the excellent visual quality of the cloaked images.
We then show in \autoref{fig:quali-utility-real_part} some qualitative samples of cloaked images. 
We can observe that when the distance budget is set as $\varepsilon$-4 (0.05), which represents a completely imperceptible perturbation, \system can achieve remarkable effectiveness performance (see matching rate in \subfig{real_hybrid_real}).
Therefore, we claim that \system provides acceptable protection for real images by much lower distance budgets and still yields good effectiveness and utility performance (see analysis of the reasons in \autoref{reason_real_low_per}).

\begin{table}[!t]
    \centering
    \caption{The quantitative utility performance of \system under Cloak v1 and v4 settings.}
    \tabcolsep 3pt
    \scalebox{0.75}
    {
    \begin{tabular}{c|c|c|c|c|c|c|c}
    \toprule
          Budget &Metric&Cloak v1& Cloak v4&Budget &Metric&Cloak v1& Cloak v4 \\
         \midrule
         \multirow{3}{*}{$\varepsilon$-0} & MSE & 1.9e-05 & 1.9e-05 &\multirow{3}{*}{$\varepsilon$-3} & MSE & 0.0003 & 0.0002\\
                                    & SSIM & 0.9968 & 0.9969&& SSIM &0.9606 & 0.967\\
                                    & PSNR & 47.205 & 47.210&& PSNR & 35.783 & 35.783\\
         \midrule
        \multirow{3}{*}{$\varepsilon$-1} & MSE & 7.1e-05 & 7.2e-05&\multirow{3}{*}{$\varepsilon$-4} & MSE & 0.0004 & 0.0004\\
                                    & SSIM & 0.9887 & 0.9887&& SSIM & 0.9422 & 0.9423\\
                                    & PSNR & 41.473 & 41.473&& PSNR & 33.983 & 33.982\\
        \midrule
        \multirow{3}{*}{$\varepsilon$-2} & MSE & 0.0002 & 0.0002\\
                                    & SSIM & 0.9764 & 0.9764\\
                                    & PSNR & 38.144 & 38.145\\
       
         \bottomrule
    \end{tabular}
    }
    \label{tab:utility_v1/4}
\end{table}
\begin{table}[!t]
    \centering
    \caption{Some visual examples of cloaked real images searched by Cloak v4 performed on StyleGANv2.}
    \scalebox{0.75}
    {
    \begin{tabular}{c|c|c|c}
    \toprule
       Target Image & $\varepsilon$-$0$ & $\varepsilon$-$2$ & $\varepsilon$-$4$\\
         \midrule
        \raisebox{-.5\height}{\includegraphics[width=0.28\linewidth]{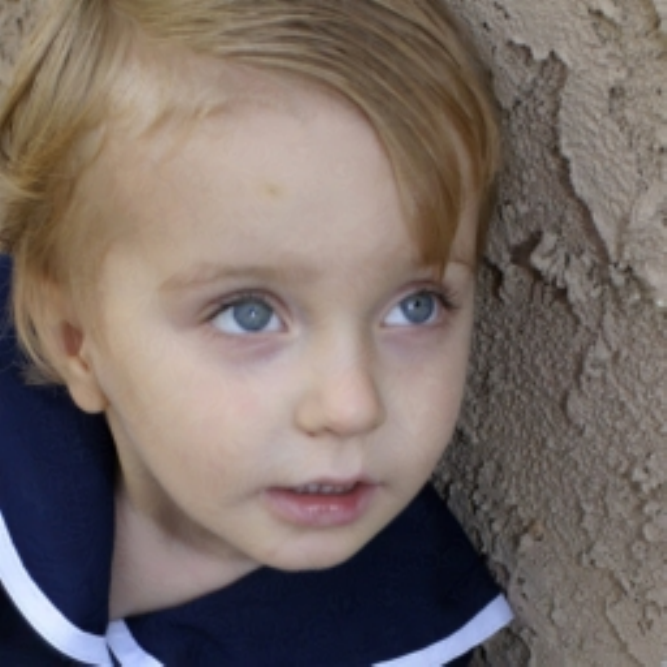}} & 
        \raisebox{-.5\height}{\includegraphics[width=0.28\linewidth]{plots/replots/6.1/utility/0.01_15.pdf}} & 
        \raisebox{-.5\height}{\includegraphics[width=0.28\linewidth]{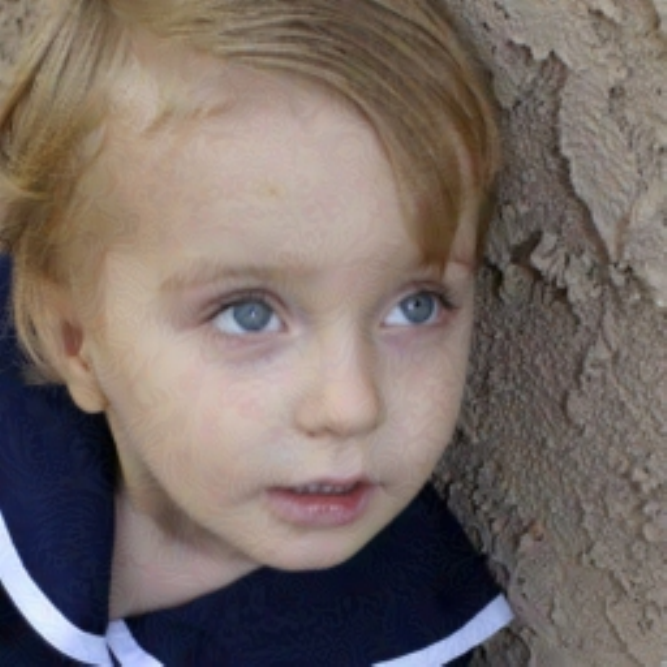}} &
        \raisebox{-.5\height}{\includegraphics[width=0.28\linewidth]{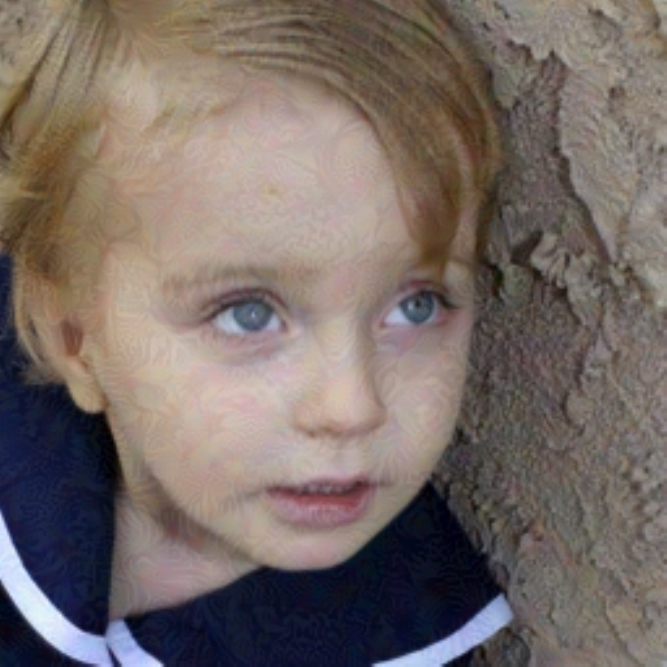}}
        \\
         \midrule
          \raisebox{-.5\height}{\includegraphics[width=0.28\linewidth]{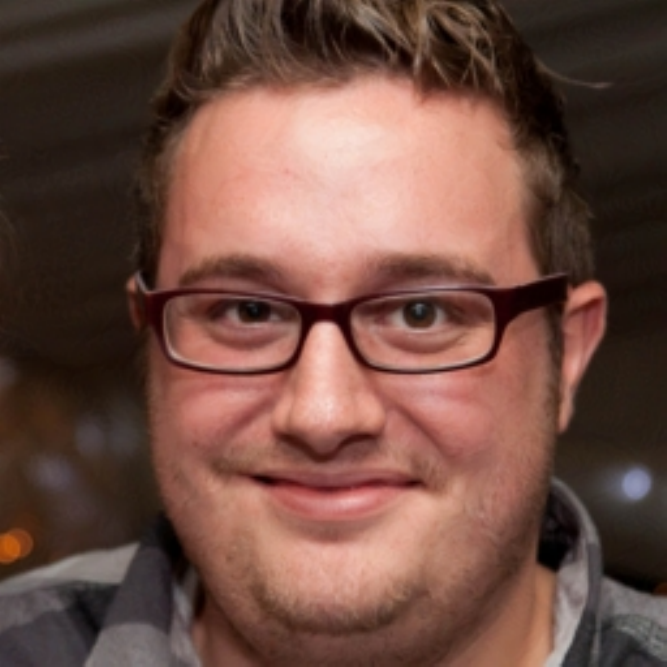}} & 
            \raisebox{-.5\height}{\includegraphics[width=0.28\linewidth]{plots/replots/6.1/utility/0.01_161.pdf}} & 
            \raisebox{-.5\height}{\includegraphics[width=0.28\linewidth]{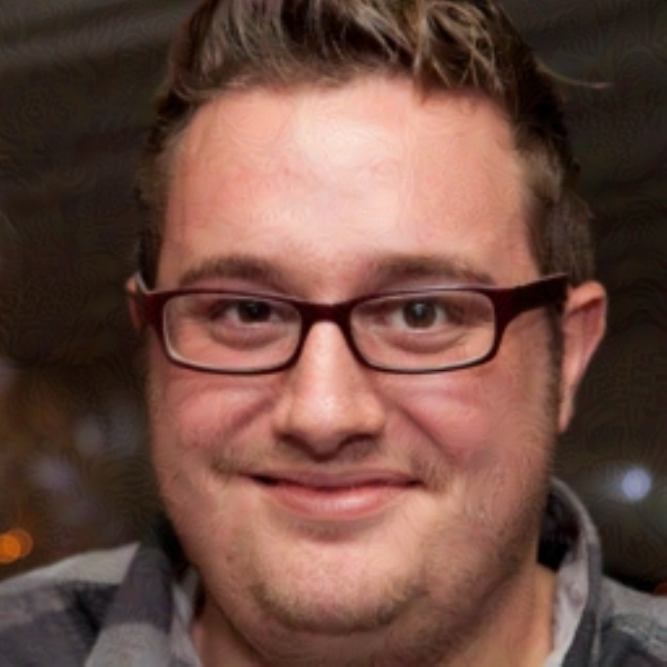}} &
            \raisebox{-.5\height}{\includegraphics[width=0.28\linewidth]{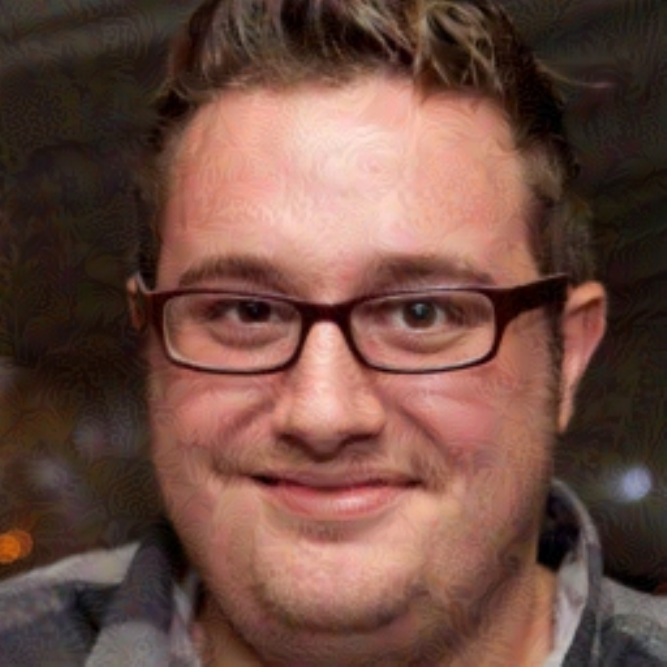}}
        \\
         \bottomrule
    \end{tabular}
    }
    \label{fig:quali-utility-real_part}
\end{table}

\begin{figure*}[t]
    \centering
    \subfloat[Cloak Overwriting\label{fig:real-adaptive-overwrite}]{
    \includegraphics[width=0.25\linewidth]{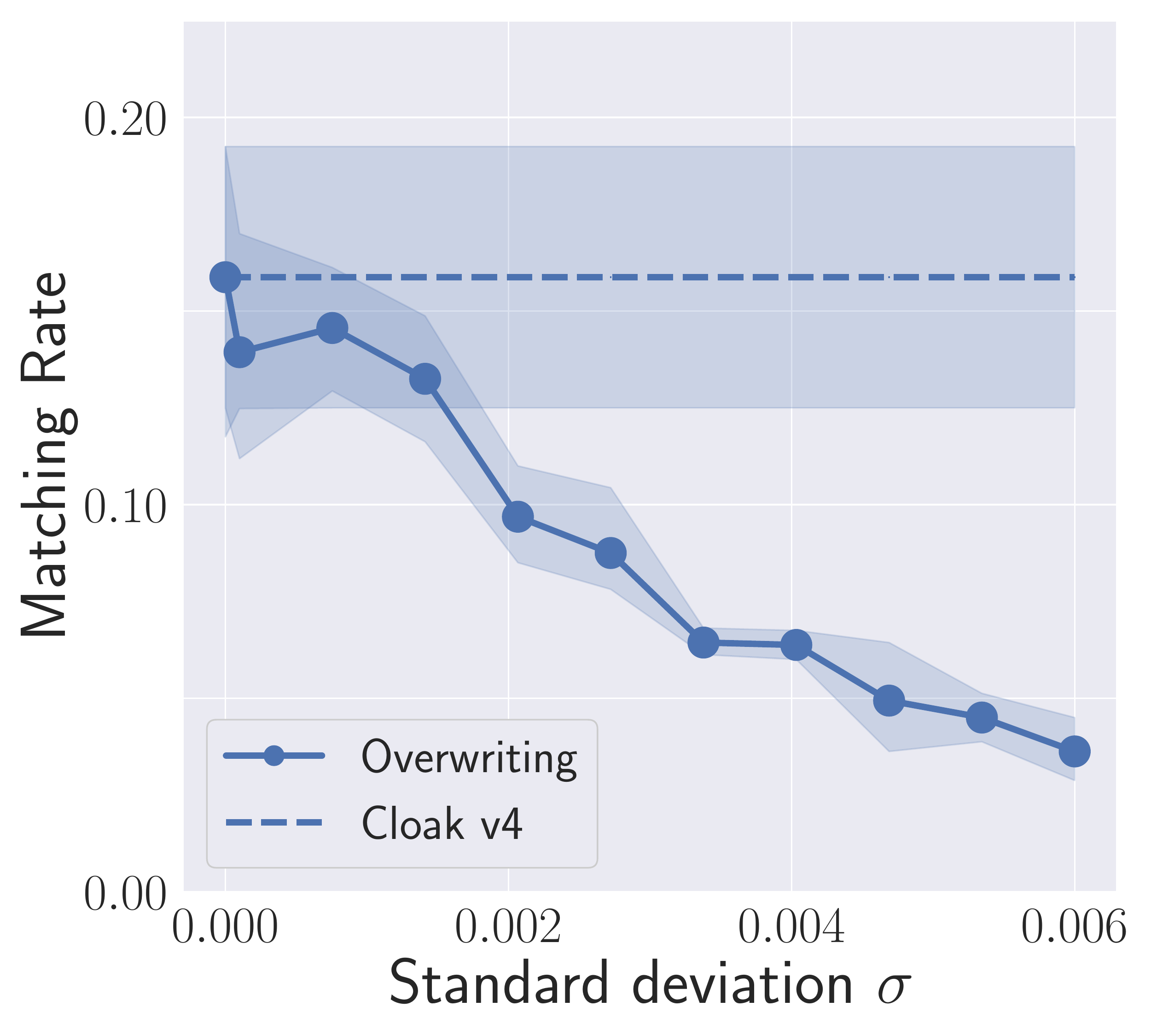}}
    \subfloat[Cloak Purification\label{fig:real-adaptive-remove}]{
    \includegraphics[width=0.25\linewidth]{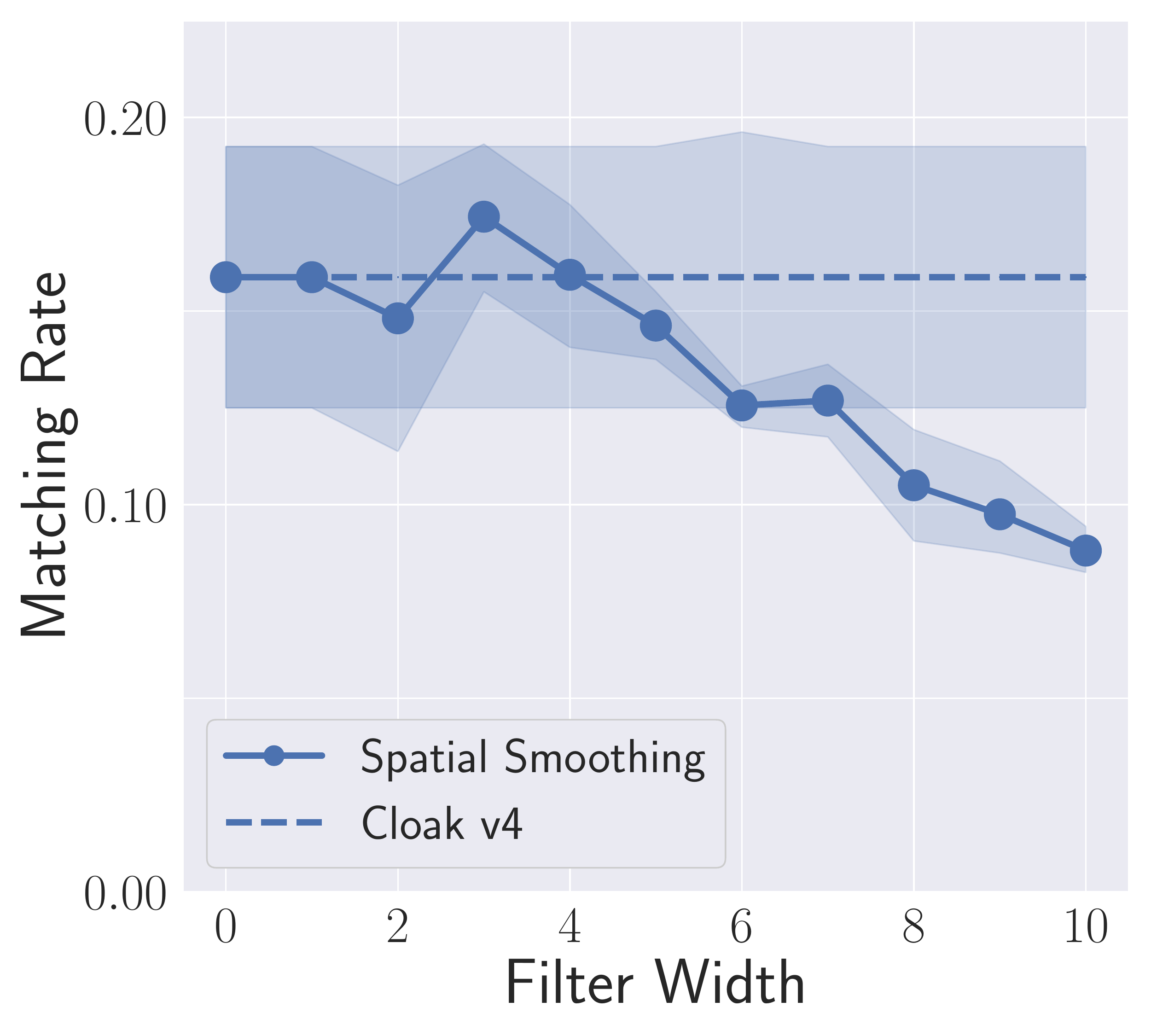}
    }
    \subfloat[More Iterations\label{fig:real-adaptive-iterations}]{
    \includegraphics[width=0.25\linewidth]{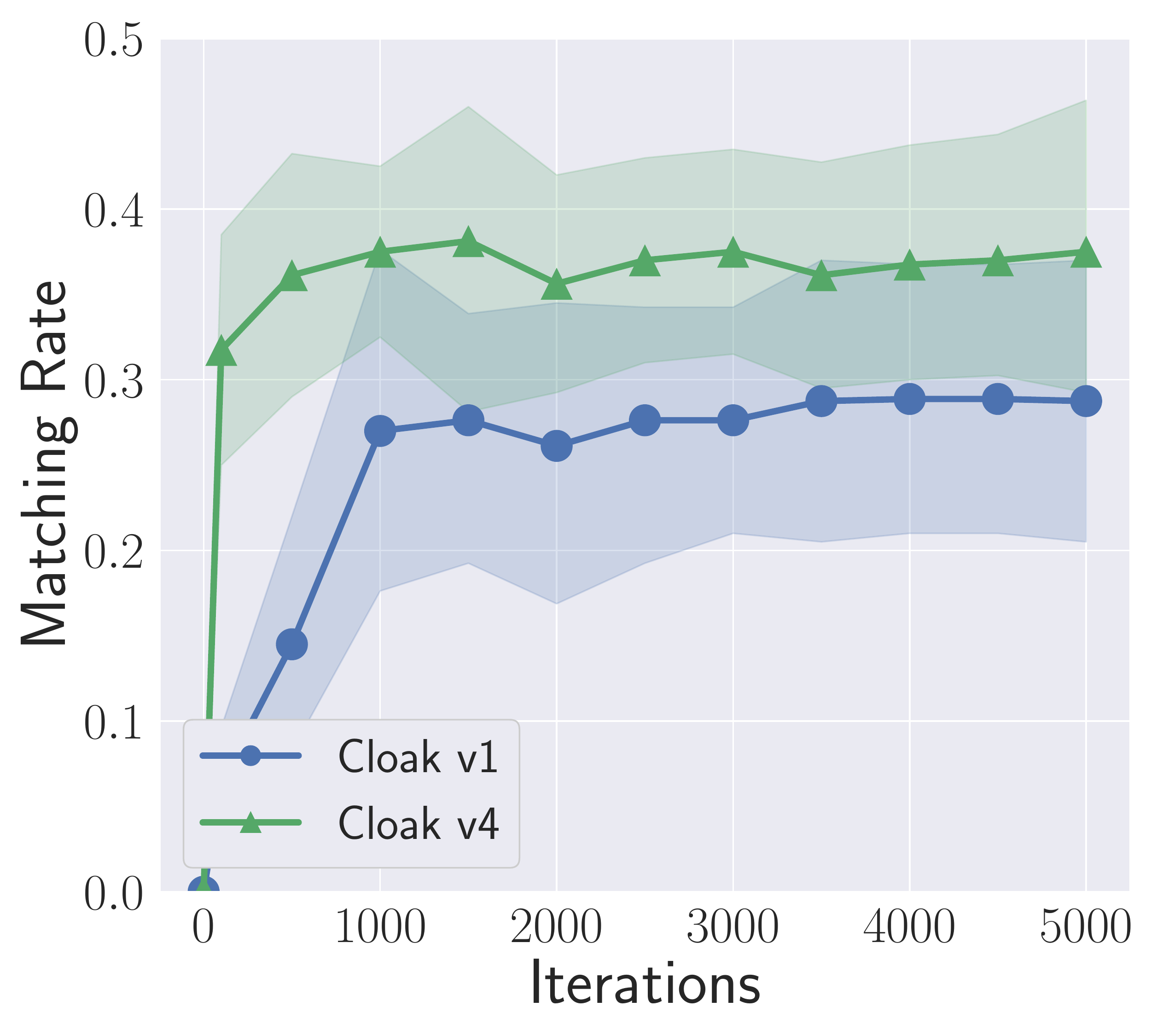}}
    \subfloat[Encoder Enhancement\label{fig:real-adaptive-encoder}]{
    \includegraphics[width=0.25\linewidth]{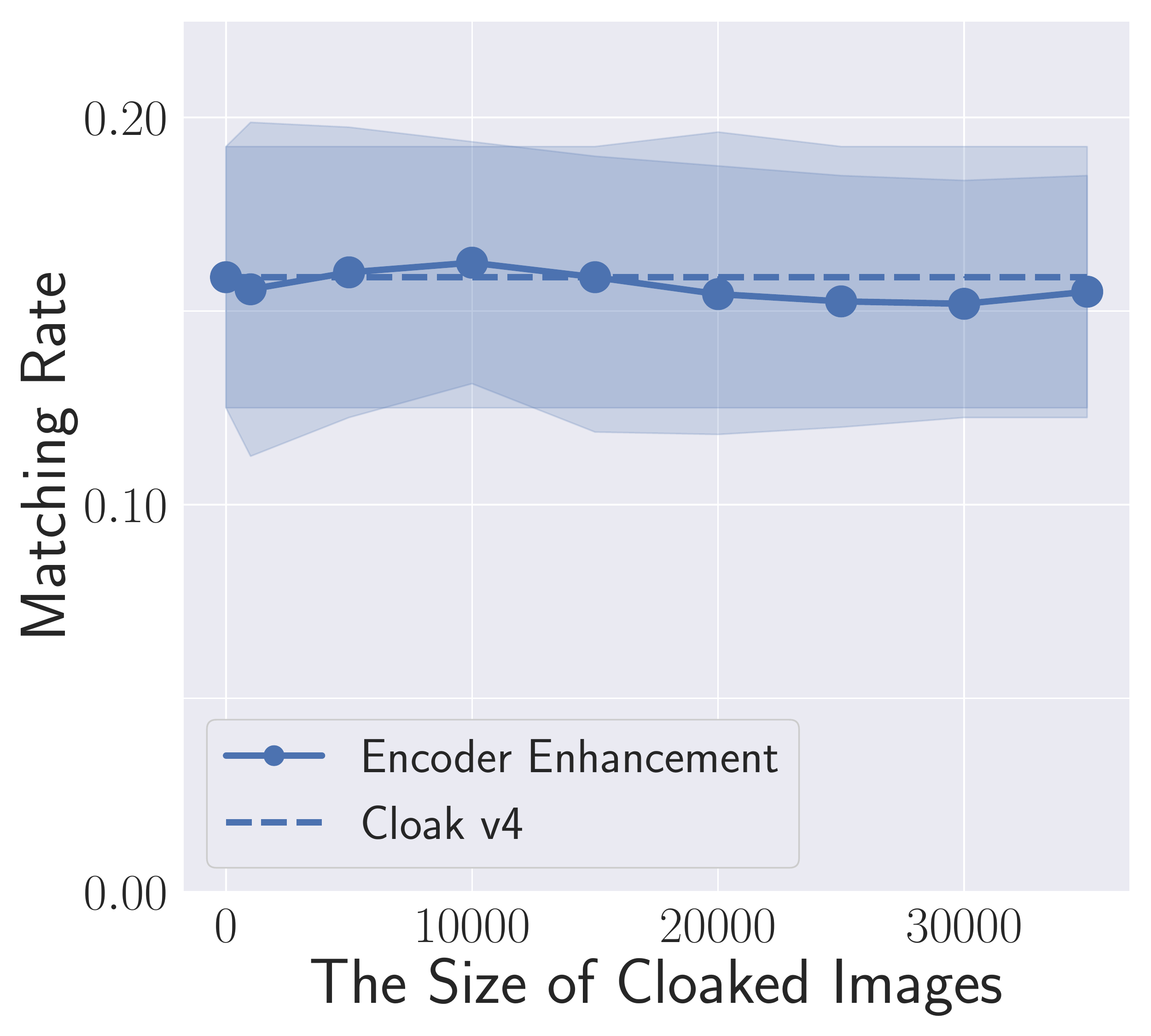}
    }
    \caption{The effectiveness performance of \system on real images under the effect of four possible adaptive adversaries.
    }
    \label{fig:real-adaptive}
\end{figure*}

\mypara{Comparison with Baselines}
We then compare \system extensively with thirteen baseline distortion methods, as shown in \autoref{tab:baseline}. 
For each baseline method, we evaluate both effectiveness and utility performance with a wide variety of different magnitude of the budget. 
More detailed descriptions of each method are presented in Appendix \autoref{tab:alldistortions}. 
\autoref{fig:real_baseline} displays the compassion between baseline methods and Cloak v1/v4, respectively (see more results of MSE/SSIM in Appendix \autoref{fig:real_baseline_appendix}). 
Thus, we can make the same observations as \system on generated images, i.e., our Cloak v1/v4 of \system achieves consistently better effectiveness (lower matching rate) and utility (lower MSE, higher SSIM, and PSNR) performance compared to all baseline methods.

\begin{figure}[t]
    \centering
    \subfloat[Optimization-based]{
    \includegraphics[width=0.5\linewidth]{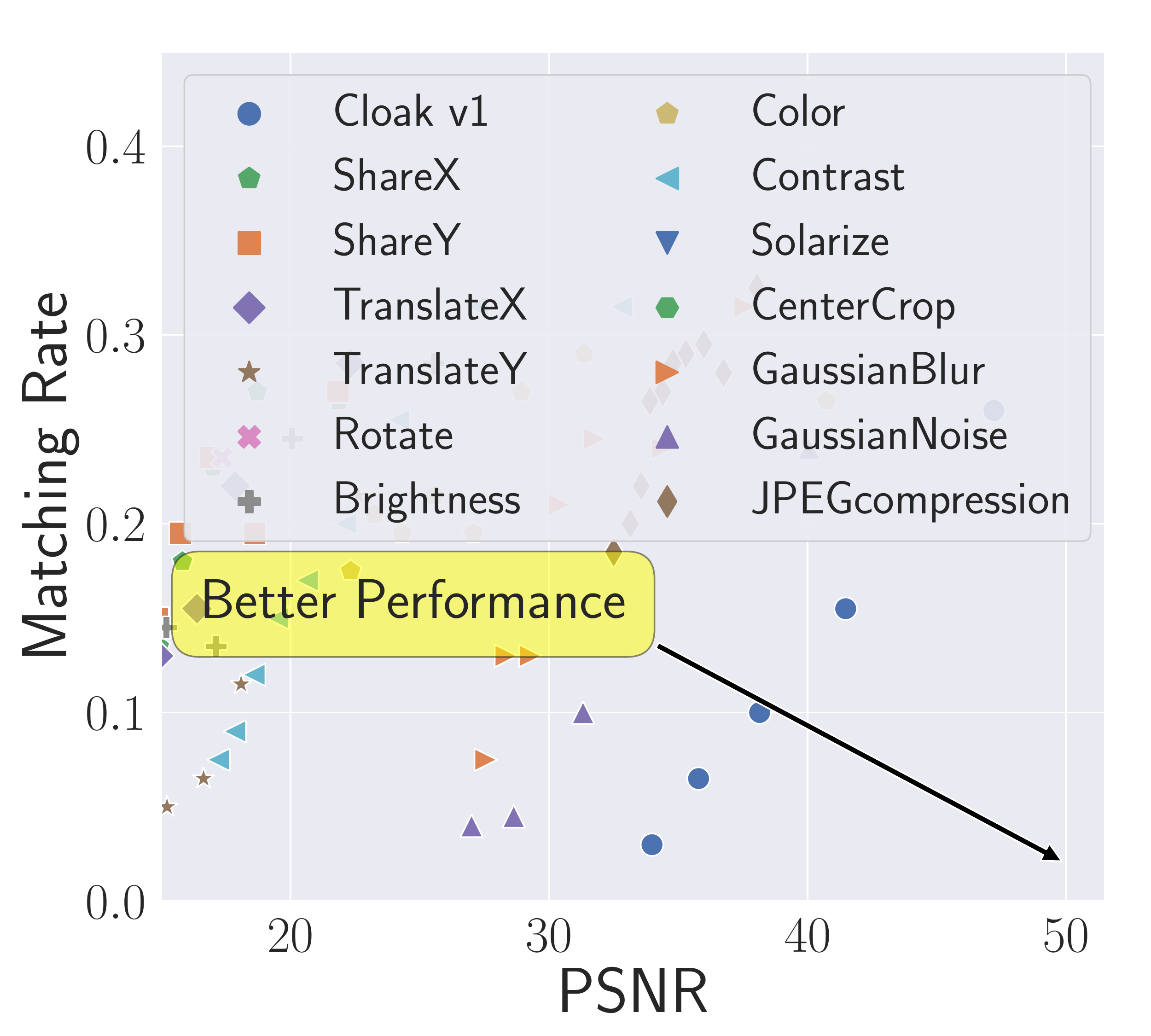}}
    \subfloat[Hybrid]{\includegraphics[width=0.5\linewidth]{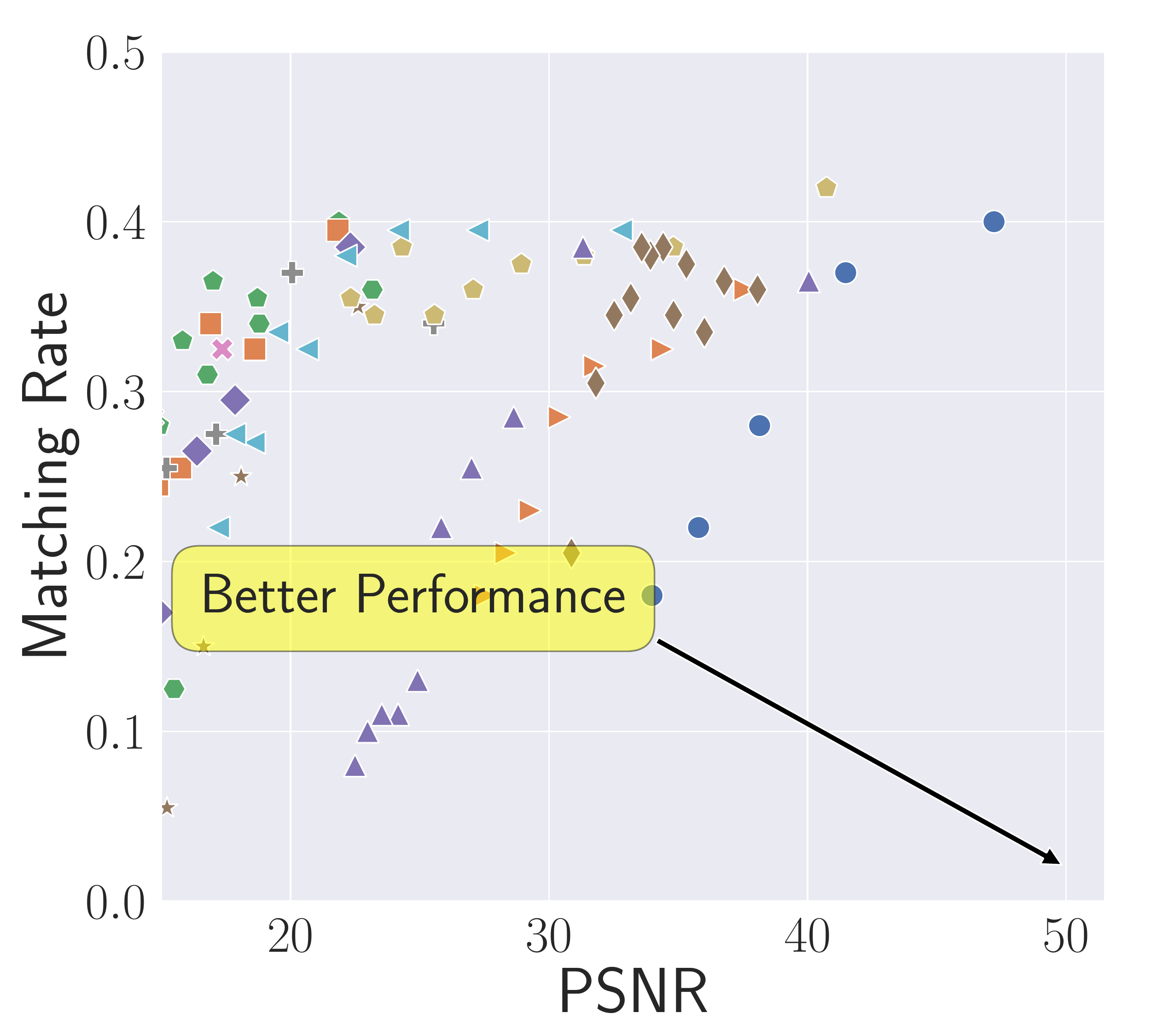}}
    \caption{Comparison between all baseline methods and Cloak v1/v4 on real images. The different points of each method represent different budgets.}
    \label{fig:real_baseline}
\end{figure}

\section{Possible Adaptive Adversary}
\label{sec:possible_ada}
Here, we explore four possible adaptive adversaries and empirically evaluate the performance of \system on real facial images. 
We conduct extensive experiments under the black-box scenario against optimization-based and hybrid inversion, i.e., Cloak v1 and Cloak v4.
Note that for the purpose of straightforward comparisons, we average the performance of \system with a varying number of distance budgets, i.e., $\varepsilon$-0/1/2/3.

\mypara{Cloak Overwriting}
This adaptive adversary aims to disturb the cloaks, i.e., the imperceptible perturbation searched by \system.
The adversary samples random noise from a Gaussian distribution $\mathcal{N}(\mu, \sigma^{2})$ to overwrite the cloaks. 

We report the matching rate by varying the standard deviation $\sigma$ (set $\mu$ as 0 for simplicity) in \subfig{fig:real-adaptive-overwrite} (see more results of Cloak v1 in Appendix \subfig{fig:real-adaptive-appendix-overwrite}).
We can observe that as the standard deviation increases, the matching rate of cloak overwriting is significantly reduced.
The reason is that the cloak overwriting actually introduces more noise in the image space on top of the imperceptible noise searched by the \system, which further jeopardizes the GAN inversion process.
These results indicate that cloak overwriting is not an applicable adaptive strategy for adversaries.

\mypara{Cloak Purification}
This adaptive adversary aims to remove or purify the cloaks searched by \system.
As aforementioned, these cloaks actually are the imperceptible noise added to the images.
Thus, we consider one of the most wide-used and easy-to-apply image noise reduction mechanisms, i.e., Spatial Smoothing~\cite{SpatialSmoothing}.
Spatial Smoothing means that pixel values are averaged with their neighboring pixel values with a low-pass filter, leading to the sharp "edges" of the image becoming blurred and the spatial correlation within the data becoming more apparent.

We report the matching rate by varying the filter widths of Spatial Spatial in \subfig{fig:real-adaptive-remove} (see more results of Cloak v1 in Appendix \subfig{fig:real-adaptive-appendix-remove}).
We can clearly observe that the matching rate increases at first and then decreases.
These results indicate that Spatial Smoothing indeed can purify the imperceptible noise added by \system to some extent.
We should also note that even the optimal setting for Spatial Smoothing can only lead to a slightly increased matching rate, and they all drop further sharply when the filter width is very large, as the Spatial Smoothing destroys the pixel space of the original image.
This observation implies that Spatial Smoothing is only a slightly effective adaptive strategy to reduce the jeopardy of \system to GAN inversions.

\mypara{More Iterations of Inversion}
This adaptive adversary has significant computational resources to perform a huge number of optimization iterations to increase the matching rate.
More specifically, we vary the number of optimization iterations from 0 to 5000 for both optimization-based and hybrid inversions.
Note that the default settings for the number of iterations are 500 and 100 for optimization-based inversion and hybrid inversion, respectively.

\subfig{fig:real-adaptive-iterations} shows the matching rate of \system under the effect of numbers of iterations.
As expected, we can find that the matching rate increases with the number of optimization iterations.
Specifically, the matching rate increases sharply up to 1000/100 iterations and continues to increase slowly afterward.
These results clearly demonstrate that more iterations of inversion indeed can reduce the jeopardy of \system to GAN inversions.
We should also note that a larger number of iterations (even up to 5000) does not lead to great effects, but is a huge cost in terms of resource usage.

\mypara{Encoder Enhancement}
We further consider another adaptive adversary where the adversary retrains the encoder to be more robust to imperceptible noise searched by \system.
More concretely, we assume that the adversary can collect a large number of cloaked images from crawler-accessible websites or social media.
We consider various numbers of cloaked images from 5k to 35k that an adversary can collect.
Note that the number of images in the full FFHQ dataset used to train StyleGANv2 is only 70k.
Then the adversary retrains the encoder by a mixed set of original clean images and collected cloaked images.

Since the encoder is only employed for hybrid inversion, we only consider here Cloak v4, the black-box setting against hybrid inversion, for evaluation.
\subfig{fig:real-adaptive-encoder} reports the matching rate under the effect of the different numbers of cloaked images collected by the adversary.
We can observe that the matching rate decreases slightly with increasing cloaked images, which means that retraining the encoder increases the jeopardy of \system to GAN inversion.
In a nutshell, encoder enhancement is not an applicable adaptive strategy for adversaries to reduce the jeopardy of \system to GAN inversions. 
We also delve into the reasons behind this observation from the perspective of the retraining algorithm in \autoref{why_encoder_fails}.

\begin{figure}[t]
    \centering
      \subfloat[Optimization-based \label{real_optim_unganable_fawkes}]{
    \includegraphics[width=0.5\linewidth]{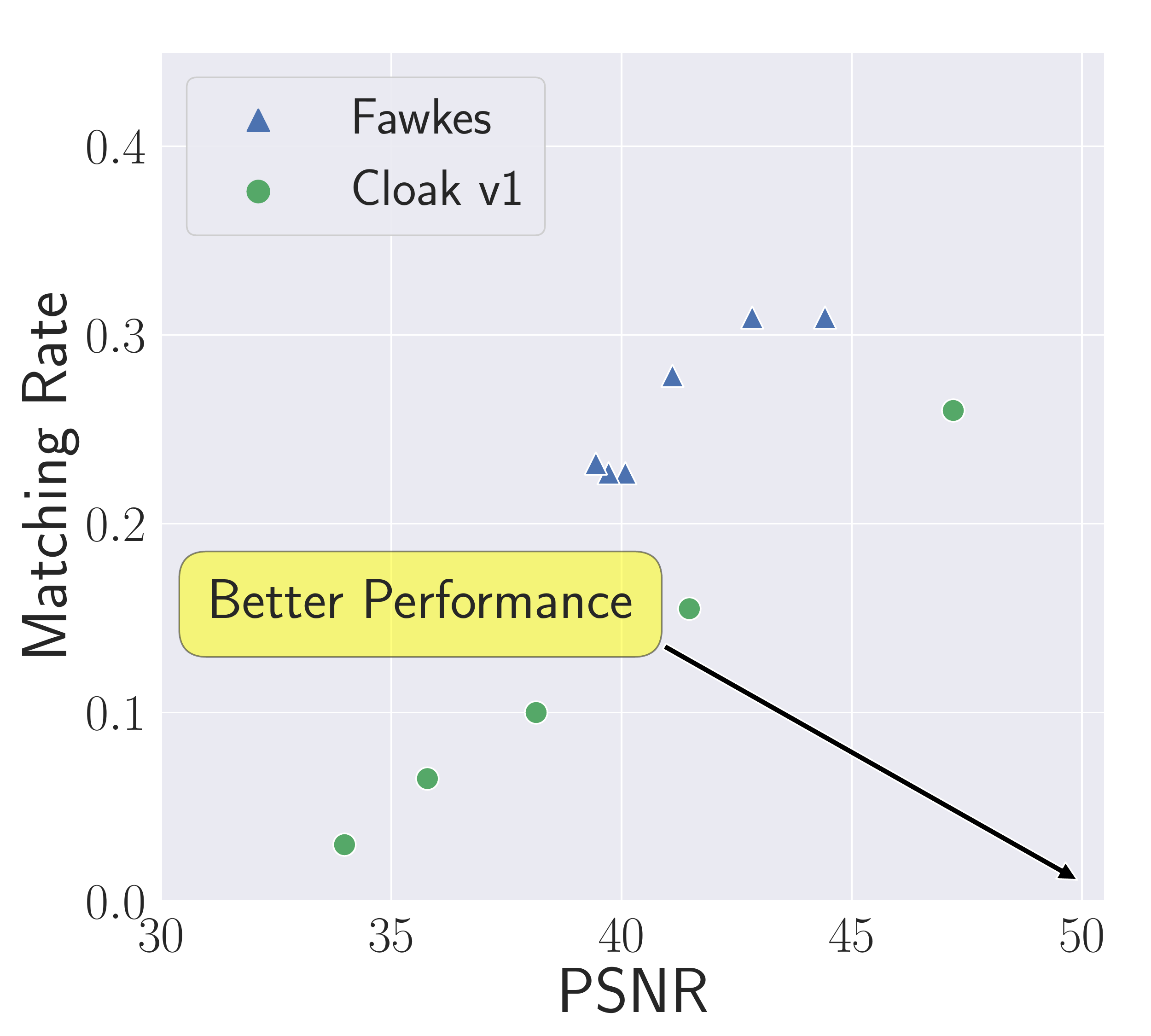}}
    \subfloat[Hybrid\label{real_hybrid_unganable_fawkes}]
    {
    \includegraphics[width=0.49\linewidth]{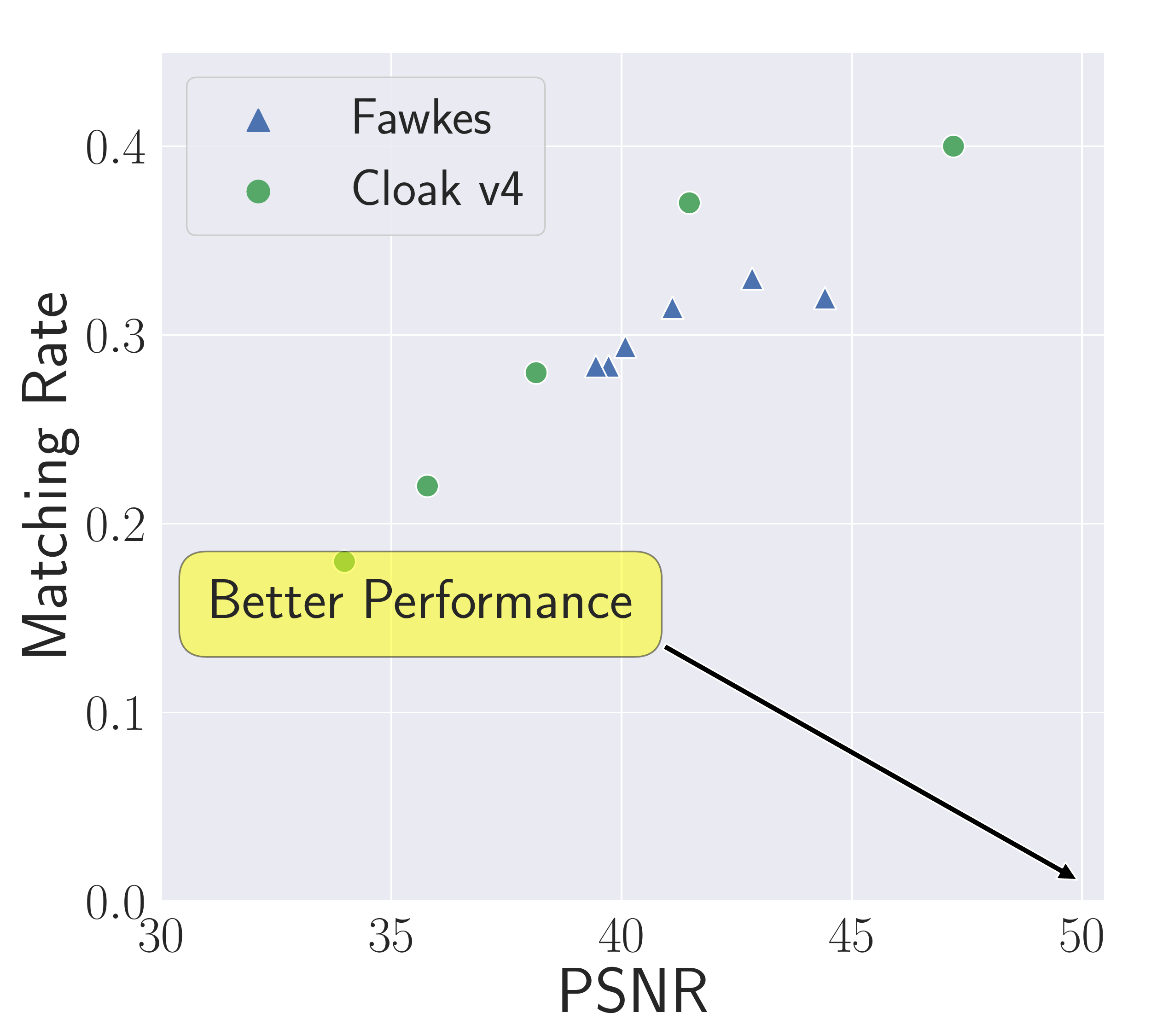}}
    \caption{Comparison between Fawkes and Cloak v1/v4 on real images. The different points of each method represent different budgets.}
    \label{fig:real_scatter_unganable_fawkes}
\end{figure}

\section{Discussion}
\label{sec:discussion}
\mypara{Comparison with Fawkes} 
Recently, the countermeasures which aim to protect faces from being stolen by recognition systems have been studied.
Fawkes~\cite{SWZLZZ20}, one of the representative works, adds pixel-level perturbations to users’ photos by altering the feature space before uploading them to the Internet. 
The functionality of unauthorized facial recognition models trained on these photos with perturbations will be deteriorated seriously.

For a convincing evaluation, we leverage the original implementation of Fawkes to protect the same real facial images as used in above evaluation.\footnote{\url{https://github.com/Shawn-Shan/fawkes}}
We set multiple different perturbation budgets to perturb these real images and evaluate the performance of Fawkes against both optimization-based and hybrid inversions.
\autoref{fig:real_scatter_unganable_fawkes} displays the comparison between Fawkes and Cloak v1/v4.
First, we can observe that Fawkes indeed can jeopardize the process of GAN inversions.
Further, we can also see that Fawkes provides much worse protection against optimization-based inversion, and similar or slightly better protection against hybrid inversion, compared to \system.
We provide a more in-depth analysis of the reasons behind these observations in \autoref{reason_fawkes}.

Here, we emphasize that except for the special black-box settings, we also propose white-box and gray-box settings, i.e., Cloak v0/v2/3.
The extensive evaluation in \autoref{fig:effectiveness-0} and \autoref{fig:effectiveness-234} shows that Cloak v0/v2/3 actually achieves better performance than Cloak v1/v4, especially in Hybrid inversion, which is naturally better than Fawkes.
That is, \system performs better than Fawkes in most cases.
More importantly, we should note that the goals of Fawkes and \system are totally different: Fawkes aims to mislead the face recognition classifiers while \system misleads the GAN inversion to prevent malicious face manipulation.

\mypara{Limitation} 
There are two major paradigms for image manipulation: GAN-inversion-based and image-translation-based. 
The latter, represented by StarGANv2~\cite{CUYH20} and AttGAN~\cite{HZKSC19}, transforms an image from the source domain to the target domain without the GAN-inversion process.
Therefore, our proposed \system is not applicable to image-translation-based manipulation, as the key idea of \system is to jeopardize the process of GAN inversion.
Moreover, we emphasize here that GAN-inversion-based and image-translation-based are two orthogonal image manipulation techniques.
Considering that the defense against the latter has been well studied~\cite{RBS20,YCTW20,HZZZY21,LYWL19}, the defense against GAN-inversion-based is still an open research problem. 
Our work is therefore well-motivated to complete this puzzle map.

Moreover, except for $\mathbf{z}$ space we consider in this work, recent works~\cite{ZSZZ20,AQW19,AQW20,TANPC21,KLAHLA20,PLXSA22,ATMGB22,WZFWC22,DTNH22} also works on $\mathbf{w}$ space, which is transformed from $\mathbf{z}$ space, leading to a better inversion performance.
We leave the in-depth exploration of more efficient \system against $\mathbf{w}$ space for future work.

\section{Conclusion}
\label{sec8:conclusion}
In this paper, we take the first step towards defending against GAN-inversion-based face manipulation by proposing \system, a system that can jeopardize the process of GAN inversion.
We consider two advanced GAN inversions: optimization-based and hybrid inversions, as well as five scenarios to comprehensively characterize the defender's background knowledge in multiple dimensions.
We extensively evaluate \system on four popular GAN models built on two benchmark face datasets of different sizes and complexity.
The results show that \system can achieve remarkable performance with respect to both effectiveness and utility. 
We further conduct a comparison of \system with thirteen image distortion methods as well as Fawkes, and the results show that \system generally outperforms all these methods.
In addition, we explore four possible adaptive adversaries against \system, and empirical evaluation shows that Spatial Smoothing and more iterations of inversion are slightly effective.

\section*{Acknowledgements}
We thank all anonymous reviewers for their constructive comments.
This work is partially funded by the Helmholtz Association within the project ``Trustworthy Federated Data Analytics'' (TFDA) (funding number ZT-I-OO1 4).

\bibliographystyle{plain}
\bibliography{normal_generated_py3}

\newpage
\appendix
\section{Face Manipulation Attack}
\label{appendx:face_edit}
\begin{table}[!t]
    \centering
    \caption{List of notations.}
    \scalebox{0.75}
    {
    \begin{tabular}{c|l}
    \toprule
        Notation & Description \\
         \midrule
         $\code$ & Latent code\\
         $\targetX$& Target image (uncloaked) \\
         $\cloakedX$  & Cloaked version of the target image $\targetX$ \\
         $\delta$   & Cloak (or perturbation) between $\targetX$ and $\cloakedX$ \\
         $\varepsilon$ & Perturbation budget\\
         $\kappa$ & Trade-off hyperparameter\\
         $I$      & GAN inversion technique\\
         $G$      & Generator \\
         $E$      & Encoder for the latent space\\
         $F$      & Feature extractor for the feature space \\
         $\mathcal{L}$  & Loss function\\
         $\advG$ & Target generator controlled by the adversary\\
         $\userG$ & Shadow generator controlled by the defender\\
         $\advE$  & Target encoder controlled by the adversary\\
         $\userE$ & Shadow encoder controlled by the defender\\
         $I_{\text{o}}$ & Optimization-based inversion\\
         $I_{\text{h}}$ & Hybrid inversion\\
         $\Lrec$ & Reconstruction loss\\
         $\Lper$ & Perceptual loss \\
         $\Lcos$ & Cosine similarity loss\\
         $\Lmse$ & MSE similarity loss\\
         \bottomrule
    \end{tabular}
    }
    \label{tab:notions}
\end{table}

\begin{figure}[t]
    \centering
    \includegraphics[width=1\columnwidth]{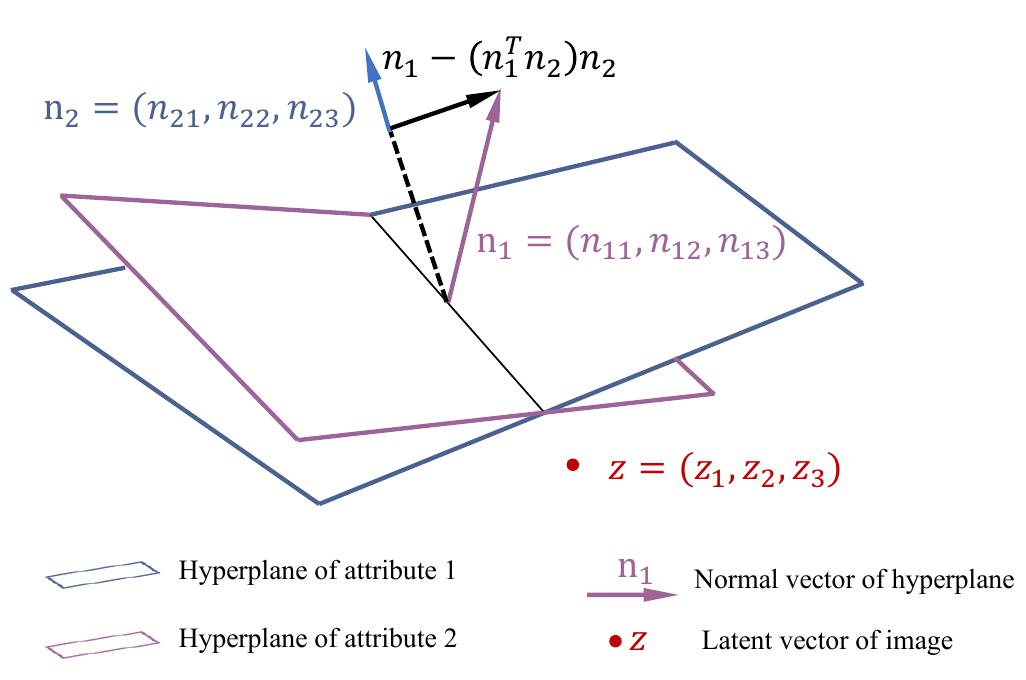}
    \caption{Illustration of the conditional manipulation in latent space.}
    \label{fig:edit_subspace}
\end{figure}

To demonstrate how face manipulation works, we consider a representative manipulation technique, namely InterFaceGAN~\cite{SGTZ20}, to perform algebraic operations in latent space.
See \autoref{fig:edit_subspace} for an illustration of the algebraic operations in latent space.

For any binary semantic (e.g, smiling \textit{v.s.} non-smiling), there is a hyperplane (determined by trained linear SVM) in the latent space serving as the separation boundary.
When the latent code walks on the same side of the hyperplane, the semantic remains the same, but when it crosses the boundary along the normal vector $\mathbf{n}$, the semantic turns into the opposite.
For multiple semantics, as shown in \autoref{fig:edit_subspace}, given two hyperplanes with normal vectors $\mathbf{n}_{1}$ and $\mathbf{n}_{2}$, there exists a projection direction $\mathbf{n}_{1}-\left(\mathbf{n}_{1}^{T} \mathbf{n}_{2}\right) \mathbf{n}_{2}$ such that moving the latent code along this new direction changes "attribute 1" without affecting "attribute 2".

\section{Algorithms of GAN inversions}
\label{appendix:inversion_algorithms}
Here, we present the algorithms of GAN inversions we consider in this paper. 
The algorithm of optimization-based inversion can be found in \autoref{algo:opti}.
The algorithm of hybrid inversion can be found in \autoref{algo:hybrid}.

\section{Algorithms of \system}
\label{appendix:unganable_algorithms}
Here, we present all the algorithms of \system. The algorithm of Cloak-0 can be found in \autoref{algo:cloak-0}. The algorithm of Cloak-1 can be found in \autoref{algo:cloak-14}. The algorithm of Cloak-2 can be found in \autoref{algo:cloak-2}. The algorithm of Cloak-3 can be found in \autoref{algo:cloak-3}. The algorithm of Cloak-4 can be found in \autoref{algo:cloak-14}.

\begin{algorithm}[!ht] 
    \caption{Latent Space Embedding for GANs of Optimization-based Inversion.}
    \label{algo:opti} 
    \KwIn{An image $\mathbf{x}$ to embed; a pre-trained generator $G_{\text{t}}(\cdot)$.}
    \KwOut{The embedded latent code $\mathbf{z}^{*}$ and the embedded image $G_{\text{t}}(\mathbf{z}^{*})$ via $F^{'}$}  
        Initialize $\mathbf{z}^{*}$=$\mathbf{z}$\;
     \For{number of optimized iterations}
    { 
$L \leftarrow L_{\text {percept }}\big(G_{\text{t}}\left(\mathbf{z}^{*}\right), \mathbf{x}\big)+\frac{1}{N}\left\|G_{\text{t}}\left(\mathbf{z}^{*}\right)-\mathbf{x}\right\|_{2}^{2}$ ; \\
$\mathbf{z}^{*} \leftarrow \mathbf{z}^{*}-\eta F^{\prime}\left(\nabla_{\mathbf{z}^{*}} L\right)$
    }
\end{algorithm} 

\begin{algorithm}[!htp] 
    \caption{Latent Space Embedding for GANs of Hybrid Inversion.}
    \label{algo:hybrid} 
    \KwIn{A image dataset $\mathcal{D}$; a encoder $E_{\text{t}}$ to train; a discriminator $D$; minibatch $m$; hyperparameter $\lambda_{\text{vgg}}$, $\lambda_{\text{adv}}$, $\lambda$; an image $\mathbf{x}$ to embed; a pre-trained generator $G_{\text{t}}(\cdot)$; a pre-trained VGG feature extraction model $F(\cdot)$.}
    \KwOut{The embedded latent code $\mathbf{z}^{*}$ and the embedded image $G_{\text{t}}(\mathbf{z}^{*})$ via $F^{'}$}  
        $\lambda_{vgg}=5 e^{-5}, \lambda_{adv}=0.1, \text { and } \gamma=10$\;
     \For{number of training epochs}
     {
        sample a minibatch of samples $\mathbf{x}^{\text {real }}$ from $\mathcal{D}$;\\
$\min _{\Theta_{E_{\text{t}}}} \mathcal{L}_{E_{\text{t}}}=\left\|\mathbf{x}^{\text {real }}-G_{\text{t}}\left(E_{\text{t}}\left(\mathbf{x}^{\text {real }}\right)\right)\right\|_{2} +\lambda_{\text{vgg}}\left\|F\left(\mathbf{x}^{\text {real }}\right)-F\left(G_{\text{t}}\left(E_{\text{t}}\left(\mathbf{x}^{\text {real }}\right)\right)\right)\right\|_{2}-\lambda_{\text {adv }} \underset{\mathbf{x}^{\text {real }} \sim \mathcal{D}}{\mathbb{E}}\left[D\left(G_{\text{t}}\left(E_{\text{t}}\left(\mathbf{x}^{\text {real }}\right)\right)\right)\right]$;\\

$\min _{\Theta_{D}} \mathcal{L}_{D} = \underset{\mathbf{x}^{\text {real }}\sim \mathcal{D}}{\mathbb{E}}[D(G_{\text{t}}(E_{\text{t}}(\mathbf{x}^{\text{real}})))]-\underset{\mathbf{x}^{\text {real }}\sim \mathcal{D}}{\mathbb{E}}[D(\mathbf{x}^{\text{real}})]+\frac{\gamma}{2} \underset{\mathbf{x}^{\text {real }}\sim \mathcal{D}}{\mathbb{E}}[\left\|\nabla_{\mathbf{x}} D\left(\mathbf{x}^{real}\right)\right\|_{2}^{2}]$;\\
     }
     Initialize $\mathbf{z}^{*}$=$\mathbf{z}$\;
     
     \For{number of optimized iterations}
    { 
 
$L \leftarrow L_{\text {percept }}\left(G_{\text{t}}\left(\mathbf{z}^{*}\right), \mathbf{x}\right)+\frac{1}{N}\left\|G_{\text{t}}\left(\mathbf{z}^{*}\right)-\mathbf{x}\right\|_{2}^{2}$ ; \\
$\mathbf{z}^{*} \leftarrow \mathbf{z}^{*}-\eta F^{\prime}\left(\nabla_{\mathbf{z}^{*}} L\right)$
    }
\end{algorithm} 

\begin{algorithm}[!htp]
    \caption{Cloaking Facial Image of Cloak-0}
    \label{algo:cloak-0} 
    \KwIn{A target image $\mathbf{x}$ to cloak; a pre-trained target generator $G_{\text{t}}(\cdot)$; a shadow encoder $E_{\text{s}}(\cdot)$; a pre-trained ResNet feature extractor $F$; cosine similarity $\mathcal{L}_{\text{cos}}(\cdot,\cdot)$; MSE similarity $\mathcal{L}_{\text{mse}}(\cdot,\cdot)$; minibatch $m$; perturbation budget $\varepsilon$; trade-off $\kappa$.}
    \KwOut{The trained shadow encoder $E_{\text{s}}$ and the cloaked image $\hat{\mathbf{x}}$.}  
    
        Initialize $\mathcal{L}_{\text{rec}}(\cdot,\cdot)=-\mathcal{L}_{\text{cos}}(\cdot,\cdot) + \mathcal{L}_{\text{mse}}(\cdot,\cdot)$\;
    
    \For{number of training iterations}
    {
    sample a minibatch of latent codes $\mathbf{z}^{'}\in \mathcal{N}(0,1)$;\\
    
    $\min _{\Theta_{E_{\text{s}}}} \mathcal{L}_{\text{rec}}\Big(E_{\text{s}}\big(G_{\text{t}}(\mathbf{z}^{'}))\big), \mathbf{z}^{'}\Big)$
    }
    Initialize $\mathbf{x}_{\text{t}}$ = \autoref{algo:opti}($\mathbf{x}$);\\
    Initialize $\delta \in \mathcal{N}(0,1)$ and $|\delta|_{\infty}<\epsilon$;\\
    Initialize $\kappa$;\\
     \For{number of optimized iterations}
    { 
$\max _{\delta} \kappa \Big( \mathcal{L}_{\text{rec}}\big(E_{\text{s}}(\mathbf{x}+\delta), \mathbf{x}_{\text{t}}\big) \Big)  +(1-\kappa )\Big( \mathcal{L}_{\text{rec}}\big(F(\mathbf{x}+\delta), F(\mathbf{x})\big) \Big) $ ; \\
clip $\delta$ for $|\delta|_{\infty}<\epsilon$;\\
clip $\mathbf{x}+\delta$ for $\mathbf{x}+\delta \in [0, 1]$;
    }
$\hat{\mathbf{x}}=\mathbf{x}+\delta$;\\
return $E_{\text{s}}, \hat{\mathbf{x}}$
\end{algorithm} 

\section{Limitations of Exiting Defenses}
\label{limitation_exiting_defense}
As face manipulation causes a great threaten to individual privacy even political security, it is of paramount importance to develop countermeasures against it. 
To mitigate this risk, many defenses have been proposed, and these defenses can be broadly divided into two categories: detection~\cite{LBZYCWG20, RCVRTN19,ANYE18,ZHMD17,MRS19,NYE19} and disrupting I2I~\cite{RBS20,YCTW20,HZZZY21,LYWL19}.
Here, I will elaborate on detection and disrupting, respectively.

\begin{algorithm}[!tp] 
    \caption{Cloaking Facial Image of Cloak-1/4}
    \label{algo:cloak-14} 
    \KwIn{A target image $\mathbf{x}$ to cloak; a pre-trained ResNet feature extractor $F$; cosine similarity $\mathcal{L}_{\text{cos}}(\cdot,\cdot)$; MSE similarity $\mathcal{L}_{\text{mse}}(\cdot,\cdot)$; perturbation budget $\varepsilon$.}
    \KwOut{The cloaked image $\hat{\mathbf{x}}$.}  
    
    Initialize $\mathcal{L}_{\text{rec}}(\cdot,\cdot)=-\mathcal{L}_{\text{cos}}(\cdot,\cdot) + \mathcal{L}_{\text{mse}}(\cdot,\cdot)$;\\
    Initialize $\delta \in \mathcal{N}(0,1)$ and $|\delta|_{\infty}<\epsilon$;\\
    
     \For{number of optimized iterations}
    { 
$\max _{\delta}  \mathcal{L}_{\text{rec}}\big(F(\mathbf{x}+\delta), F(\mathbf{x})\big)  $ ; \\
clip $\delta$ for $|\delta|_{\infty}<\epsilon$;\\
clip $\mathbf{x}+\delta$ for $\mathbf{x}+\delta \in [0, 1]$;
    }
$\hat{\mathbf{x}}=\mathbf{x}+\delta$;\\
return $\hat{\mathbf{x}}$
\end{algorithm} 

\begin{algorithm}[!tp] 
    \caption{Cloaking Facial Image of Cloak-2}
    \label{algo:cloak-2} 
    \KwIn{A target image $\mathbf{x}$ to cloak; a pre-trained target encoder $E_{\text{t}}(\cdot)$; a pre-trained ResNet feature extractor $F$; cosine similarity $\mathcal{L}_{\text{cos}}(\cdot,\cdot)$; MSE similarity $\mathcal{L}_{\text{mse}}(\cdot,\cdot)$; perturbation budget $\varepsilon$; trade-off $\kappa$.}
    \KwOut{The cloaked image $\hat{\mathbf{x}}$.}  
    
    Initialize $\mathcal{L}_{\text{rec}}(\cdot,\cdot)=-\mathcal{L}_{\text{cos}}(\cdot,\cdot) + \mathcal{L}_{\text{mse}}(\cdot,\cdot)$\;
    Initialize $\delta \in \mathcal{N}(0,1)$ and $|\delta|_{\infty}<\epsilon$;\\
    Initialize $\kappa$;\\
     \For{number of optimized iterations}
    { 
$\max _{\delta} \kappa \Big( -\mathcal{L}_{\text{rec}}\big(E_{\text{t}}(\mathbf{x}+\delta), 0\big) \Big)  +(1-\kappa )\Big( \mathcal{L}_{\text{rec}}\big(F(\mathbf{x}+\delta), F(\mathbf{x})\big) \Big) $ ; \\
clip $\delta$ for $|\delta|_{\infty}<\epsilon$;\\
clip $\mathbf{x}+\delta$ for $\mathbf{x}+\delta \in [0, 1]$;
    }
$\hat{\mathbf{x}}=\mathbf{x}+\delta$;\\
return $\hat{\mathbf{x}}$
\end{algorithm} 

\begin{algorithm}[!t] 
    \caption{Cloaking Facial Image of Cloak-3}
    \label{algo:cloak-3} 
    \KwIn{A target image $\mathbf{x}$ to cloak; a pre-trained target encoder $E_{\text{t}}(\cdot)$; a shadow encoder $E_{\text{s}}$; a shadow generator $G_{\text{s}}$; a pre-trained ResNet feature extractor $F$; cosine similarity $\mathcal{L}_{\text{cos}}(\cdot,\cdot)$; MSE similarity $\mathcal{L}_{\text{mse}}(\cdot,\cdot)$; perturbation budget $\varepsilon$; trade-off $\kappa$.}
    \KwOut{The trained shadow encoder $E_{\text{s}}$, the trained shadow generator $G_{\text{s}}$ and the cloaked image $\hat{\mathbf{x}}$.}
    
    Initialize $\mathcal{L}_{\text{rec}}(\cdot,\cdot)=-\mathcal{L}_{\text{cos}}(\cdot,\cdot) + \mathcal{L}_{\text{mse}}(\cdot,\cdot)$\;
    \For{number of training iterations}{
    
    sample a minibatch of latent codes $\mathbf{z}^{'}\in \mathcal{N}(0,1)$;\\
    
    $\min _{\Theta_{E_{\text{s}}}} \mathcal{L}_{\text{rec}}\Big(E_{\text{s}}\big(G_{\text{s}}(\mathbf{z}^{'}))\big), \mathbf{z}^{'}\Big)$;\\
    
    $\max _{\Theta_{G_{\text{s}}}} \mathcal{L}_{\text{rec}}\Big(E_{\text{s}}\big(G_{\text{s}}(\mathbf{z}^{'}))\big), \mathbf{z}^{'}\Big)$;\\
    }
    Initialize $\delta \in \mathcal{N}(0,1)$ and $|\delta|_{\infty}<\epsilon$;\\
    Initialize $\kappa$;\\
     \For{number of optimized iterations}
    { 
$\max _{\delta} \kappa \Big( -\mathcal{L}_{\text{rec}}\big(E_{\text{s}}(\mathbf{x}+\delta), 0\big) \Big)  +(1-\kappa )\Big( \mathcal{L}_{\text{rec}}\big(F(\mathbf{x}+\delta), F(\mathbf{x})\big) \Big) $ ; \\
clip $\delta$ for $|\delta|_{\infty}<\epsilon$;\\
clip $\mathbf{x}+\delta$ for $\mathbf{x}+\delta \in [0, 1]$;
    }
$\hat{\mathbf{x}}=\mathbf{x}+\delta$;\\
return $E_{\text{s}}, G_{\text{s}},\hat{\mathbf{x}}$
\end{algorithm} 

\mypara{Detection}
This type of defense is designed in a passive manner to detect whether face images have been tampered with after wide propagation.
That is, the adversary has completed the manipulation of the target individual's face, and the manipulated face image has already spread online, i.e., the harm to the individual has already been caused.
In contrast, \system defends against GAN-inversion-based face manipulation in a proactive manner.
Concretely, \system focuses on defending GAN inversion, an essential step for face manipulation. 
\system can jeopardize the GAN inversion process, thus the following latent code manipulation step will not achieve the ideal result.
Thus, \system protects the individual from malicious face manipulation in a proactive manner.

\mypara{Disrupting}
There are two major paradigms for image manipulation: GAN-inversion-based and image-translation-based. 
Disrupting is only applicable for the latter manipulation technique.
Typically, image-translation-based leverages an end-to-end network to transform an image from the source domain to the target domain.
Disrupting mitigate image-translation-based face manipulation by spoofing the end-to-end network.
However, there is still no approach to defend against GAN-inversion-based face manipulation. 
Here, we emphasize that GAN-inversion-based and image-translation-based are two orthogonal image manipulation techniques.
Compared to the latter, GAN-inversion-based enjoys unique advantages: on-manifold manipulation, gradual manipulation effect, image interpolation, etc.
Considering that the defense against the latter has been well studied~\cite{RBS20,YCTW20,HZZZY21,LYWL19}, the defense against GAN-inversion-based is still an open research problem. 
Our work is therefore well-motivated to complete this puzzle map.

\section{GAN Models and Datasets}
\label{appendix:GAN_dataset}
\mypara{DCGAN} 
DCGAN~\cite{RMC16}  uses convolutions in the discriminator and fractional-strided convolutions in the generator.

\mypara{WGAN} 
WGAN~\cite{GAADC17} minimizes the Wasserste in distance between the generated and real data distributions, which offers more model stability and makes the training process easier.

\mypara{StyleGANv1/v2} 
StyleGANv1~\cite{KLA19} implicitly learns hierarchical latent styles for image generation. The StyleGAN generator takes per-block incorporation of style vectors (defined by a mapping network) and stochastic variation (provided by the noise layers) as inputs, instead of samples from the latent space, to generate a synthetic image. The StyleGANv2~\cite{KLAHLA20} further improves the image quality by proposing weight demodulation, path length regularization, redesigning generator, and removing progressive growing.

\mypara{CelebA} 
CelebA~\cite{LLWT15} is a large-scale face attributes dataset consisting of 200K celebrity images with 40 attribute annotations each. 

\mypara{FFHQ} 
Flickr-Faces-HQ (FFHQ)~\cite{KLA19,KLAHLA20} is a high-quality image dataset of human faces crawled from Flickr, which consists pixels and contains considerable variation in terms of age, of 70,000 high-quality human face images of 1024 × 1024 ethnicity, and image background.

\section{Metrics of Utility}
\label{appendix:metrics}
\mypara{MSE} 
Mean squared error (MSE) is the most commonly used loss function for reconstruction. More concretely, for two images $I$ and $\hat{I}$ with $N$ pixels, the MSE is given by
\begin{equation}
\operatorname{MSE}(I, \hat{I})=\frac{1}{N} \sum_{i=0}^{N}\left(I-\hat{I}_{i}\right)^{2}
\end{equation}

\begin{table}[!t]
    \centering
    \caption{The exact settings of $\varepsilon$ and $\kappa$.
    These settings may vary in terms of the target images and other factors.}
    \tabcolsep 3pt
    \scalebox{0.75}
    {
    \begin{tabular}{c|c|c|c|c|c|c}
    \toprule
          \multirow{2}{*}{Budget} &\multirow{2}{*}{Cloaks}&\multirow{2}{*}{$\varepsilon$}& \multicolumn{4}{c}{$\kappa$} \\
          &&&DCGAN & WGAN & StyleGANv1 & StyleGANv2 \\
         \midrule
         \multirow{3}{*}{$\varepsilon$-0} & Cloak v0 & 0.01 & 0.7 & 0&0.2&1\\
                                    & Cloak v2 & 0.01 & 0.1&0.4&0.5&0.4\\
                                    & Cloak v3 & 0.01 & 0.2&0.1&0.5&1\\
         \midrule
        \multirow{3}{*}{$\varepsilon$-1} & Cloak v0 & 0.0167 & 0& 0.7&0.2&0.4\\
                                    & Cloak v2 & 0.02 & 0.5&0.1&1&1\\
                                    & Cloak v3 & 0.02 & 0.8&0.9&0&0.1\\
        \midrule
        \multirow{3}{*}{$\varepsilon$-2} & Cloak v0 & 0.0233 & 0.3& 0.4&0.9&0.6\\
                                    & Cloak v2 & 0.03 & 0&0.1&1&1\\
                                    & Cloak v3 & 0.03 & 0.9&0.7&0&1\\
        \midrule
        \multirow{3}{*}{$\varepsilon$-3} & Cloak v0 & 0.03 & 0.2& 0.6&0.9&0\\
                                    & Cloak v2 & 0.04 & 0.8&0.7&1&1\\
                                    & Cloak v3 & 0.04 & 0.7&0.3&0.6&0.3\\
        \midrule
        \multirow{3}{*}{$\varepsilon$-4} & Cloak v0 & 0.0367 & 0.2& 0.8&0.5&0.3\\
                                    & Cloak v2 & 0.05 & 0.3&0.3&1&1\\
                                    & Cloak v3 & 0.05 & 0.9&0.1&0&0.6\\
        \midrule
         \multirow{3}{*}{$\varepsilon$-5} & Cloak v0 & 0.0433 & 0.3& 0.5&0&0.7\\
                                    & Cloak v2 & 0.06 & 0.2&0.1&1&1\\
                                    & Cloak v3 & 0.06 & 0.0&0.3&00.6&1\\
         \midrule
       \multirow{3}{*}{$\varepsilon$-6} & Cloak v0 & 0.05 & 0.4& 0.2&1&0.2\\
                                    & Cloak v2 & 0.07 & 0.8&0&1&1\\
                                    & Cloak v3 & 0.07 & 0.8&0.6&0.9&0.7\\
        \midrule
       \multirow{3}{*}{$\varepsilon$-7} & Cloak v0 & 0.0567 & 0.2& 0.2&1&1\\
                                    & Cloak v2 & 0.08 & 0.6&0.1&1&1\\
                                    & Cloak v3 & 0.08 & 0.7&0&1&0.8\\
        \midrule
        \multirow{3}{*}{$\varepsilon$-8} & Cloak v0 & 0.0633 & 0.6& 0&1&1\\
                                    & Cloak v2 & 0.09 & 0.3&0.5&1&1\\
                                    & Cloak v3 & 0.09 & 0.1&0&1&0.9\\
        \midrule
        \multirow{3}{*}{$\varepsilon$-9} & Cloak v0 & 0.07 & 0.6& 0.5&1&1\\
                                    & Cloak v2 & 0.1 & 0.2&0.9&1&1\\
                                    & Cloak v3 & 0.1 & 0.3&0&0.9&0.9\\
         \bottomrule
    \end{tabular}
    }
    \label{tab:eps_kappa}
\end{table}

\begin{table*}[!t]
    \centering
    \caption{List of all image distortions.}
    \scalebox{0.75}
{
    \begin{tabular}{c|l}
    \toprule
        Notation & Description \\
         \midrule
         ShearX(Y) & Shear the image along the horizontal (vertical) axis with rate \emph{magnitude}. \\
         \hline
         TranslateX(Y) & Translate the image in the horizontal (vertical) direction by \emph{magnitude} number of pixels.\\
         \hline
         Rotate &  Rotate the image \emph{magnitude} degrees. \\
         \hline
         Brightness & \tabincell{l}{ Adjust the brightness of the image. A \emph{magnitude}=0 gives a black image, whereas \emph{magnitude}=1 \\gives the original image. } \\
         \hline
         Color & \tabincell{l}{Adjust the color balance of the image, in a manner similar to the controls on a colour TV set. \\A \emph{magnitude}=0 gives a black \& white image, whereas \emph{magnitude}=1 gives the original image.}   \\
         \hline
         Contrast & Maximize the image contrast, by making the darkest pixel black and lightest pixel white. \\
         \hline
         Solarize & Invert all pixels above a threshold value of \emph{magnitude}\\
         \hline
         CenterCrop & Crop the center part of the original image by \emph{magnitude} and resize it to the original size.\\
         \hline
         GaussianBlur & Blur the original image by a Gaussian function, typically to reduce image noise and reduce detail.\\
         \hline
         GaussianNoise & Add Gaussian-distributed additive noise on the original image.\\
         \hline
         JPEGCompression & \tabincell{l}{JPEGCompression uses the DCT (Discrete Cosine Transform) method for coding transformation.\\ It allows a trade-off between storage size and the degree of compression can be adjusted.}\\
         \bottomrule
    \end{tabular}
    }
    \label{tab:alldistortions}
\end{table*}

\mypara{SSIM} 
Structural Similarity (SSIM) measures the structural similarity between images based on independent comparisons in terms of luminance, contrast, and structures. For two images $I$ and $\hat{I}$ with $N$ pixels, the SSIM is given by 
\begin{equation}
\operatorname{SSIM}(I, \hat{I})=\left[\mathcal{C}_{l}(I, \hat{I})\right]^{\alpha}\left[\mathcal{C}_{c}(I, \hat{I})\right]^{\beta}\left[\mathcal{C}_{s}(I, \hat{I})\right]^{\gamma}
\end{equation}
where $\alpha, \beta, \gamma$ are control parameters for adjusting the relative importance. The details of these terms can be found in 

\mypara{PSNR} 
Peak Signal-to-Noise Ratio (PSNR) is one of the most widely-used criteria to measure the quality of reconstruction. The PSNR between the ground truth image $I$ and the reconstruction $\hat{I}$ is defined by the maximum possible pixel value of the image (denoted as $L$) and the mean squared error (MSE) between image:
\begin{equation}
\operatorname{PSNR}(I, \hat{I})=10 \cdot \log _{10}\left(\frac{L^{2}}{\frac{1}{N} \sum_{i=1}^{N}(I(i)-\hat{I}(i))^{2}}\right)
\end{equation}

\section{Detailed Assumptions for \system}
\label{assumption_defender}
Recall that GAN inversion actually works in three spaces, i.e., the image space, the feature space, and the latent space (see \autoref{sec3:intuition}).
These observations motivate our \system that aims to maximize deviations in both latent and feature spaces by adding imperceptible noise in image space.
Therefore, we elaborate on the assumptions from the latent space and feature space, respectively.
Note that, in the following discussion, ``original image''  represents the image the defender aims to protect and ``cloaked image'' represents the protected original image to which the defender has added imperceptible noise.

First, for optimization-based inversion, we consider two different scenarios to characterize the defender’s background knowledge.

\mypara{White-Box Against Optimization-based (Cloak v0)}
\begin{itemize}
    \item \mypara{Latent Space} The defender aims to maximize deviation between the cloaked image and its original image in the latent space.
    In other words, the defender needs to obtain the cloaked latent code and original latent code, and maximize deviation between them.
    
    To obtain the original latent code, we assume that the defender has white-box access to the target generator and knows the adversary's inversion technique (i.e., optimization-based inversion) to invert the original image to obtain its exact latent code. 
    
    To obtain the cloaked latent code, the defender needs to build an end-to-end model, namely shadow encoder to map the cloaked image of each iteration to its approximate latent code.
    The reason is that the defender iteratively optimize the cloaked image with respect to maximizing latent deviation.
    Thus, the defender needs to compute the gradient through backpropagation performed on an end-to-end encoder (see \autoref{optim_method} (\textbf{Cloak v0}) for how to build such an encoder), which is intractable through optimization-based inversion to achieve the above goal.

    Finally, the defender iteratively optimizes the cloaked image by maximizing latent deviation between the cloaked latent code and original latent code.
    Note that the original latent code is always fixed throughout the optimization process.
    
    \item \mypara{Feature Space} The defender also aims to maximize deviation between the cloaked image and its original image in the feature space.
    Thus, the defender needs to obtain the cloaked feature and original feature, and maximize deviation between them.
    To this end, we assume that the defender has access to a feature extractor to map the image to its feature.
    In this work, we adopt the easy-to-download, widely-used, and pre-trained ResNet-18 as the feature extractor. 
\end{itemize}

\mypara{Black-Box Against Optimization-based (Cloak v1)}
In this scenario, we assume the defender has no knowledge of the target generator or inversion technique, and the defender can only produce significant alterations to images’ feature space.
\begin{itemize}
\item \mypara{Feature Space} 
The defender can only access the feature extractor to map the image to its features.
\end{itemize}

For hybrid inversion, we consider three different scenarios to characterize a defender's background knowledge. 

\mypara{White-Box Against Hybrid Inversion (Cloak v2)} 
\begin{itemize}
\item \mypara{Latent Space}
Note that hybrid inversion leverages a learned predictive encoder (called target encoder) to map the image directly to the latent space and uses it as initialization for the subsequent optimization stage.
Furthermore, \autoref{fig:distri} shows the trend of perceptual and MSE loss of the subsequent optimization stage, respectively.
We can clearly observe that the loss trend remains constant when the initialization is set to zero, which means that it is difficult to converge when the initial latent code is set to zero.
This observation suggests a new perspective on the latent deviation – misleading the target encoder to provide zero initialization, or close to zero. 
In other words, our defense’s goal against hybrid inversion should be to force the output of the target encoder to zero. 
This is actually a special case of maximizing latent deviation, which provides the movement direction of the cloaked image in the latent space, i.e., towards zero.

To force the output of the target encoder to zero, we assume that the defender has full knowledge of the target encoder.
Thus the defender can feed the image to the target encoder to obtain its latent code and iteratively update the image by minimizing the deviation between the target encoder's output and zero.
\item \mypara{Feature Space} 
Same as in previous scenarios, we assume that the defender has access to a feature extractor to map the image to its feature.
Thus the defender can maximize deviation between the cloaked image and its original image in the feature space.
\end{itemize}

\mypara{Grey-Box (Cloak v3)}
In this scenario, the goal of the defender is the same as the Cloak v2, i.e., minimizing the deviation between the output of the target encoder (accepts cloaked images as input) and zero in the latent space and maximizing the deviation between the cloaked images and its original image in the feature space.
Here, we relax the assumption that the defender has complete knowledge of the target encoder. 

\begin{itemize}
\item \mypara{Latent Space}
In particular, we assume that the defender can send many queries to the target encoder and train a shadow encoder to mimic the behavior of the target encoder, and relies on the shadow encoder to act as the target encoder (see \autoref{sec:hybrid_method} (\textbf{Cloak v3}) for how to train the shadow encoder). 
Then the defender can feed the image to the shadow encoder to obtain its latent code and iteratively update the image by minimizing the deviation between the shadow encoder's output and zero.

\item \mypara{Feature Space} 
Same as in previous scenarios, we assume that the defender has access to a feature extractor to map the image to its feature.
Thus the defender can maximize deviation between the cloaked image and its original image in the feature space.
\end{itemize}

\mypara{Black-Box (Cloak v4)}
Here, we assume the defender has no knowledge of the adversary's generator or encoder. 
Here, the defender only has access to a feature extractor $F$.
Then the defender can maximize the deviation between the cloaked image and its original image in the feature space.

\section{Threshold Choosing}
\label{appendix:threshold}
Here, we focus on the threshold choosing to determine whether the original and reconstructed images belong to same person. More specifically, we select 1000 images that could be successfully inverted to the latent space. Then, we feed all the pairs of original and reconstructed images to FaceNet and obtain the L2 distances. We then sort all L2 distances from low to high. We use the dichotomy method to manually check the distance around which the identity of the reconstructed and original image changes significantly, and this distance is the threshold value we need.

\section{Why Real Images Requiring Lower Perturbation}
\label{reason_real_low_per}
As we discussed in \autoref{eval_real}, protecting real images requires a lower distance budget.
The reason behind this observation is that the generated images are generated by each GAN model, i.e., the GAN model has seen the latent code before, and therefore, it is easier for the GAN to invert these generated images into their latent code.
In contrast, the GAN model never knows what the latent code of the real image looks like, making it more difficult to obtain the latent code of the real image, and thus a much lower distance budget is sufficient to protect the real image.

\section{Why Encoder Enhancement Fails}
\label{why_encoder_fails}
The algorithm of retraining the encoder is same as the original one (see Appendix \autoref{algo:hybrid} for more details):
\begin{equation}\label{retrain_eq}
\begin{aligned}
\min _{\Theta_{E}} \mathcal{L}_{E}=\left\|\mathbf{x}-G\left(E\left(\mathbf{x}\right)\right)\right\|_{2} &+\lambda_{v g g}\left\|F\left(\mathbf{x}\right)-F\left(G\left(E\left(\mathbf{x}\right)\right)\right)\right\|_{2} \\
&-\lambda_{a d v} \underset{\mathbf{x} \sim P_{\text {data }}}{\mathbb{E}}\left[D\left(G\left(E\left(\mathbf{x}\right)\right)\right)\right]
\end{aligned}
\end{equation}
where $G$ and $D$ represent the generator and discriminator, the two main components of StyleGANv2, and $F$ represents the feature extractor VGG. 

To delve into the reasons behind the above observation, we need to scrutinize the above \autoref{retrain_eq} again.
Considering when the encoder $E$ accept $\mathbf{x}^{\text{cloaked}}$ as inputs for retraining, then the first two items of above formula are formalized as follows:
\begin{equation}\label{retrain_eq_2}
\begin{array}{r}
\left\|\mathbf{x}^{\text {cloaked }}-G\left(E\left(\mathbf{x}^{\text {cloaked }}\right)\right)\right\|_{2} \\
\left\|F\left(\mathbf{x}^{\text {cloaked }}\right)-F\left(G\left(E\left(\mathbf{x}^{\text {cloaked }}\right)\right)\right)\right\|_{2}
\end{array}
\end{equation}
We can clearly see that these two items actually optimize the encoder so that the generator reconstructs the cloaked image, which is completely contrary to the original optimization goal of reconstructing a clean image.
In other words, these newly collected cloaked images act as a poisoned set for retraining the encoder, thus leading to increased jeopardy of \system to GAN inversion.

Referring to the objective function setting in adversarial training to make a classifier robust to adversarial examples, thus \autoref{retrain_eq_2} can be formalized as follows:
\begin{equation}
\begin{array}{r}
\left\|\mathbf{x}^{\text {corresponding clean }}-G\left(E\left(\mathbf{x}^{\text {cloaked }}\right)\right)\right\|_{2} \\
\left\|F\left(\mathbf{x}^{\text {corresponding clean }}\right)-F\left(G\left(E\left(\mathbf{x}^{\text {cloaked }}\right)\right)\right)\right\|_{2}
\end{array}
\end{equation}
The newly designed objective functions optimize the encoder so that the generator reconstructs the corresponding clean images, i.e., the encoder bears the robustness to remove or purify the noise added to the image.
However, it is not realistic to use the corresponding clean images as ground truth labels to guide the optimization.
In other words, the adversary does not need to retrain the encoder if he/she has already collected a clean image that he/she intends to invert and manipulate.

\section{Why Fawkes Behaves This Way}
\label{reason_fawkes}
\mypara{Why Fawkes Performs Worse Against Optimization-based Inversion}
Recall that Fawkes changes the features of one face towards another face, while \system changes the features of one face away from itself.
However, in the Fawkes, the two faces inevitably have some common attributes, for example, they may have similar hairstyle or similar skin color or the same gender.
Besides, as shown in \autoref{fig:edit_subspace}, the latent vector of a face remains unchanged if its' attributes are already the target attributes.
Similarly, in the feature space, since two faces inevitably share some common attributes, the protected face is limited in some direction, i.e., it cannot move alone in the direction of the common attributes.
While in \system, there are no restrictions on the protected face, which means it can move in any direction and as far away from itself as possible.
In a nutshell, since the objective functions of Fawkes and \system are different, the cloaked faces of Fawkes are relatively closer to the original faces in feature space, thus leading to its worse performance compared to \system.

\begin{figure}[t]
    \centering
    \subfloat[DCGAN]{\includegraphics[width=0.5\linewidth]{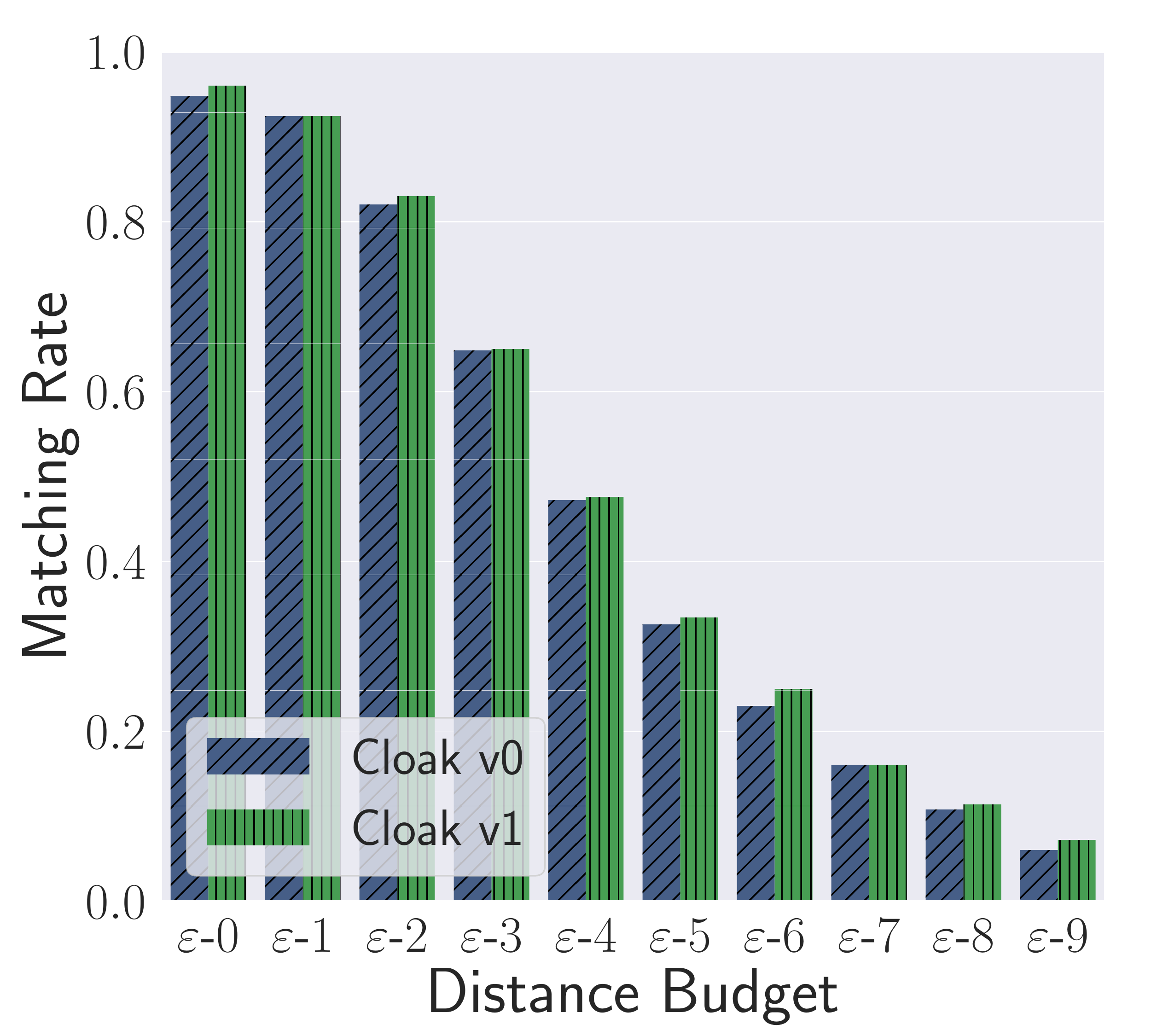}}
    \subfloat[WGAN]{\includegraphics[width=0.5\linewidth]{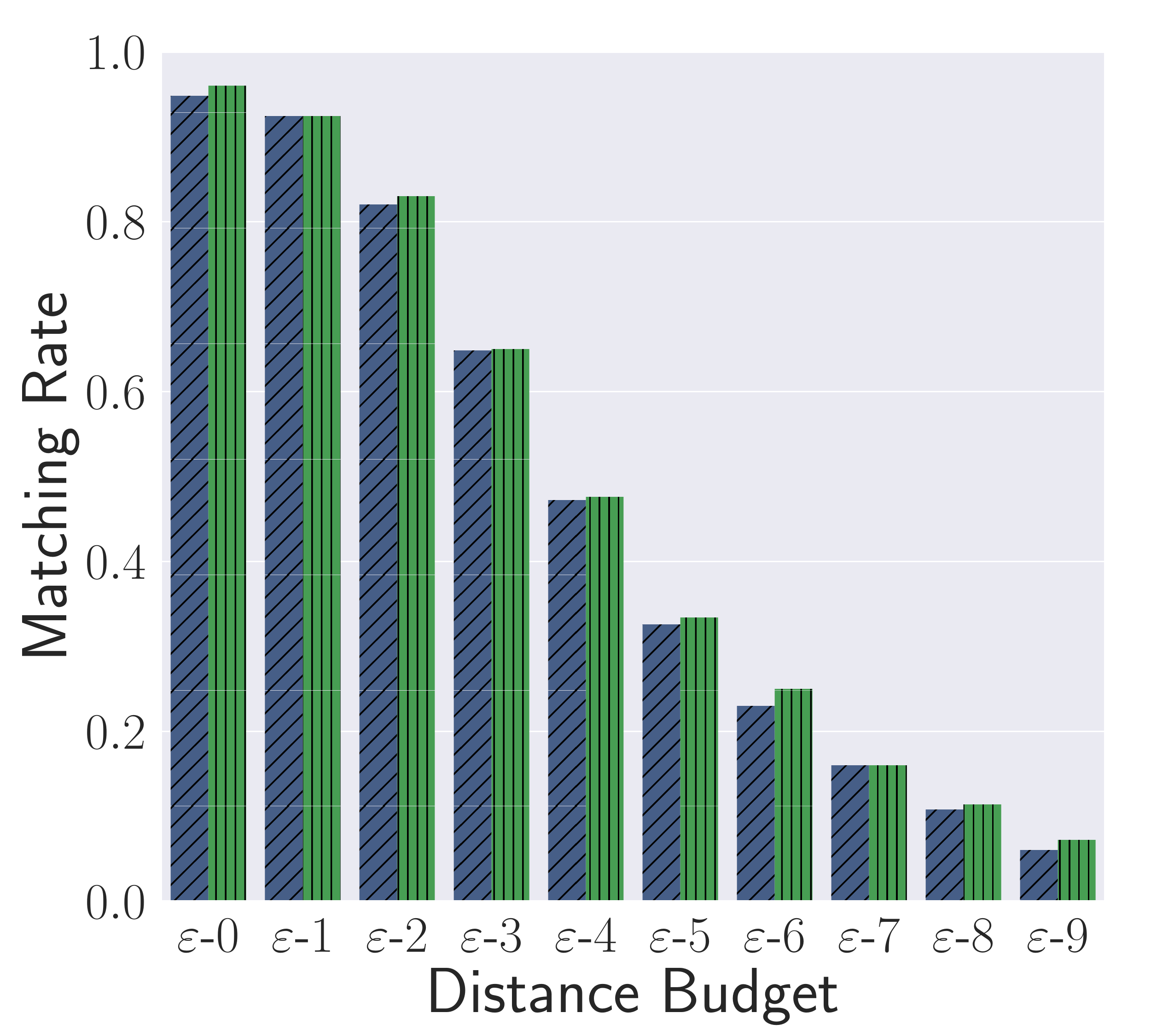}}
    
    \subfloat[DCGAN]{\includegraphics[width=0.5\linewidth]{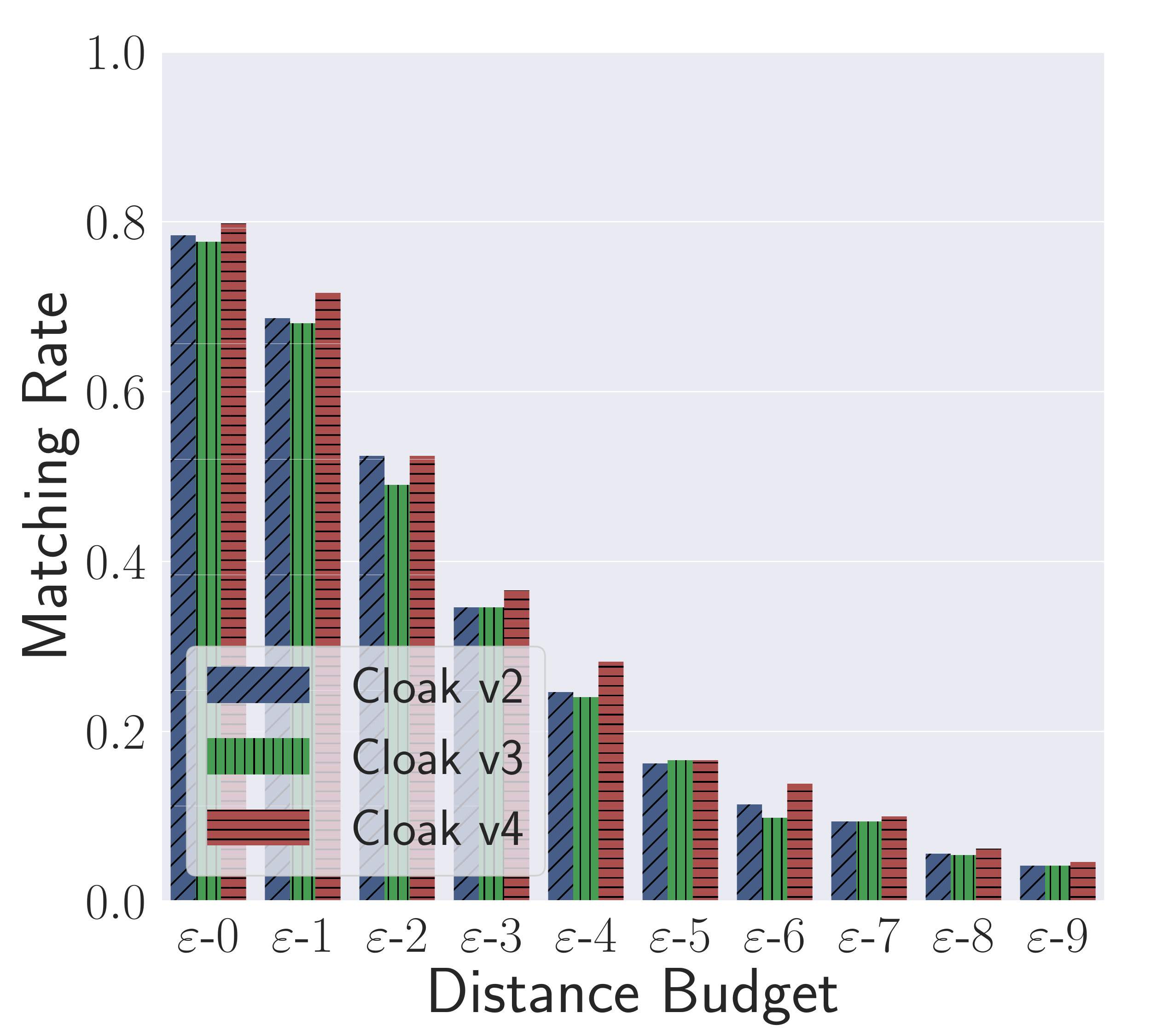}}
    \subfloat[WGAN]{\includegraphics[width=0.5\linewidth]{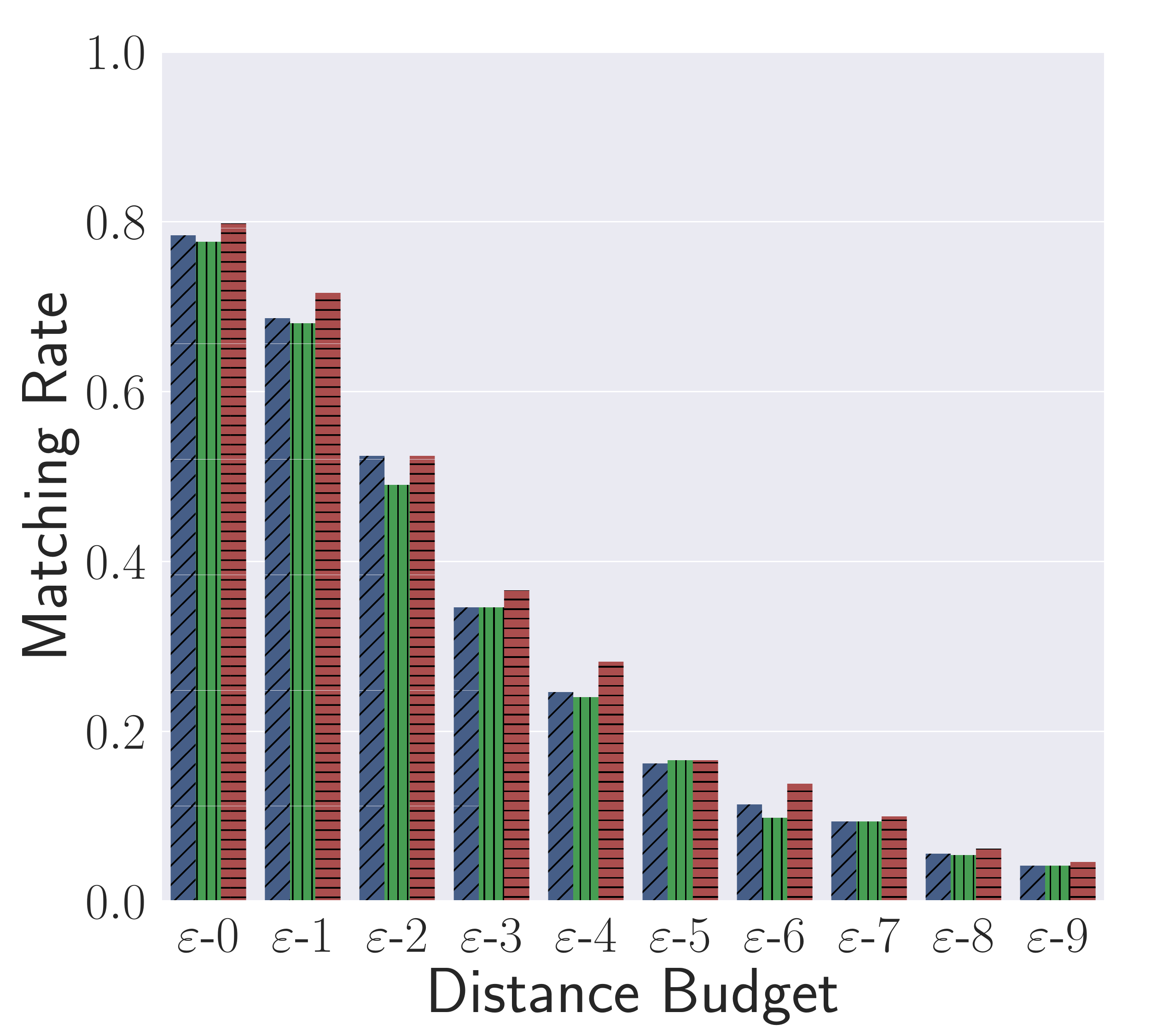}}
    \caption{The effectiveness performance of \system on DCGAN and WGAN.}
    \label{fig:effectiveness-appendix}
\end{figure}

\begin{figure}[t]
    \centering
    \subfloat[Cloak Overwriting\label{fig:real-adaptive-appendix-overwrite}]{
    \includegraphics[width=0.5\linewidth]{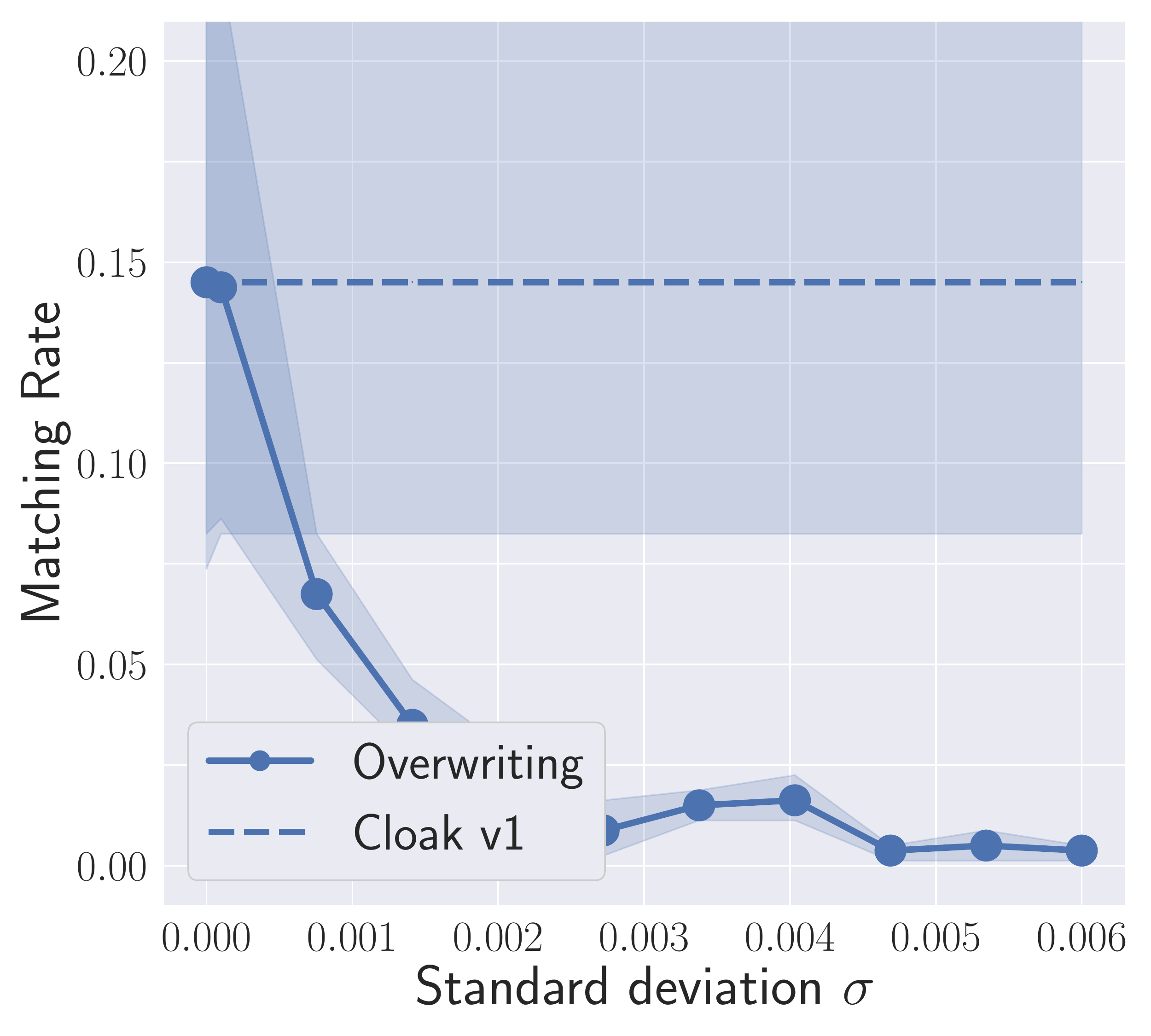}}
    \subfloat[Cloak Purification\label{fig:real-adaptive-appendix-remove}]{
    \includegraphics[width=0.5\linewidth]{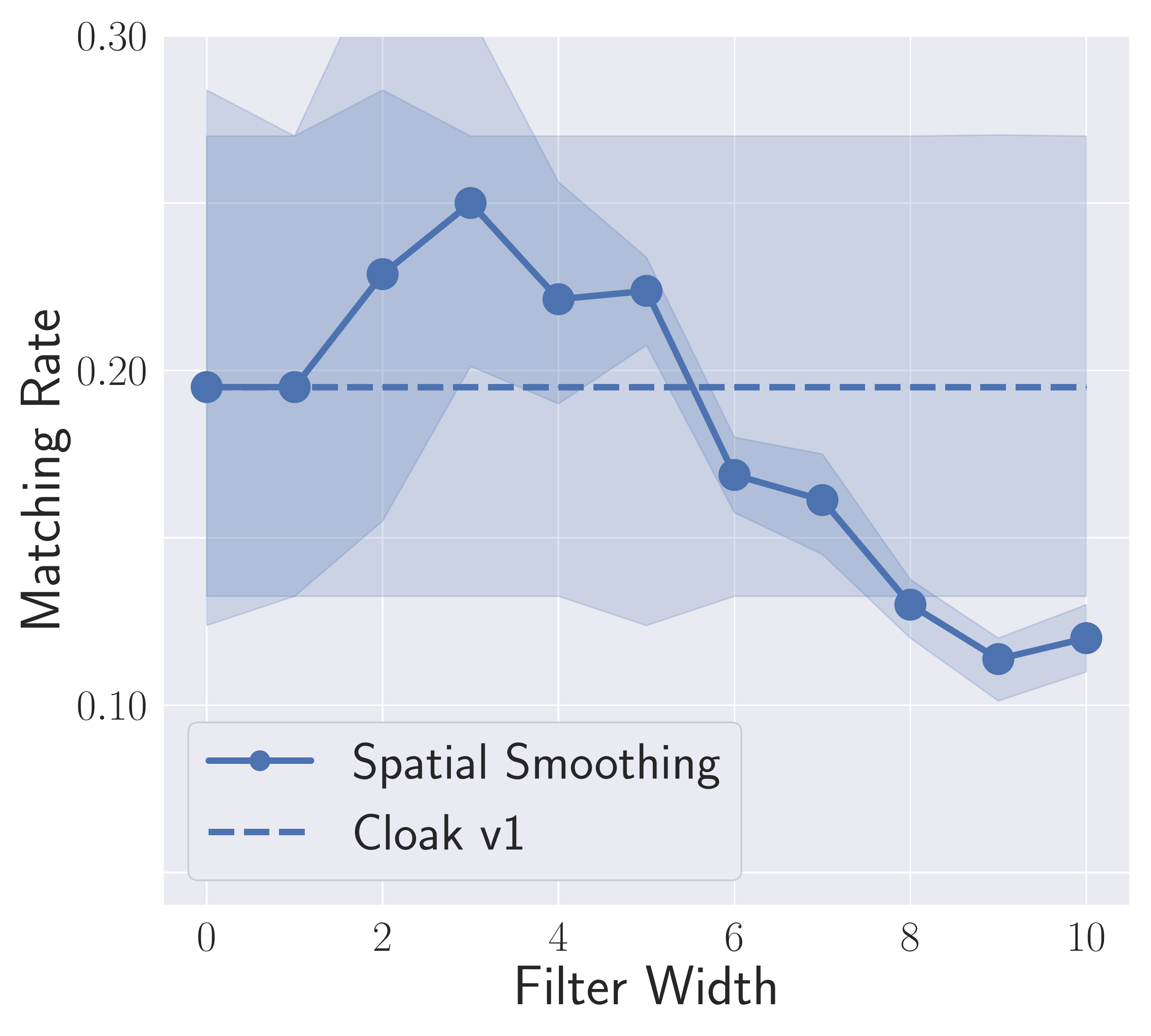}
    }
    \caption{The effectiveness performance of \system under the effect of two possible adaptive adversaries, i.e., Cloak Overwriting, Cloak Purification.
    }
    \label{fig:real-adaptive-appendix}
\end{figure}

\mypara{Why Fawkes Performs Similarly Against Hybrid Inversion}
Hybrid inversion leverages a learned predictive encoder to provide better initialization that can lead to better initial and final optimization performance, as shown in \autoref{fig:distri}.
In addition, as specified by \autoref{retrain_eq}, the encoder is trained based on clean images and is trained to reconstruct clean images.
Thus, the encoder is insensitive to imperceptible noise added by Fawkes or \system (both Fawkes and \system only mislead a feature extractor) and can map their cloaked images to similarly distributed and easy-to-optimized initial latent codes, ultimately leading to a good state of convergence in subsequent optimization-based inversion stage.
As for the slight variance between Fawkes and \system, we believe it is due to the feature extractor used differently of them.
Fawkes adopts InceptionResNet V2~\cite{SIVA17} and DenseNet-121~\cite{HLMW17}, which are much larger than ResNet-18 adopted in \system, thus leading to very slight better protection.
Note that, here we only compare Cloak v4 with Fawkes, our Cloak v2/3 shown in \autoref{fig:effectiveness-234} can achieves much better performance than Cloak v4, which is naturally better than Fawkes.

\newpage
\begin{figure*}[t]
    \centering
    \subfloat[DCGAN]{\includegraphics[width=0.26\linewidth]{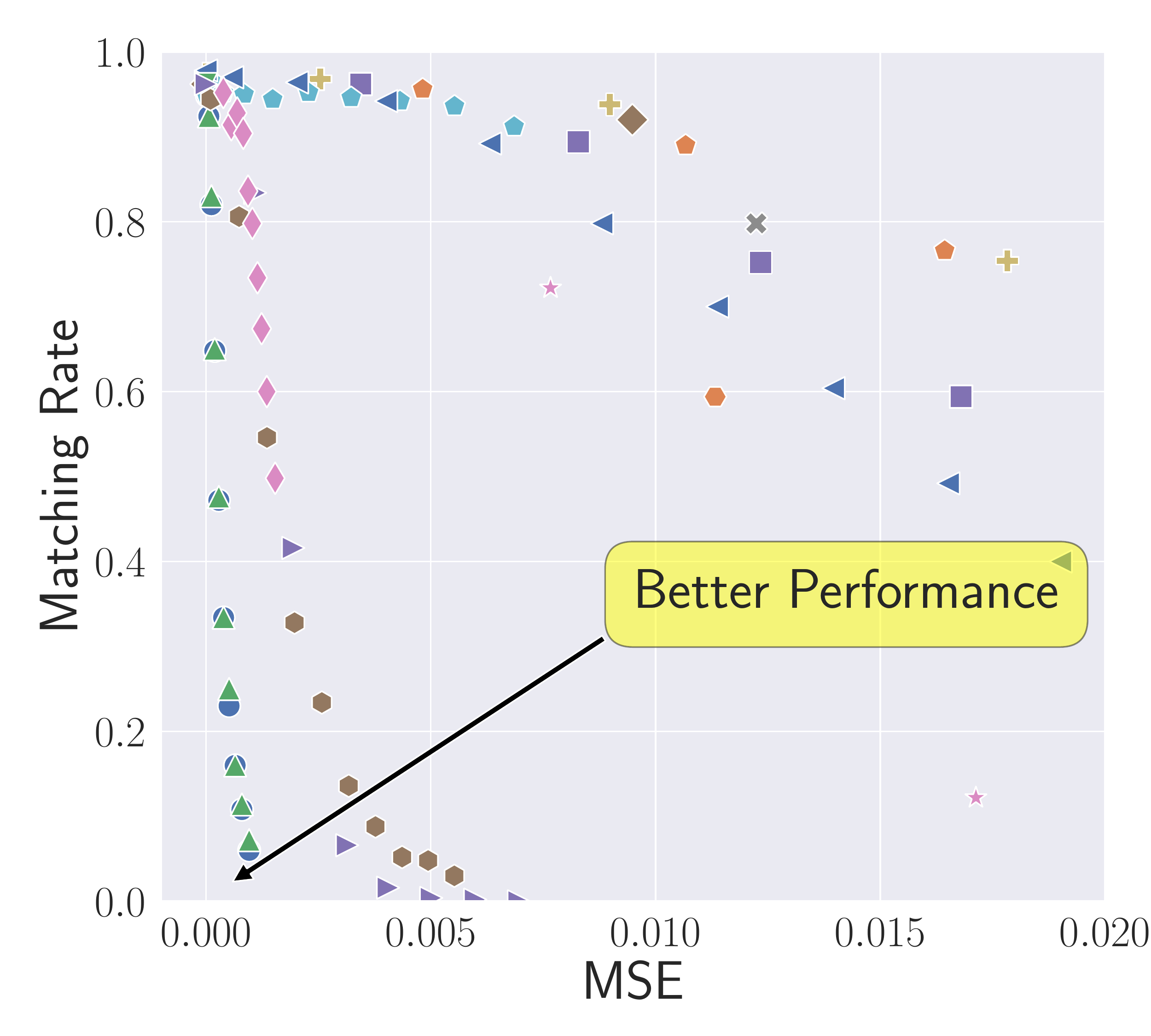}
\includegraphics[width=0.26\linewidth]{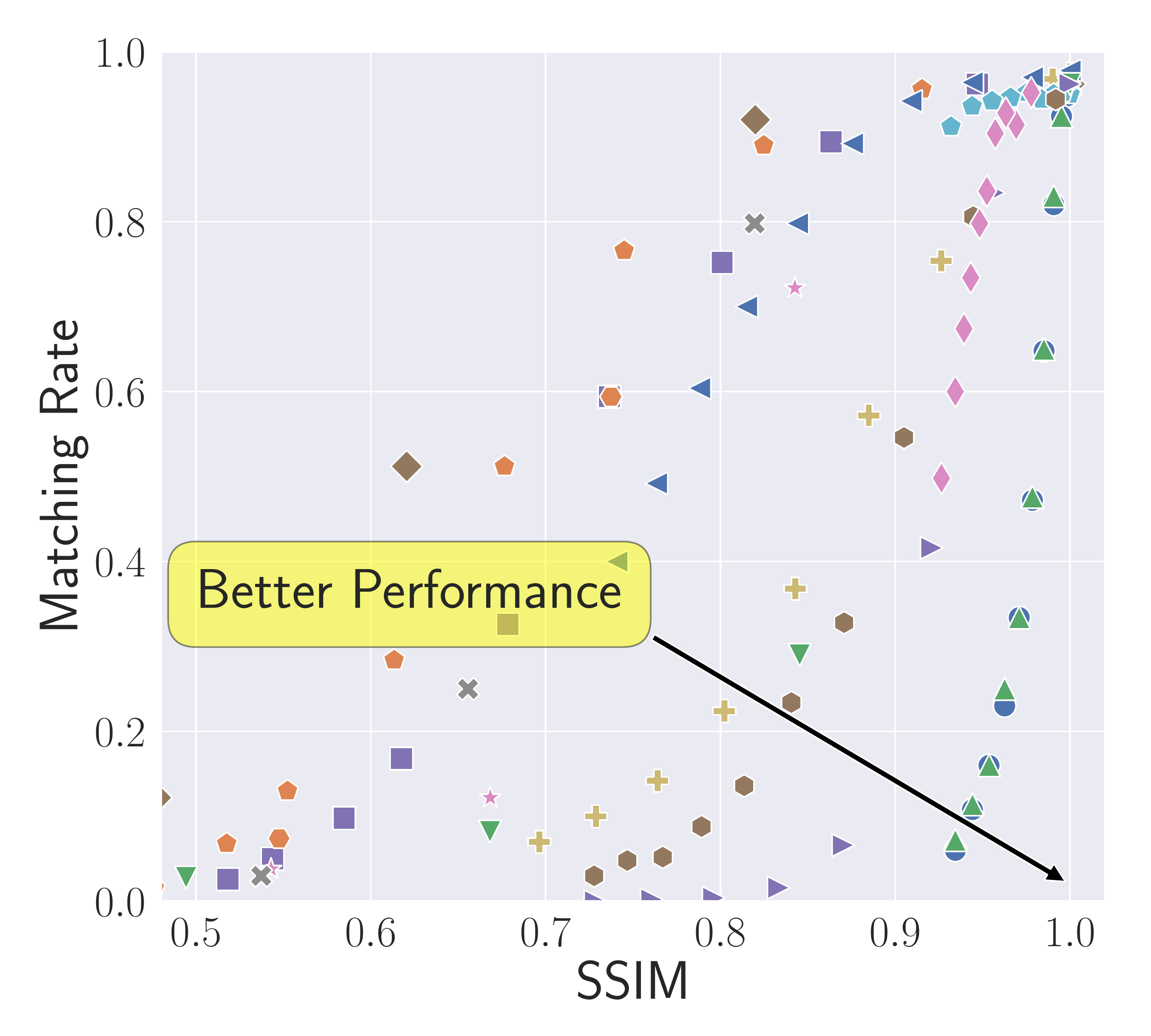}
\includegraphics[width=0.26\linewidth]{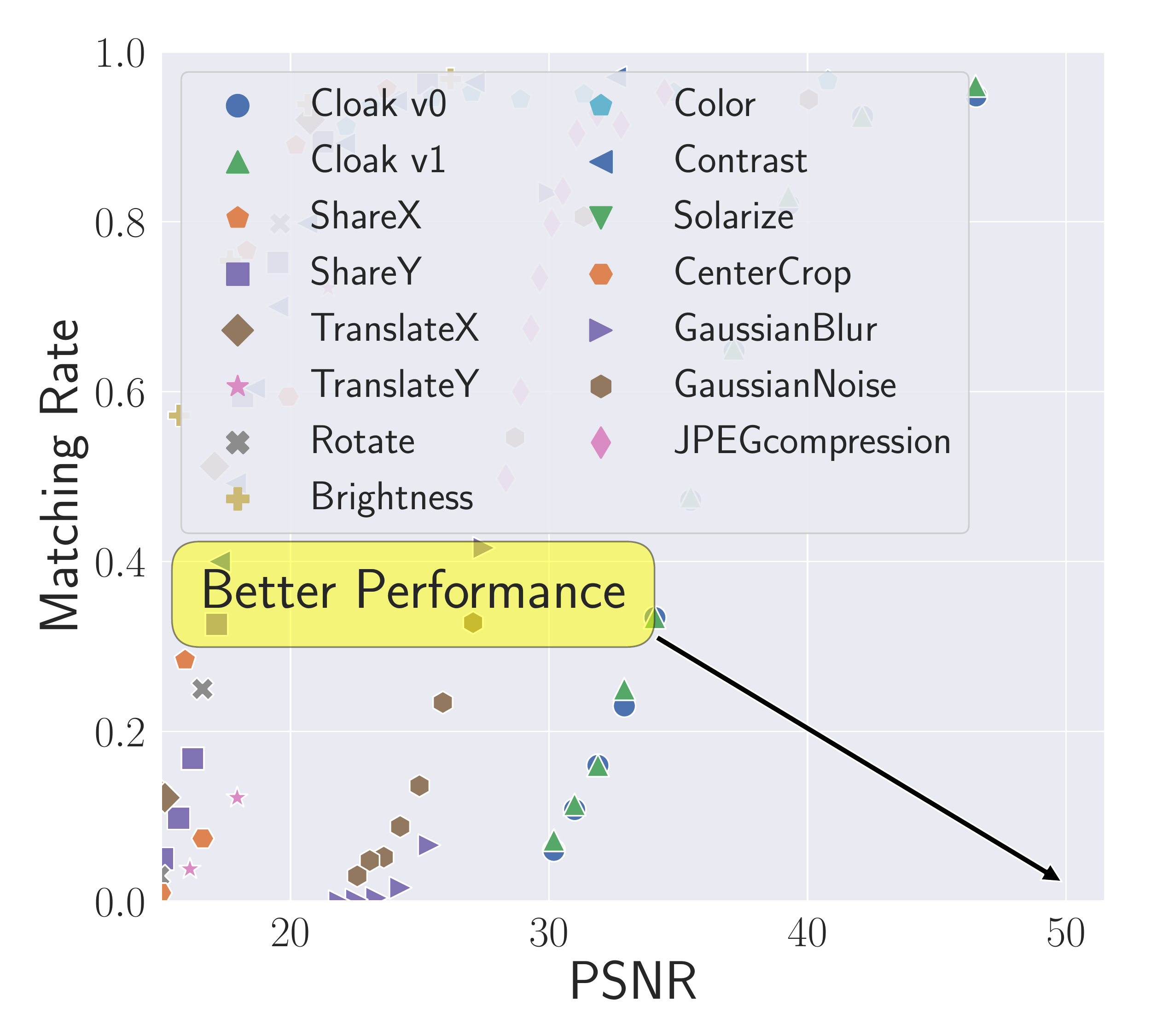}}

    \subfloat[WGAN]{\includegraphics[width=0.26\linewidth]{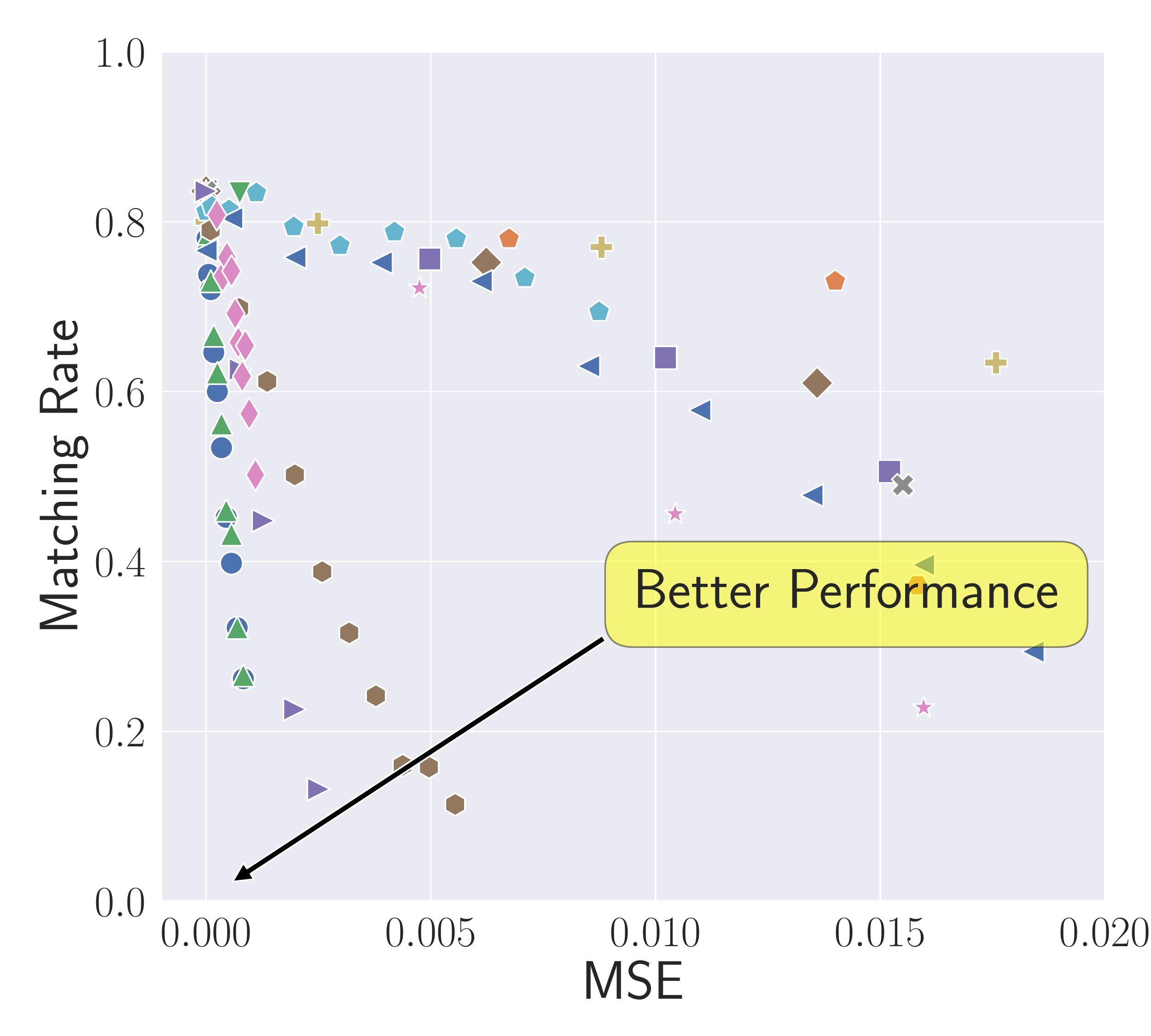}
\includegraphics[width=0.26\linewidth]{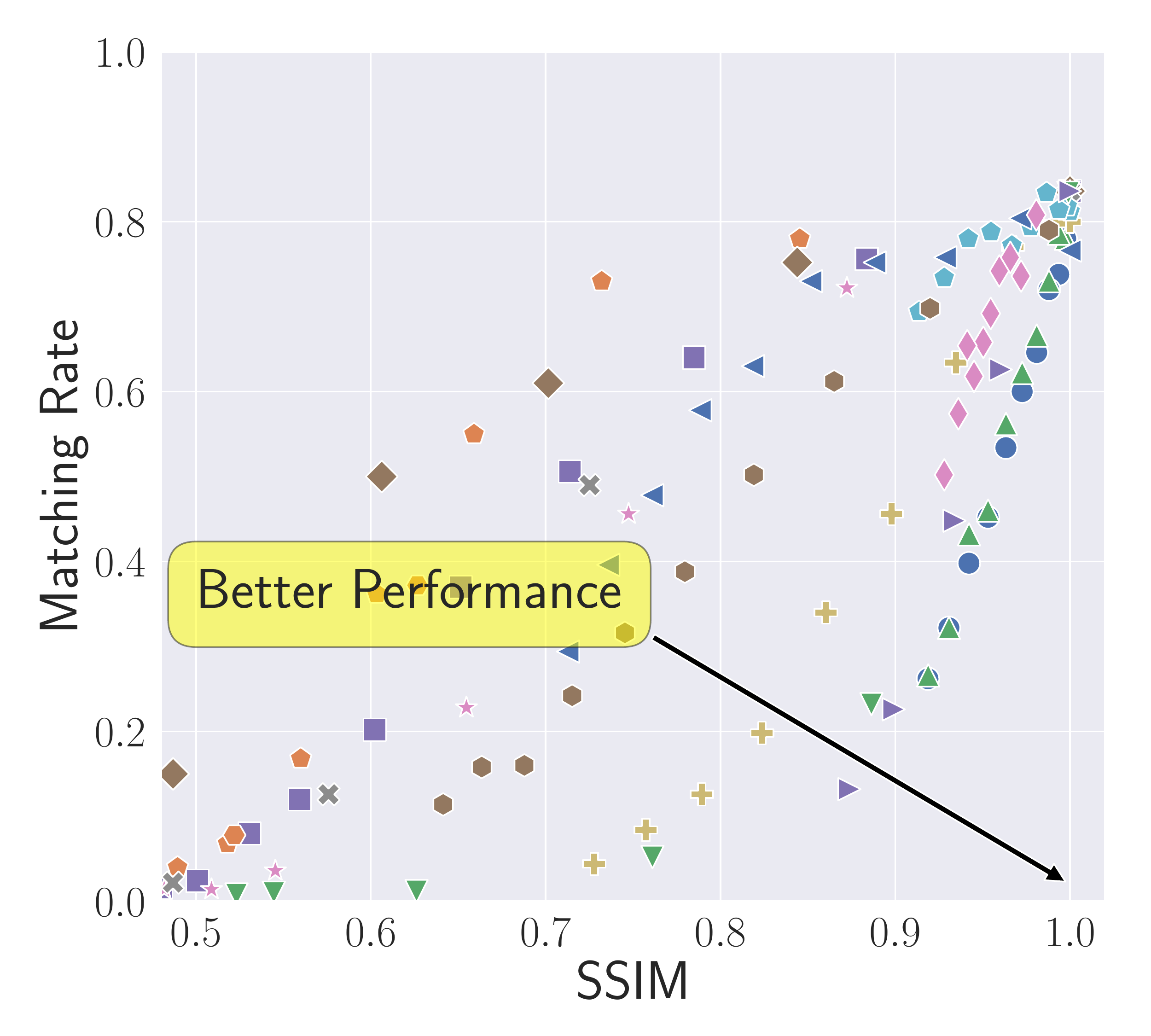}
\includegraphics[width=0.26\linewidth]{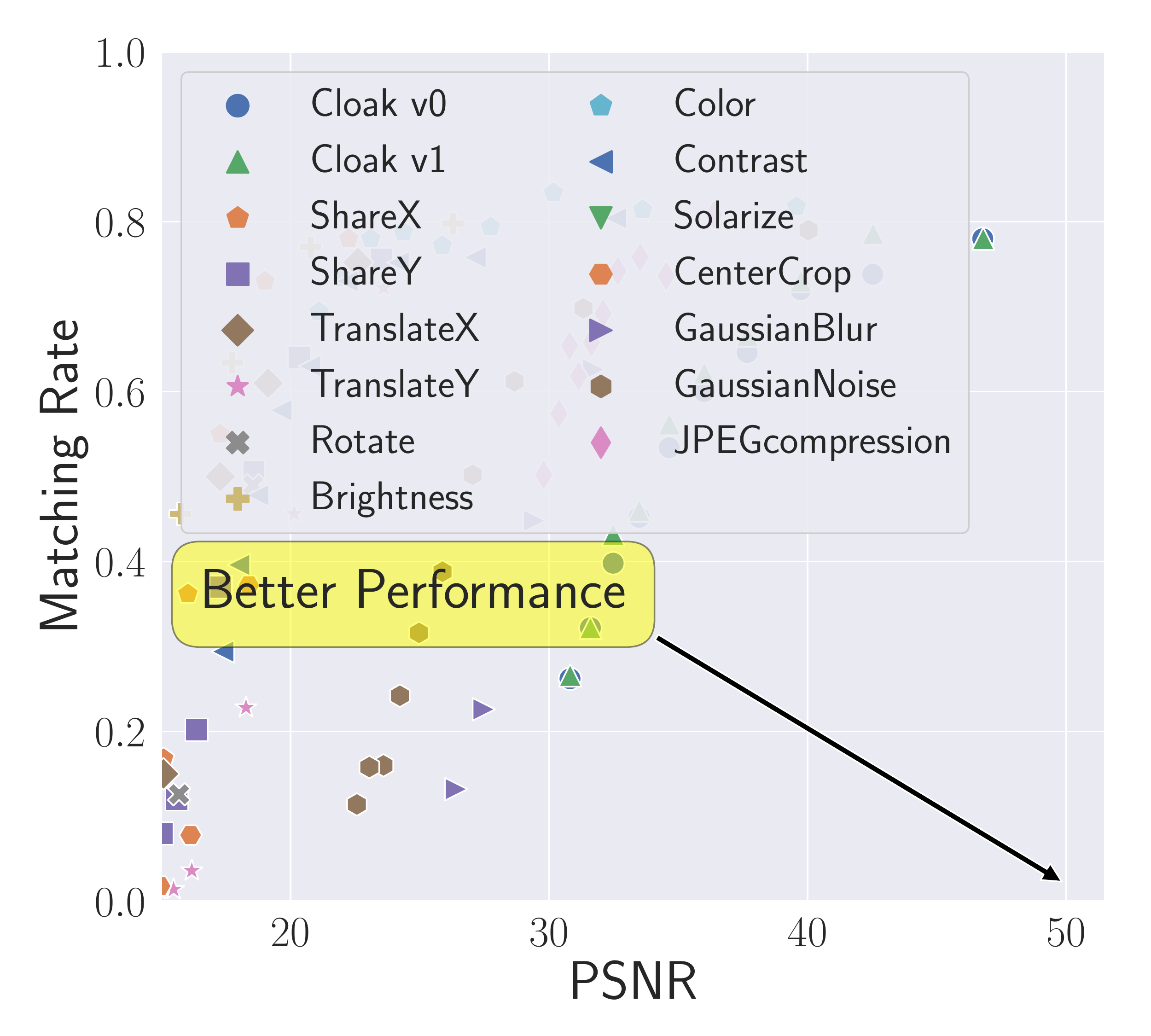}}

    \subfloat[StyleGANv1]{\includegraphics[width=0.26\linewidth]{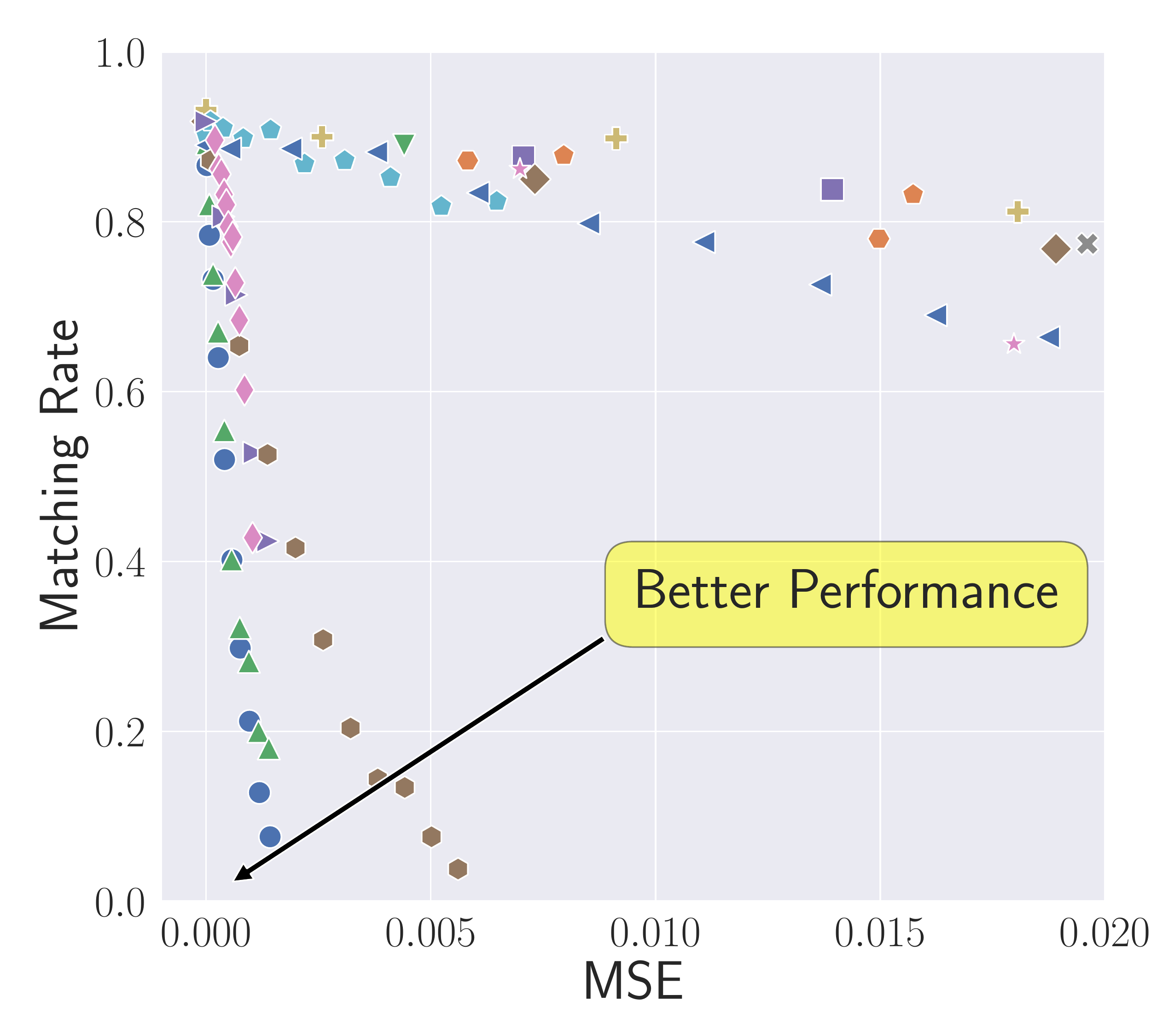}
\includegraphics[width=0.26\linewidth]{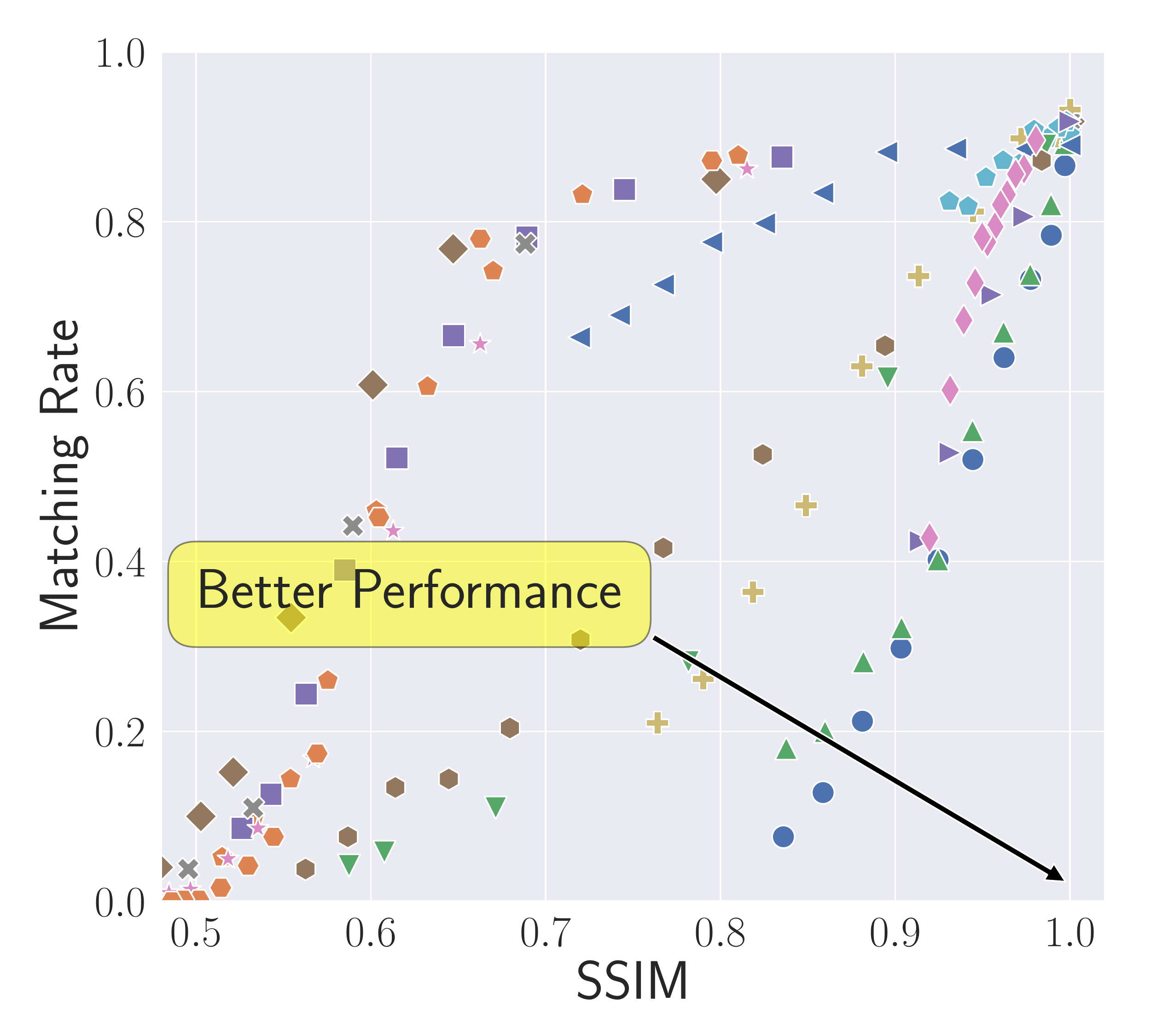}
\includegraphics[width=0.26\linewidth]{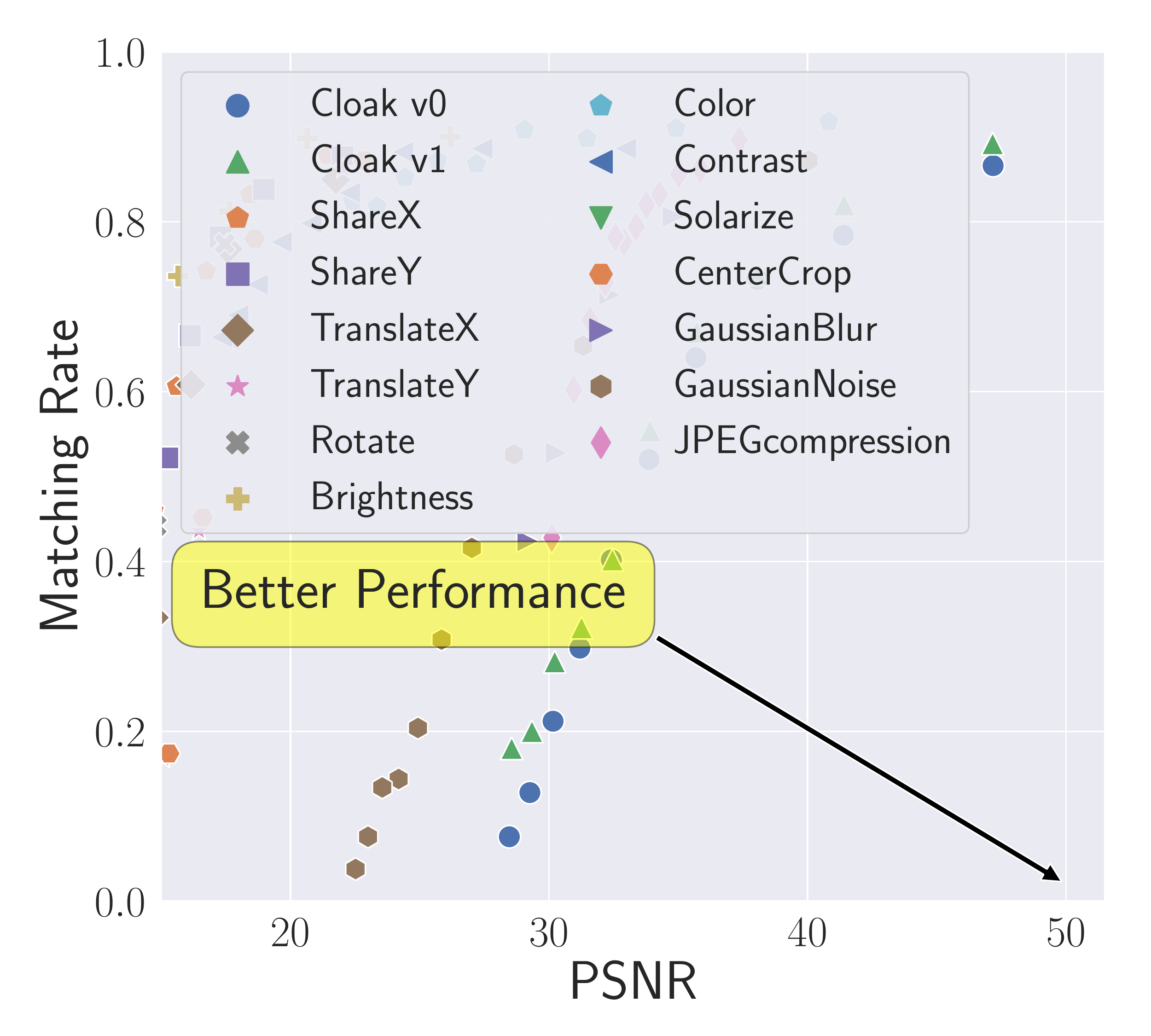}}
    \caption{Comparison between all baseline methods and Cloak v0/v1 on generated images. The different points of each method represent different budgets.}
   \label{appfig:scatter_opti}
\end{figure*}
\begin{figure*}[t]
    \centering
   \subfloat[DCGAN]{\includegraphics[width=0.26\linewidth]{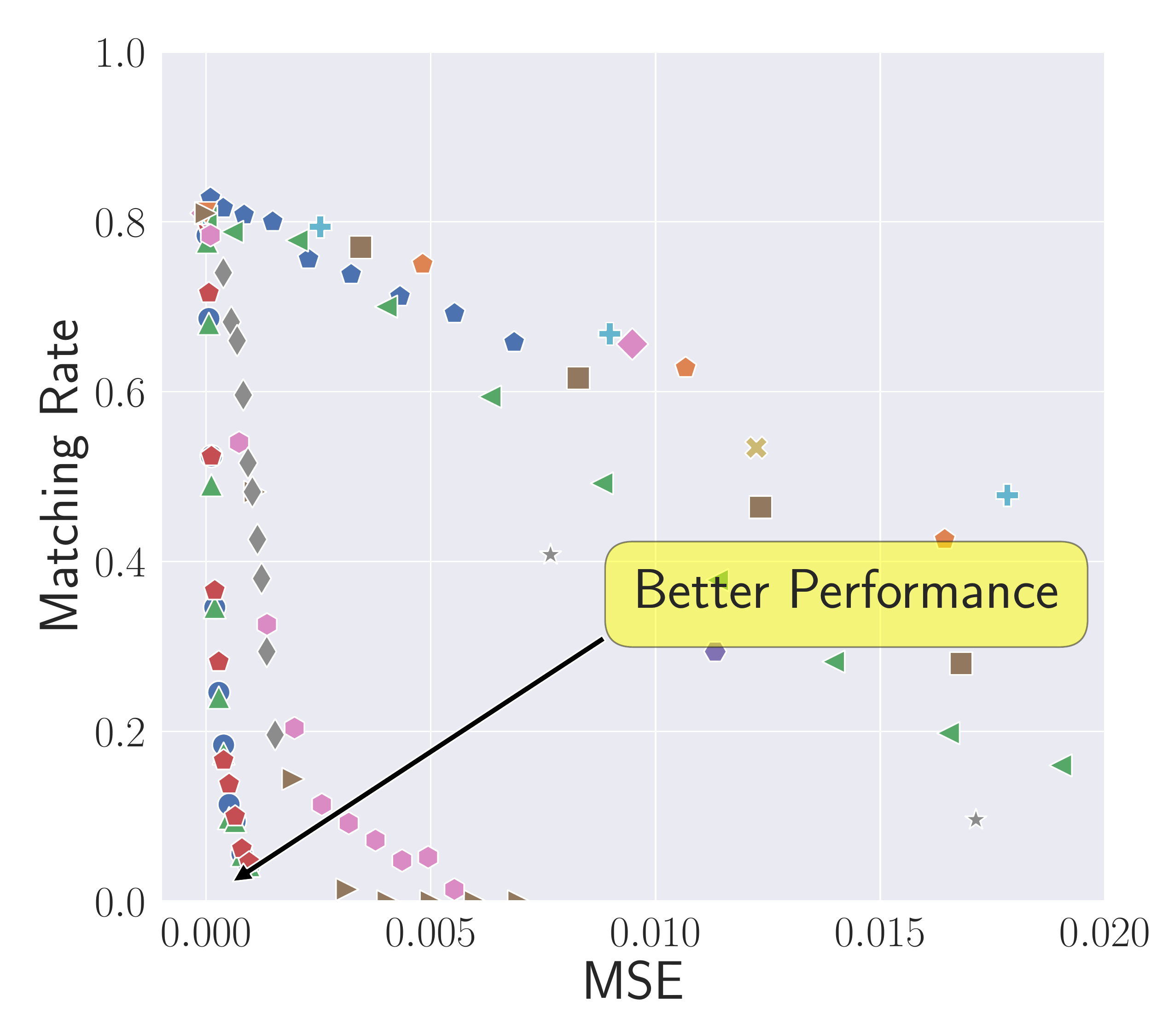}
\includegraphics[width=0.26\linewidth]{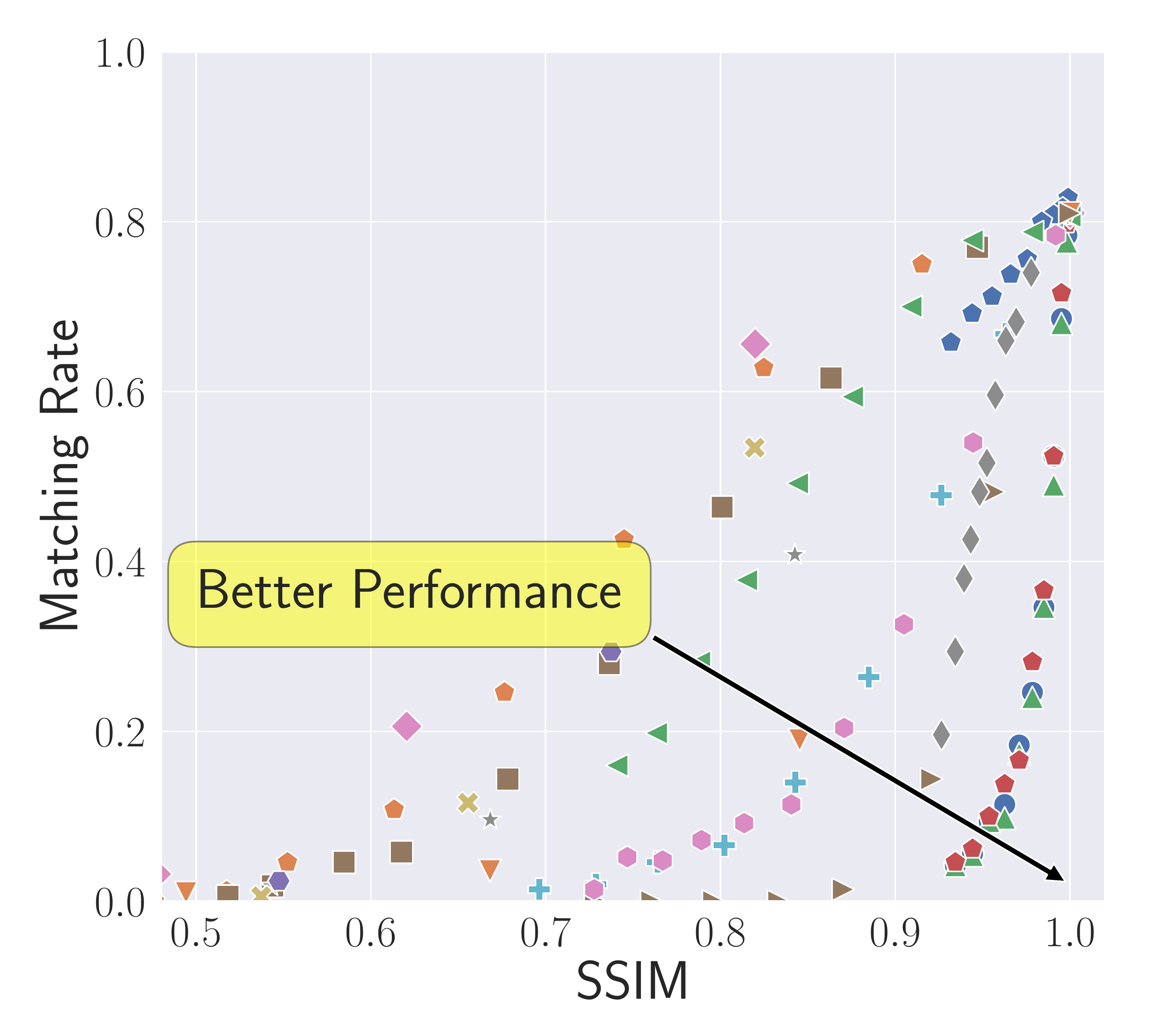}
\includegraphics[width=0.26\linewidth]{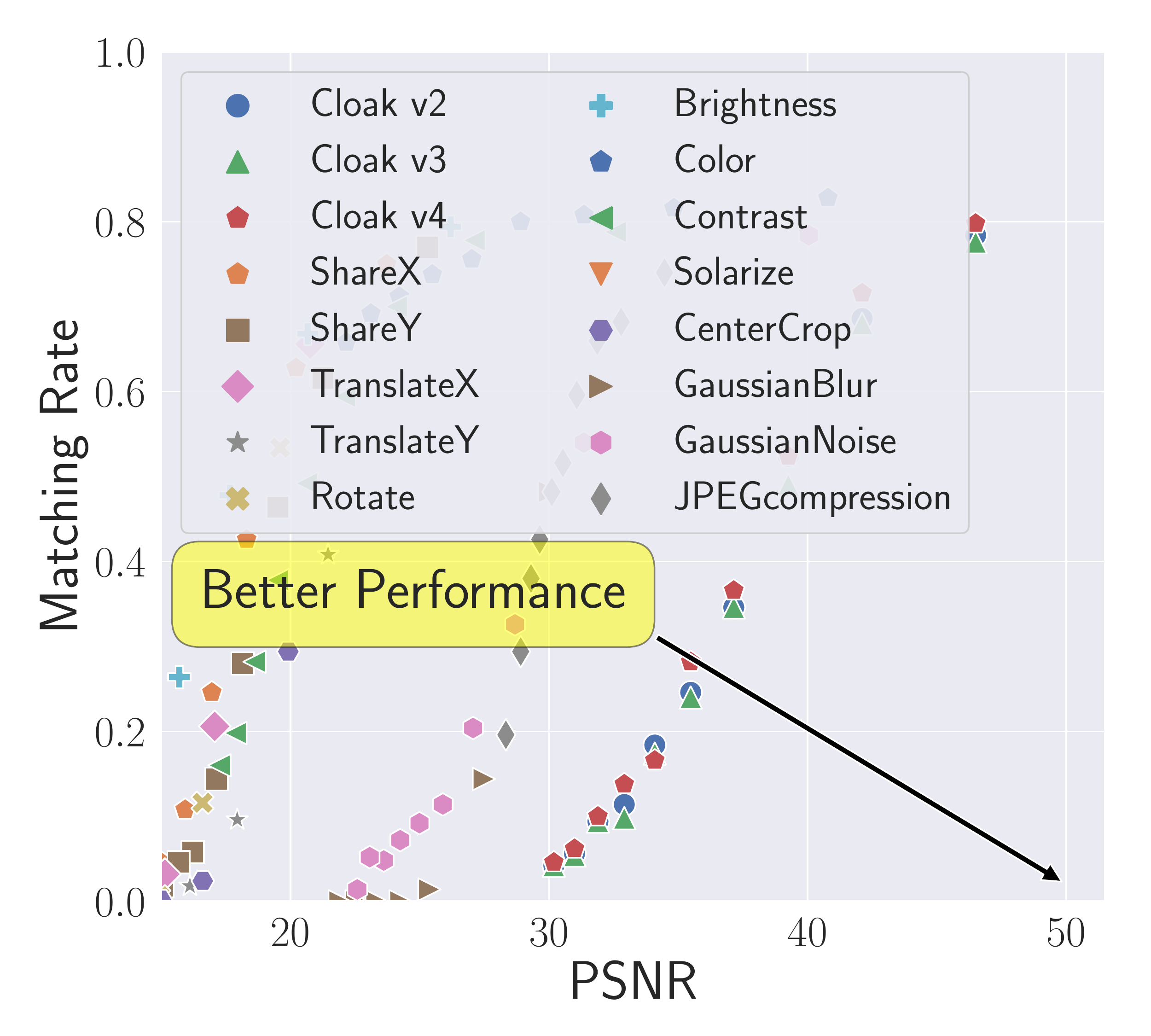}}

    \subfloat[WGAN]{\includegraphics[width=0.26\linewidth]{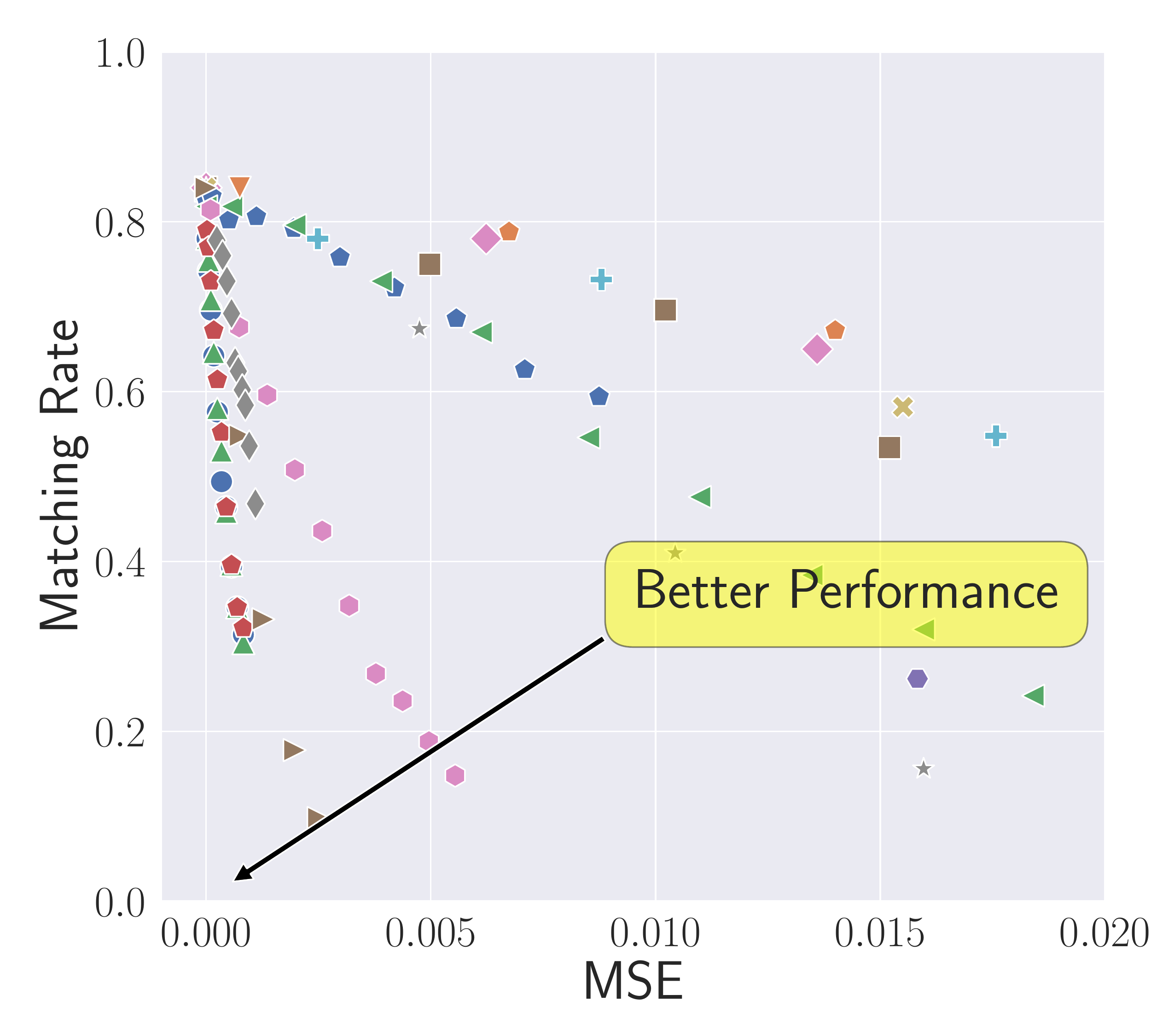}
\includegraphics[width=0.26\linewidth]{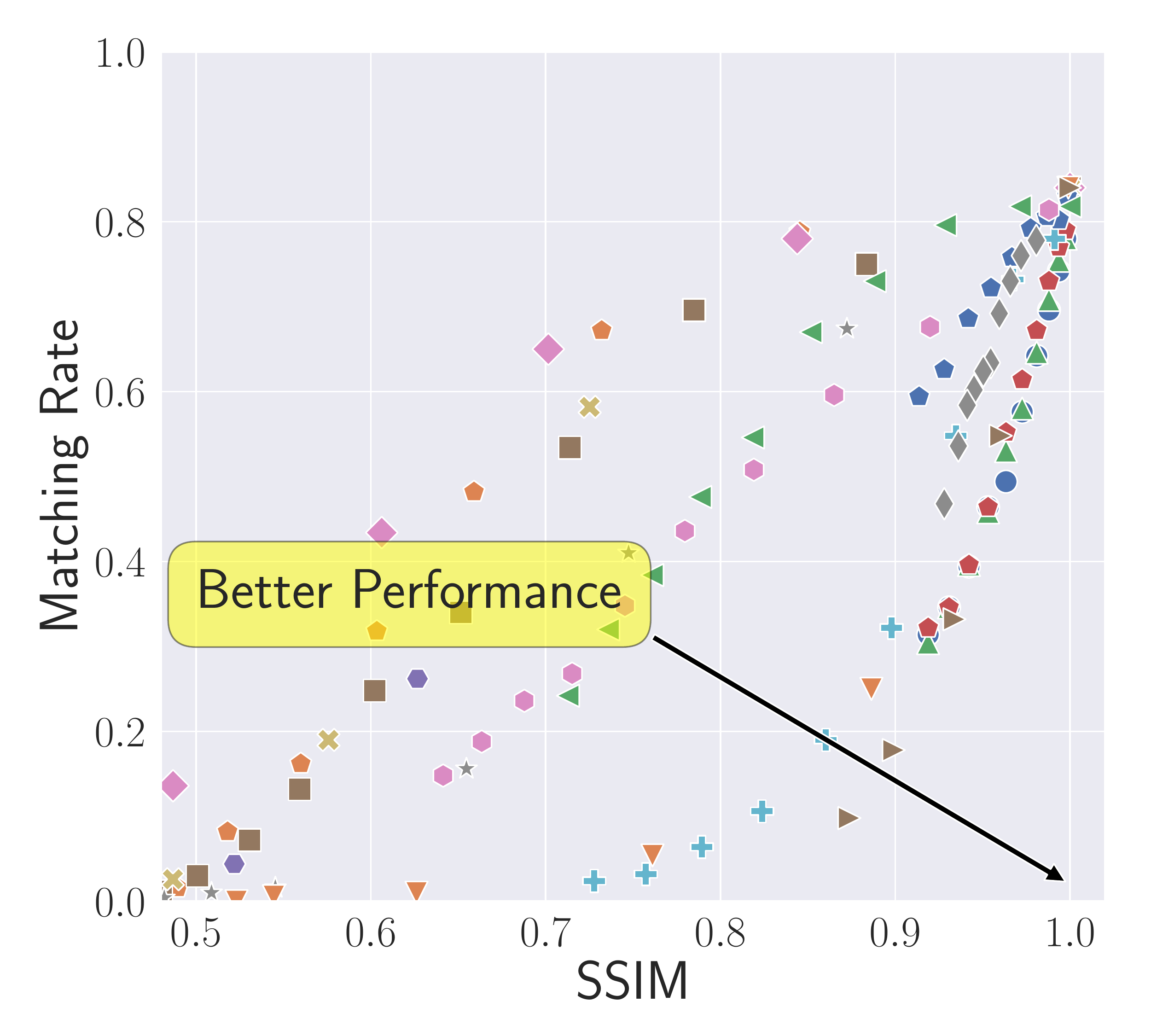}
\includegraphics[width=0.26\linewidth]{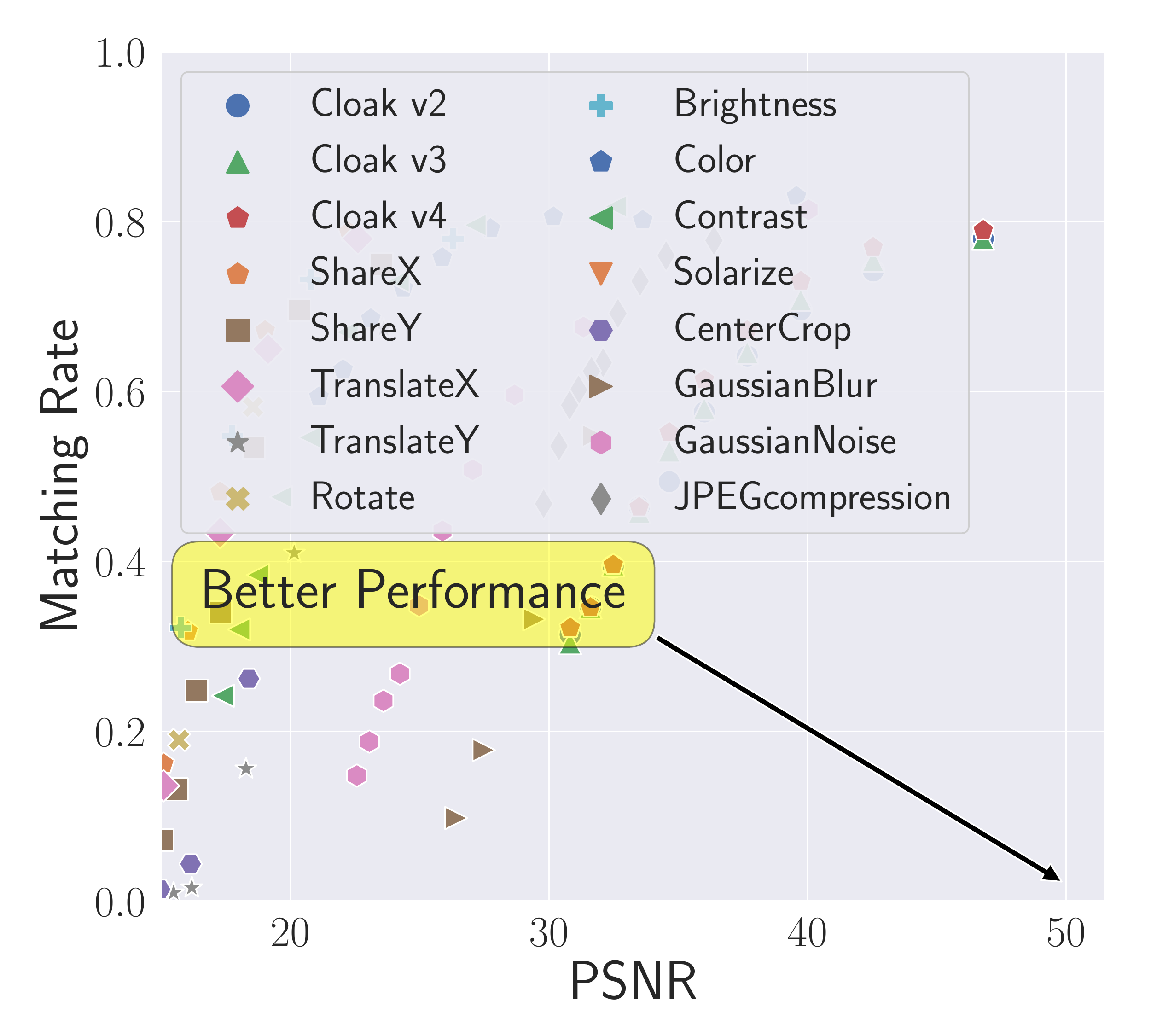}}

    \subfloat[StyleGANv1]{\includegraphics[width=0.26\linewidth]{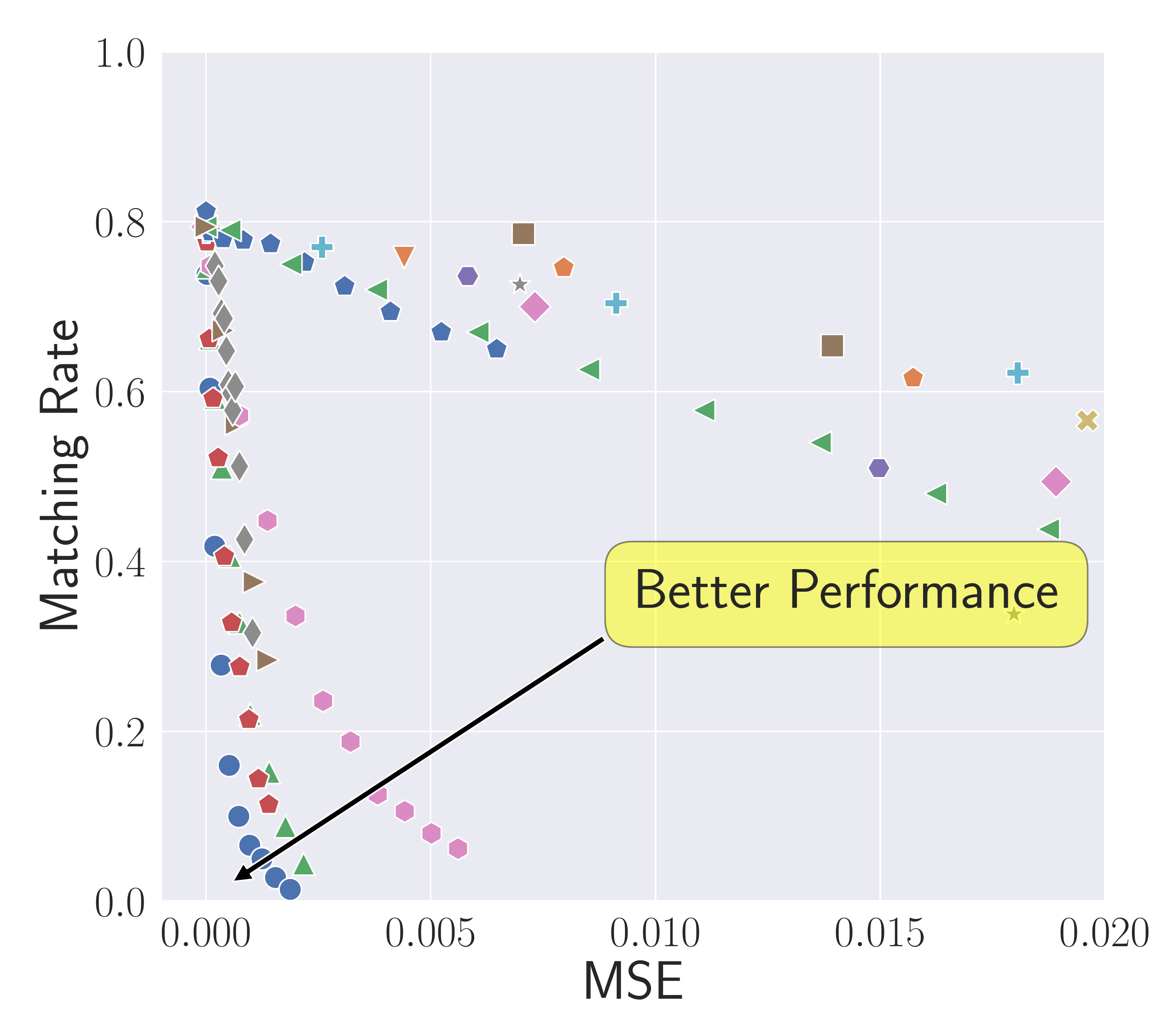}
\includegraphics[width=0.26\linewidth]{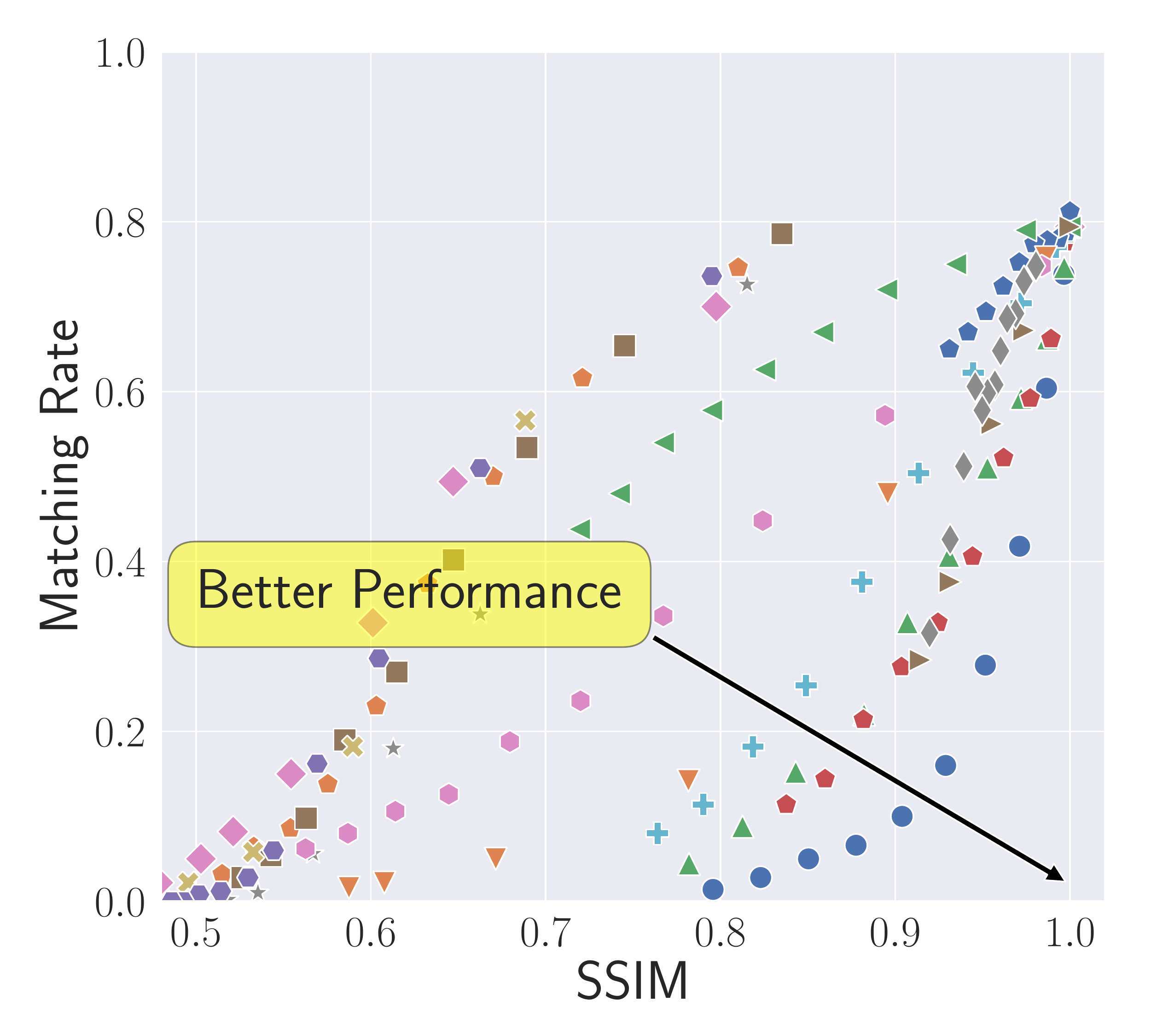}
\includegraphics[width=0.26\linewidth]{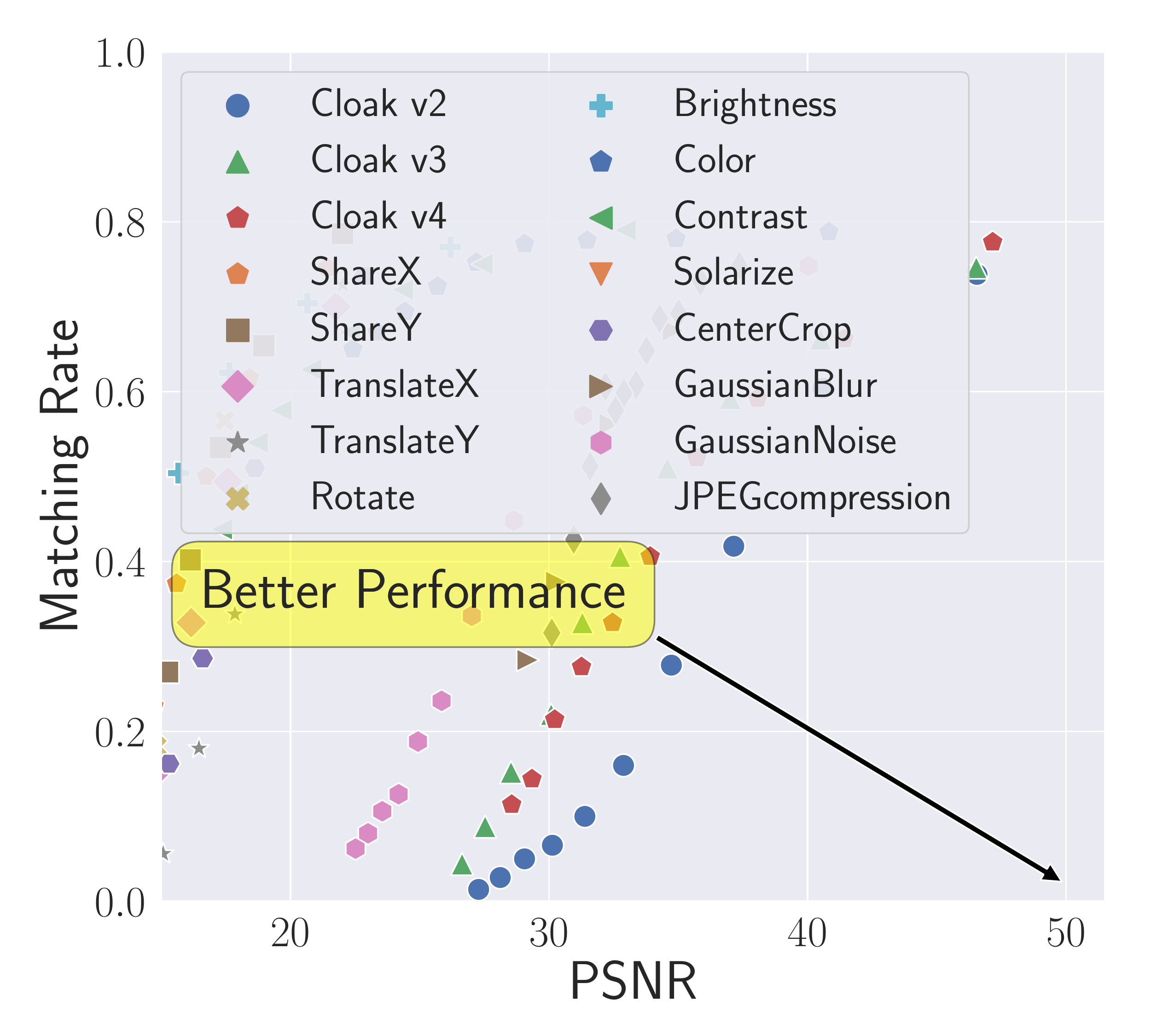}}
    \caption{Comparison between all baseline methods and Cloak v2/v3/v4 on generated images. The different points of each method represent different budgets.}
    \label{appfig:scatter_hybrid}
\end{figure*}

\begin{table*}[!t]
    \centering
    \caption{Some visual examples of cloaked real images searched by Cloak v4 performed on StyleGANv2 under different perturbation budgets.}
    \scalebox{0.75}
    {
    \begin{tabular}{c|c|c|c|c|c}
    \toprule
       Target Image & $\varepsilon$-$0$ & $\varepsilon$-$1$ & $\varepsilon$-$2$& $\varepsilon$-$3$ & $\varepsilon$-$4$\\
         \midrule
        \raisebox{-.5\height}{\includegraphics[width=0.19\linewidth]{plots/replots/6.1/utility/0.01_15.pdf}} & 
        \raisebox{-.5\height}{\includegraphics[width=0.19\linewidth]{plots/replots/6.1/utility/0.01_15.pdf}} & 
        \raisebox{-.5\height}{\includegraphics[width=0.19\linewidth]{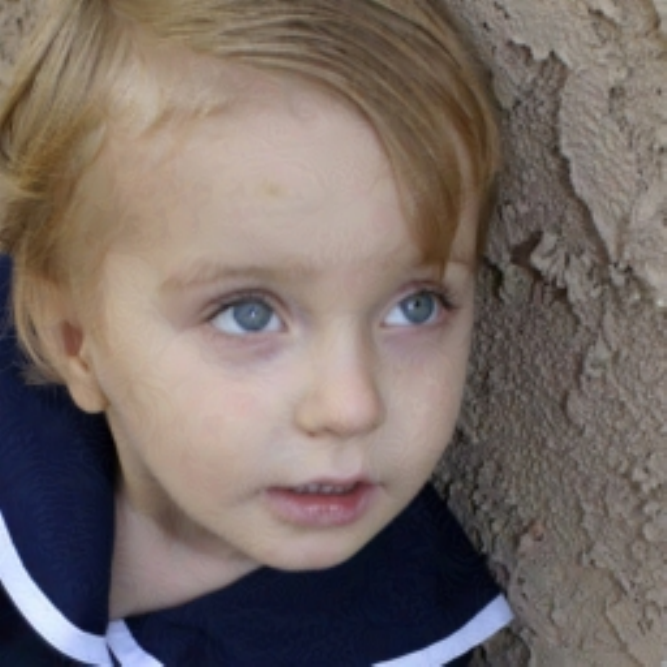}} & \raisebox{-.5\height}{\includegraphics[width=0.19\linewidth]{plots/replots/6.1/utility/0.03_15.pdf}} &
        \raisebox{-.5\height}{\includegraphics[width=0.19\linewidth]{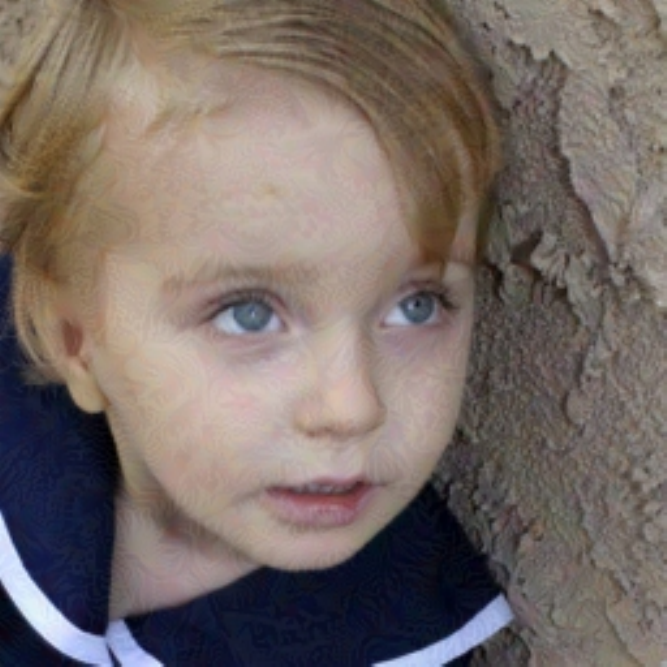}} &
        \raisebox{-.5\height}{\includegraphics[width=0.19\linewidth]{plots/replots/6.1/utility/0.05_15.pdf}}
        \\
         \midrule
          \raisebox{-.5\height}{\includegraphics[width=0.19\linewidth]{plots/replots/6.1/utility/0.01_161.pdf}} & 
            \raisebox{-.5\height}{\includegraphics[width=0.19\linewidth]{plots/replots/6.1/utility/0.01_161.pdf}} & 
            \raisebox{-.5\height}{\includegraphics[width=0.19\linewidth]{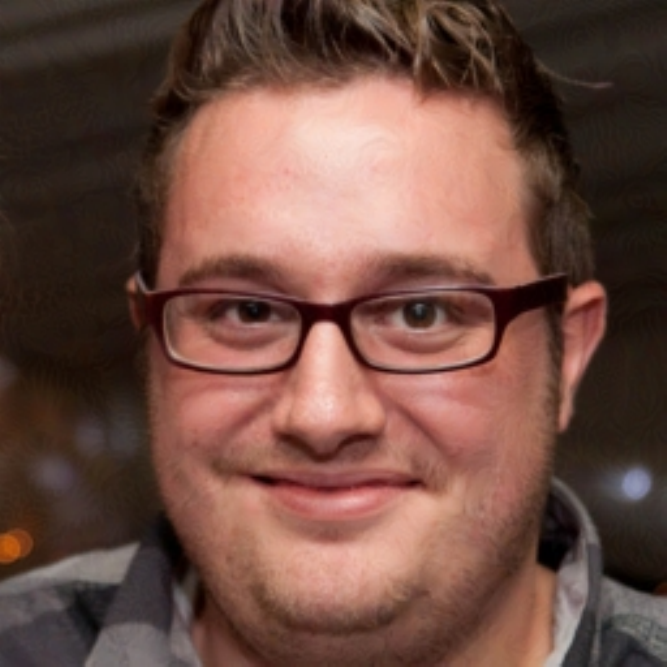}} & \raisebox{-.5\height}{\includegraphics[width=0.19\linewidth]{plots/replots/6.1/utility/0.03_161.pdf}} &
            \raisebox{-.5\height}{\includegraphics[width=0.19\linewidth]{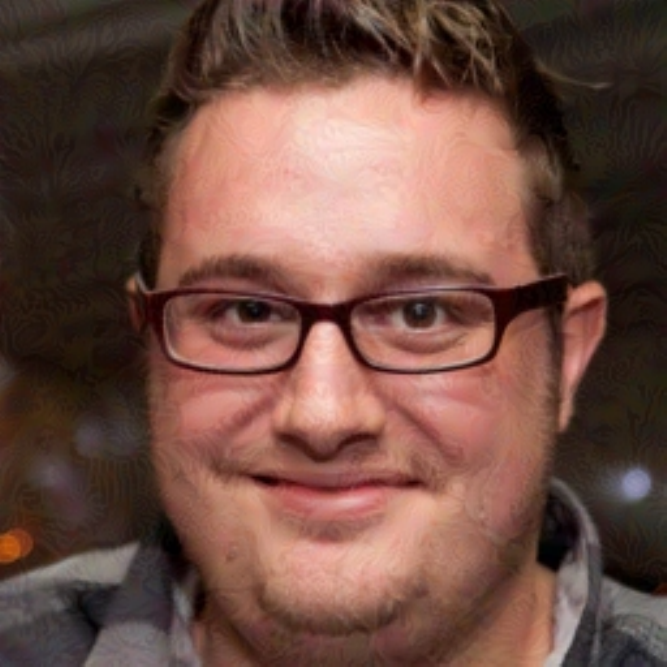}} &
            \raisebox{-.5\height}{\includegraphics[width=0.19\linewidth]{plots/replots/6.1/utility/0.05_161.pdf}}
        \\
         \bottomrule
    \end{tabular}
    }
    \label{fig:quali-utility-real}
\end{table*}

\begin{figure*}[t]
    \centering
    \subfloat[Optimization-based\label{fig:opt-dcgan}]{
    \includegraphics[width=0.25\linewidth]{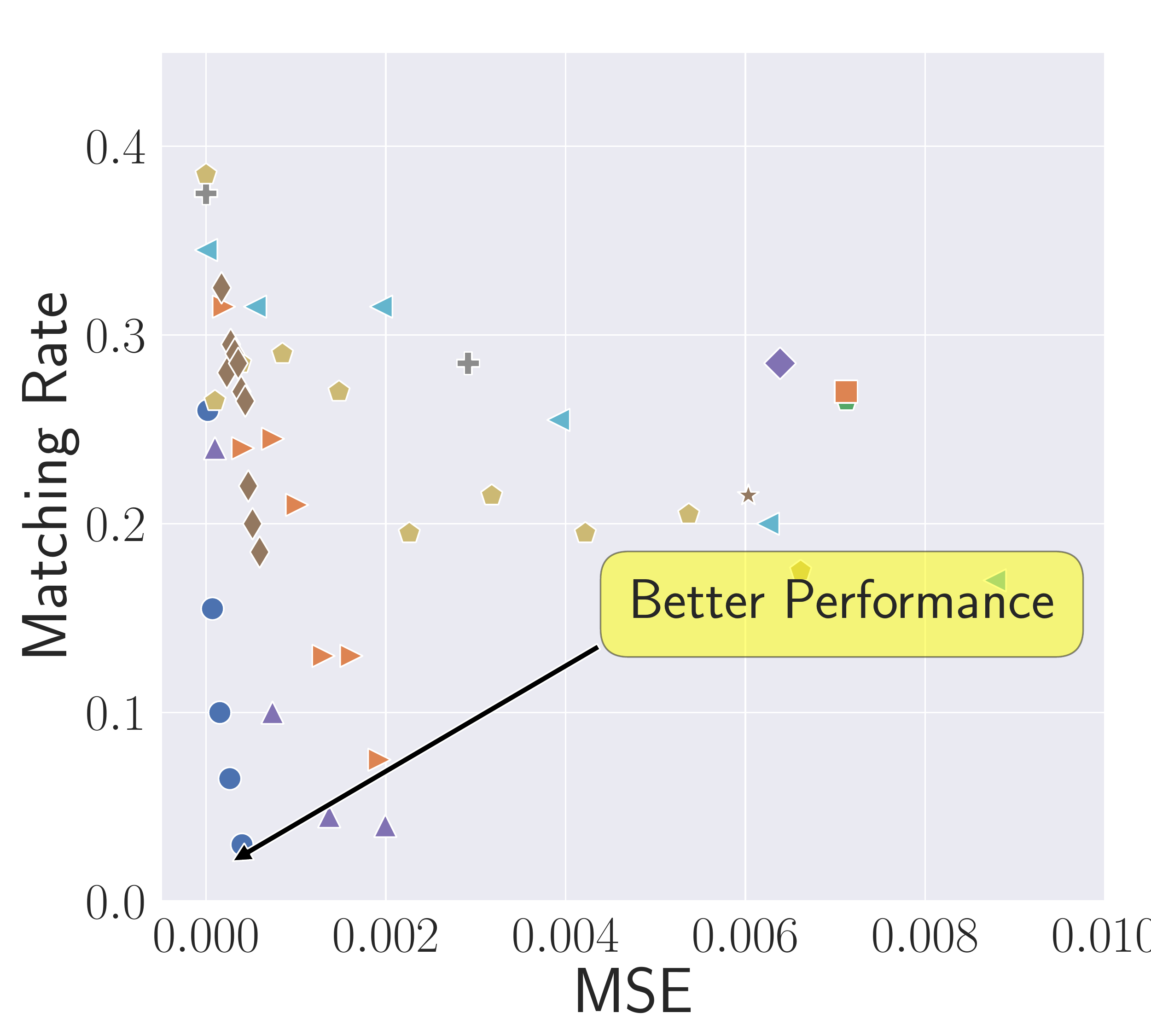}
    \includegraphics[width=0.25\linewidth]{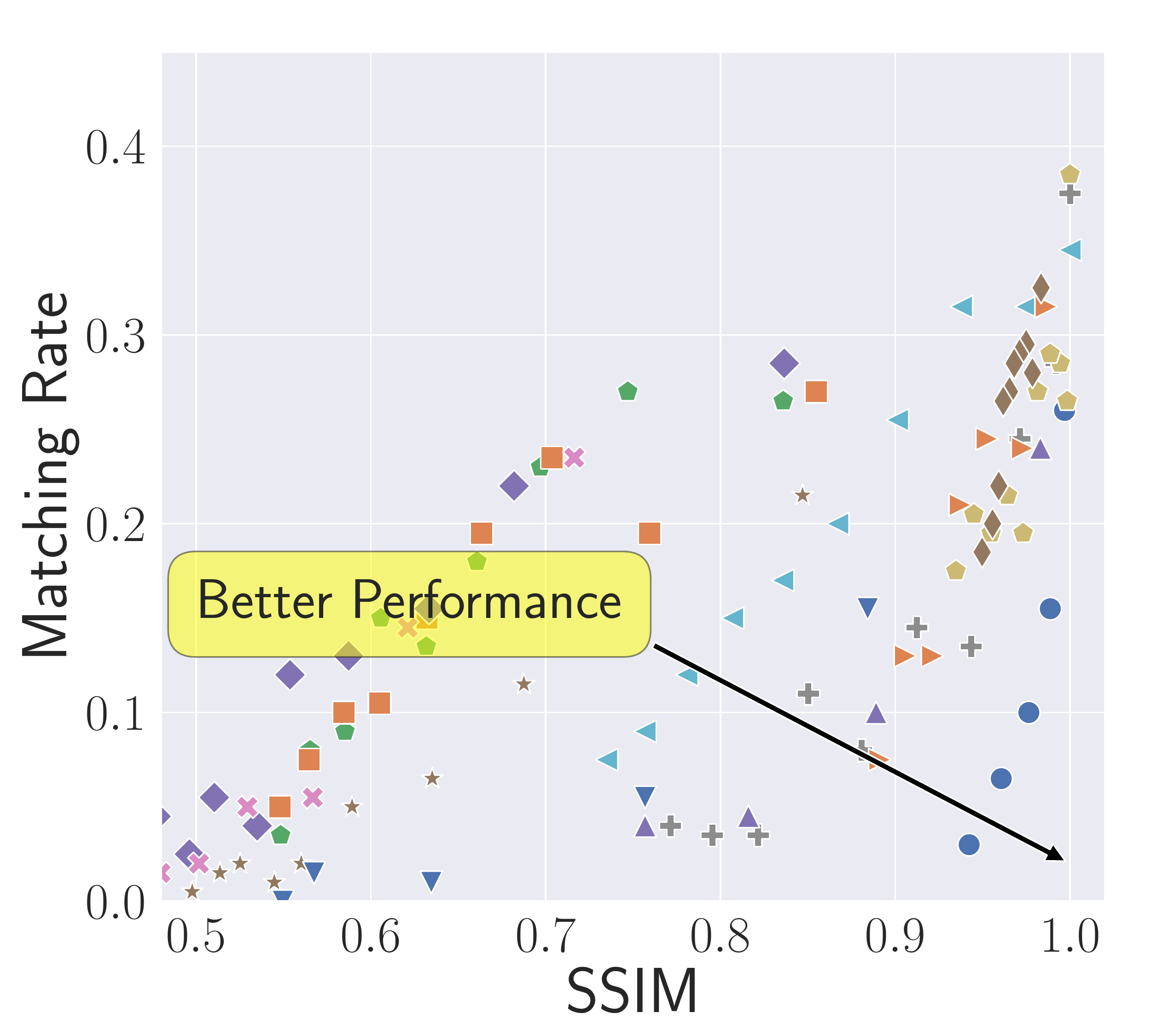}}
    \subfloat[Hybrid\label{fig:hybrid-dcgan}]{
    \includegraphics[width=0.25\linewidth]{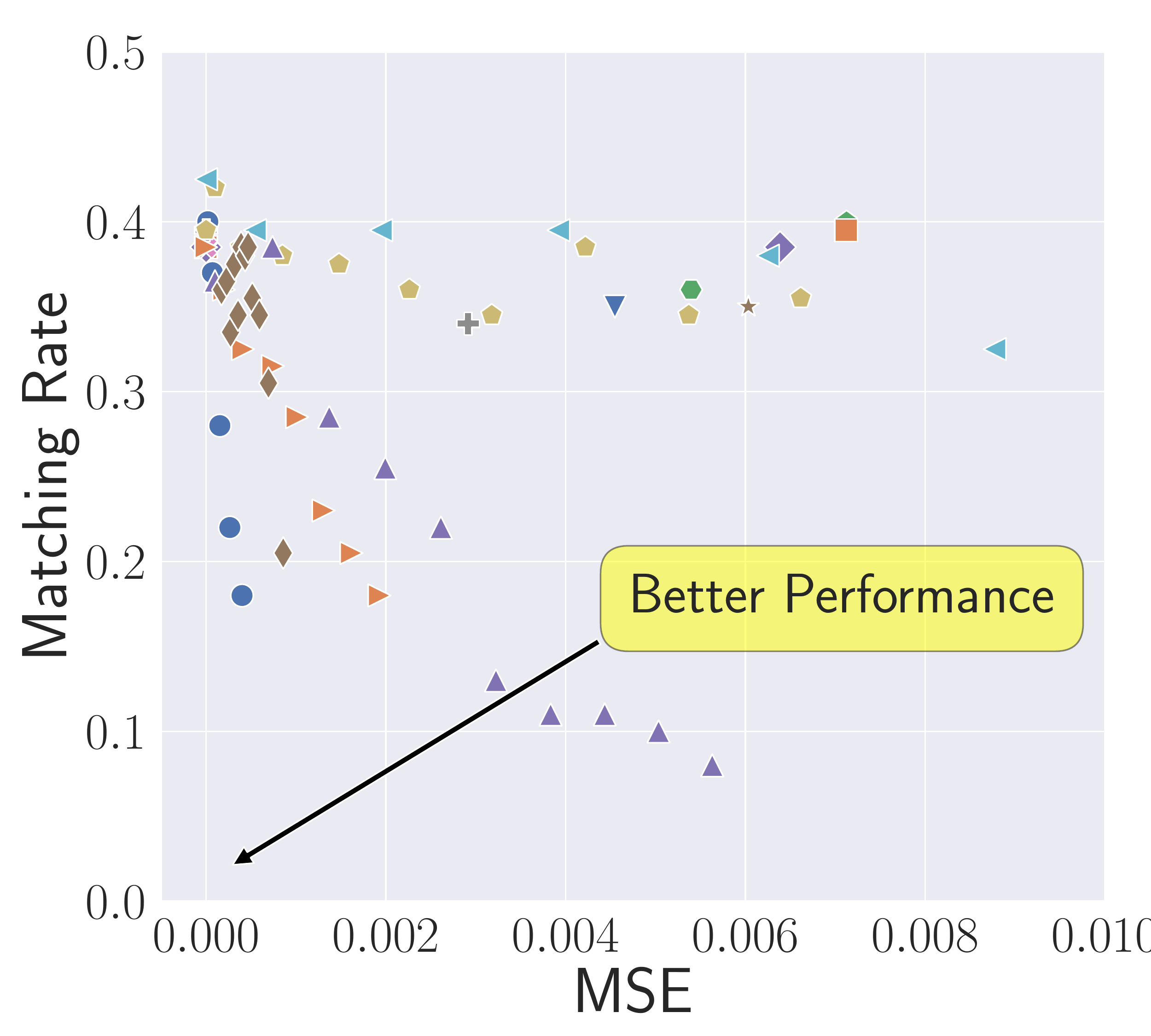}
    \includegraphics[width=0.25\linewidth]{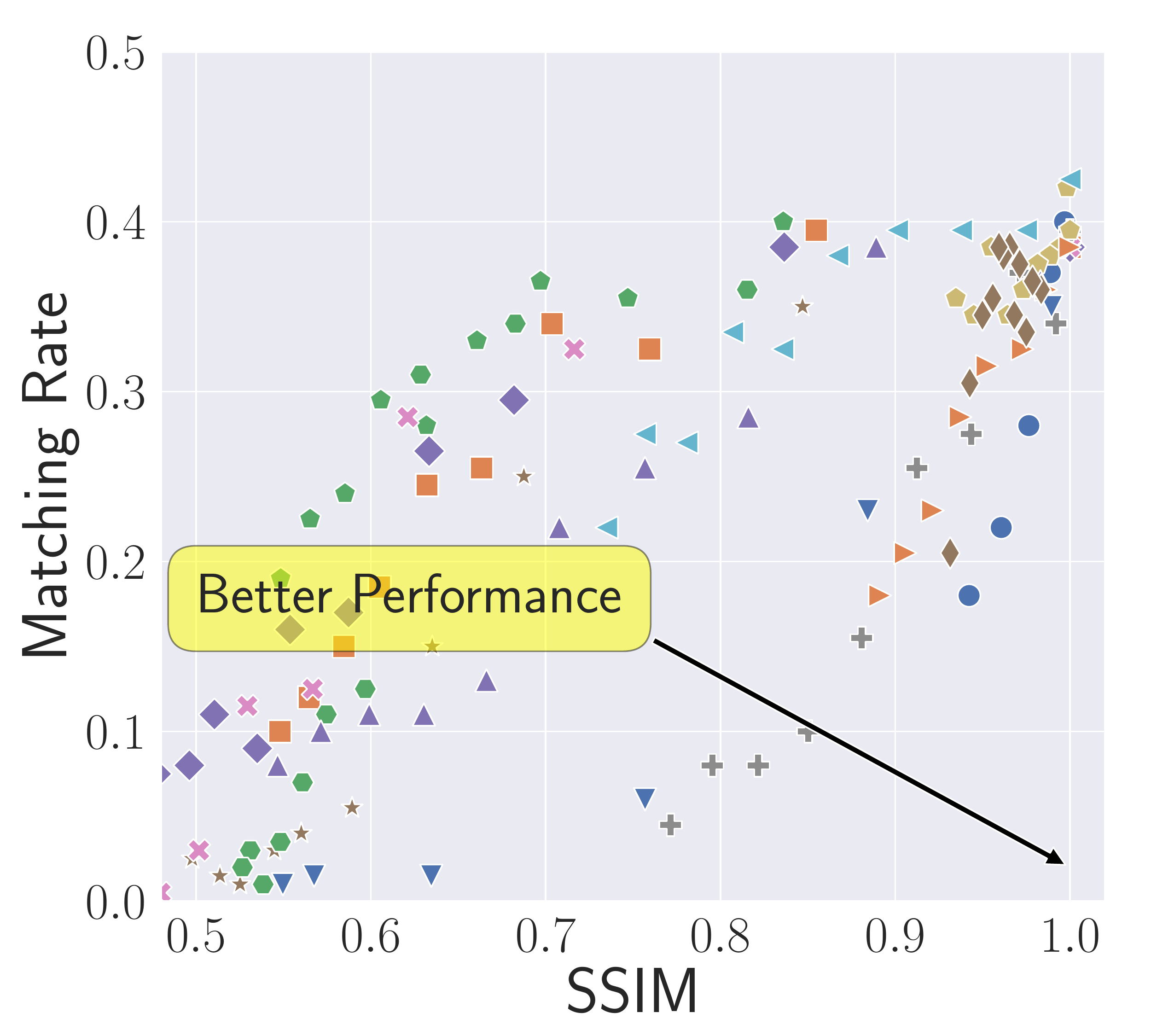}}
    \caption{Comparison between all baseline methods and Cloak v1/v4 on real images. The different points of each method represent different budgets.}
    \label{fig:real_baseline_appendix}
\end{figure*}

\end{document}